\documentclass[preprint]{imsart}
\pubyear{2021}
\volume{TBA}
\issue{TBA}
\firstpage{1}
\lastpage{1}

\usepackage{amsthm}
\usepackage{amsmath}
\usepackage{natbib}
\usepackage[colorlinks,citecolor=blue,urlcolor=blue,filecolor=blue,backref=page]{hyperref}
\usepackage{graphicx}

\startlocaldefs
\numberwithin{equation}{section}
\theoremstyle{plain}

\endlocaldefs
\usepackage[utf8]{inputenc}
\usepackage{float} 
\usepackage{amsmath,bm}
\usepackage{graphicx}
\usepackage{array,xcolor}
\usepackage[toc,page]{appendix}
\usepackage[nottoc]{tocbibind}
\usepackage{amssymb, bbm,hyperref}
\usepackage{comment}

\DeclareMathOperator{\Var}{\mathrm{Var}}

\begin{document}

\begin{frontmatter}
\title{Variance matrix priors for Dirichlet process mixture models with Gaussian kernels.}
\runtitle{Variance matrix priors}

\begin{aug}
\author{\fnms{Wei} \snm{Jing$^{1}$}\thanksref{}\ead[label=e1]{wj4@st-andrews.ac.uk}},
\author{\fnms{Michail} \snm{Papathomas$^{1}$}\thanksref{}\ead[label=e2]{M.Papathomas@st-andrews.ac.uk}}
\and
\author{\fnms{Silvia} \snm{Liverani$^{2}$}\thanksref{}%
\ead[label=e3]{s.liverani@qmul.ac.uk}%
\ead[label=u1,url]{}
}

\runauthor{W. Jing et al.}

\address[addr1]{ School of Mathematics and Statistics, University of St Andrews, UK
    \printead{e1} 
    \printead{e2}
}

\address[addr2]{School of Mathematical Sciences, Queen Mary University of London, London, UK and The Alan Turing Institute, The British Library, London, UK
    \printead{e3}%
}


\end{aug}


\begin{abstract}
The Dirichlet Process Mixture Model (DPMM) is a Bayesian non-parametric approach widely used for density estimation and clustering. In this manuscript, we study the choice of prior for the variance or precision matrix when Gaussian kernels are adopted. Typically, in the relevant literature, the assessment of mixture models is done by considering observations in a space of only a handful of dimensions. Instead, we are concerned with more realistic problems of higher dimensionality, in a space of up to 20 dimensions. We observe that the choice of prior is increasingly important as the dimensionality of the problem increases. After identifying certain undesirable properties of standard  priors in problems of higher dimensionality, we review and implement possible alternative priors. The most promising priors are identified, as well as other factors that affect the convergence of MCMC samplers. Our results show that the choice of prior is critical for deriving reliable posterior inferences. This manuscript offers a thorough overview and comparative investigation into possible priors, with detailed guidelines for their implementation. Although our work focuses on the use of the DPMM in clustering, it is also applicable to density estimation.  
\end{abstract}



\end{frontmatter}

\section{Introduction}

The Dirichlet Process Mixture Model (DPMM) is a popular non-parametric Bayesian modelling approach, widely used for density estimation and clustering. In this manuscript, we focus on clustering applications, although the results are also applicable to density estimation. The DPMM allows for model based clustering, where the mixture distribution is the likelihood for each vector observation. For continuous data, the kernel distribution in the mixture likelihood is typically set to be the multivariate Gaussian density. Under the Bayesian framework, when the variance matrix of the Gaussian kernels is unknown, a prior should be specified for it. Due to computational simplicity, the inverse Wishart distribution is often chosen. 

Based on our simulation studies, the DPMM struggles to uncover clearly distinct clusters under this choice, even when the dataset consists of just a handful of variables and the normality assumption is correct. Adding hyperpriors on the parameters of the inverse Wishart distribution does not improve the performance of the model to a satisfactory degree, as we demonstrate in Section 4. 

In this manuscript, we consider problems in a space of up to 20 dimensions. This is a higher dimensionality than usually tested in simulation studies, relevant to more realistic problems. It can be viewed as high dimensionality in particular within the context of clustering,  when unknown covariance matrices are to be estimated for a selection of clusters, some potentially of small size. 

Some standard approaches for tackling the problem of dimensionality, listed in \cite{chandra2020escaping}, involve strong assumptions on the structure of the covariance matrix. One assumption, for instance, is that variables are conditionally independent given the cluster allocation. This presumes clusters of specific shape and orientation, and the DPMM can perform badly when this assumption is violated; see also Section 4 and the Supplemental material Section S5.1.

We observe that the choice of prior for the variance matrix of the Gaussian kernel is increasingly important as the dimensionality as well as the number of target clusters increases. Using simulated and real data, we demonstrate that a sparsity inducing prior is the most promising one, in terms of effecting MCMC convergence and identifying true clusters. 

When the DPMM model is adopted, the number of components in the mixture is set to be infinite. The likelihood for each data point $X_{i}$ in $\mathbb{R}^{J}$ is,  
\begin{equation} \label{eq:1}
X_{i}|\bm{\psi},\bm{\Theta} \sim \sum_{c=1}^{\infty} \psi_{c} f(X_{i}|\bm{\Theta}_{c}),
\end{equation} 
where $X_{i}=[X_{i,1}, X_{i,1},...,X_{i,J}]^{T}$ is the data vector for subject $i$, after observing $J$ variables. Here,  $f(X_{i}|\bm{\Theta}_{c})$ represents a family of distributions with cluster specific parameters $\bm{\Theta}_{c}$ for component $c$. Also, $\psi_{c}$ denotes the latent probability with which $X_{i}$ belongs to component $c$, with $\sum_{c=1}^{\infty} \psi_{c}=1$. To simplify the likelihood and subsequent computations, an auxiliary vector $\bm{Z}=[Z_{1}, Z_{2},..., Z_{n}]$ is often introduced. Conditionally on $\bm{Z}$, the infinite mixture likelihood becomes,
\begin{equation} \label{eq:2}
X_{i}|Z_{i}, \boldsymbol{\Theta} \sim
f(X_{i}|\boldsymbol{\Theta}_{Z_{i}}).
\end{equation} Conditionally on $\bm{\psi}$, the probability mass function of $Z_{i}$ is,
\begin{equation*}
p(Z_{i}=c|\boldsymbol{\psi})=\psi_{c}.
\end{equation*}

The core part of the DPMM is the Dirichlet process (DP). Denote by $\boldsymbol{\tilde{\Theta}}=(\boldsymbol{\tilde{\Theta}}_1,...,\boldsymbol{\tilde{\Theta}}_{n})$, the vector of cluster specific parameters for subjects $i=1,...,n$. Elements of this vector can be identical, so that if subjects $i$ and $i^{'}$ belong to cluster $c$, then $\boldsymbol{\tilde{\Theta}}_{i}=\boldsymbol{\tilde{\Theta}}_{i^{'}}=\boldsymbol{\Theta}_{c}$. 
The DP can be represented with the stick-breaking construction \cite{sethuraman1994constructive} given below. \begin{equation} \label{eq:3}
\begin{aligned}
& G=\sum_{c=1}^{\infty} \psi_{c}\delta_{\boldsymbol{\Theta}_{c}} \\
& \psi_{c}=V_{c} \prod_{l<c} (1-V_{l}), \ \ \psi_{1}=V_{1} \\
& V_{c} | \alpha \sim Beta(1, \alpha) \\
& \boldsymbol{\Theta}_{c} | G_{0} \sim G_{0},
\end{aligned}
\end{equation} 
where $G_{0}$ denotes the base distribution defined on the parameter space of $\bm{\Theta}_{c}$ and $\alpha$ is the concentration parameter.
As the realization of the DP is almost surely discrete \citep{blackwell1973discreteness}, the DPMM is introduced so that the distribution of the $X_i$ can be absolutely continuous \citep{lo1984class}, namely 
\begin{equation}
\begin{aligned} \label{eq:4}
& X_{i}|\boldsymbol{\tilde{\Theta}}_{i} \sim f(X_{i}|\boldsymbol{\tilde{\Theta}}_{i}) \\
& \boldsymbol{\tilde{\Theta}}_{i}|G \sim G \\
& G|G_{0}, \alpha \sim DP(G_{0},\alpha),
\end{aligned}
\end{equation}
where the $\bm{\tilde{\Theta}}$ parameters are drawn using (\ref{eq:3}). 

Based on (\ref{eq:2}), (\ref{eq:3}) and (\ref{eq:4}), the joint distribution of the data and parameters becomes
\begin{equation} 
\begin{aligned} \label{eq:5}
p(X_{i}, \boldsymbol{\Theta},\boldsymbol{\psi}|\alpha, G_{0}, Z_{i})
& =  f(X_{i}|\boldsymbol{\Theta}_{Z_{i}}) 
p(\boldsymbol{\Theta})p(\boldsymbol{\psi}) \\
& = f(X_{i}|\boldsymbol{\Theta}_{Z_{i}}) 
\Big\{ \prod_{c=1}^{\infty} p(\boldsymbol{\Theta}_{c})p(\psi_{c}) \Big\}.
\end{aligned}
\end{equation}

For datasets that contain only continuous variables, the typical choice for the component distribution is the multivariate normal density, namely,
\begin{equation} \label{eq:6}
f(X_{i}|\boldsymbol{\Theta}_{Z_{i}})=
(2\pi)^{-\frac{J}{2}}
|\Sigma_{Z_{i}}|^{-\frac{1}{2}}\exp\bigg\{-\frac{1}{2}(X_{i}-\mu_{Z_{i}})^{T} \Sigma_{Z_{i}}^{-1}(X_{i}-\mu_{Z_{i}})\bigg\},
\end{equation} 
where $\mu_{Z_{i}}$ and $\Sigma_{Z_{i}}$ are the mean vector and the variance-covariance matrix for cluster $Z_{i}$ respectively. In this case, $\boldsymbol{\Theta}_{Z_{i}}=(\mu_{Z_{i}},\Sigma_{Z_{i}})$. 
One commonly used setting for the base distribution $G_{0}$ is,  
\begin{equation} \label{eq:7}
G_{0} \equiv \emph{N}_{J}(\mu_{c};\mu_{0},\Sigma_{0}) \times InvWishart_{J}(\Sigma_{c};R_{0},\kappa_{0}),
\end{equation} 
where $\emph{N}_{J}$ denotes the multivariate normal distribution of dimension $J$ (with mean $\mu_{0}$ and covariance matrix $\Sigma_{0}$); $InvWishart_{J}$ represents the Inverse Wishart distribution (IW) of dimension $J$ (with mean matrix $R_{0}$ and scale parameter $\kappa_{0}$). Note that $\mu_{c}$ and $\Sigma_{c}$ are independent a priori. 

Despite being commonly employed as the prior for a variance-covariance matrix, the Inverse Wishart distribution has received a great amount of criticism in the literature. \cite{gelman2006prior} stated that the IW allocates little mass for variances in the region near zero. \cite{o1994bayesian} pointed out that the IW uses only one degree of freedom ($\kappa_{0}$) to model the variability and dependence between all the parameters in the matrix. Consequently, a great amount of dependency among the entries of correlations and variances is introduced. More specifically, the larger the variances, the larger the correlations in absolute value \citep{tokuda2011visualizing}. Lastly, the IW is sensitive to the choice of the hyperparameter values \citep{hennig2015handbook}.

We illustrate the problem of using the IW in the context of the DPMM with two simulation examples, with specifications given in Table \ref{tab:sim1} (well-separated clusters, with small cluster-specific variances within $\Sigma_{c}^{true}$, and large sample size in relation to the number of model parameters) and Table \ref{tab:sim2} (clusters are closer, with larger cluster-specific variances and small sample size). Five clusters are simulated from the multivariate normal distribution. In the Supplementary material, Section S7, we include reduced dimensionality plots of the generated true clusters for one dataset for each simulation scenario. 

\begin{table}[t!] 
	\centering
	\caption{Simulated data I with $J=6$. The  $\mbox{Var}_{c}^{true}$ and $\rho_{c}^{true}$ columns show the identical diagonal elements, and identical correlations in $\Sigma_{c}^{true}$ respectively.}
	\begin{tabular}{  m{2.0cm} m{4.2cm} m{1.0cm} m{1.0cm} m{1cm} } 
		\hline
		Cluster No. & $\mu_{c}^{true}$ & $\mbox{Var}_{c}^{true}$ & $\rho_{c}^{true}$ & $n_{c}$ \\ \hline
		1 & (5, 35, 75, 5, 5, 5) & 1 & 0 & 100 \\ 
		2 & (35, 5, 5, 5, 5, 5) & 1 & 0 & 100 \\ 
		3 & (5, 75, 5, 5, 5, 35) & 1 & 0 & 100 \\ 
		4 & (5, 5, 35, 5, 5, 75) & 1 & 0 & 100 \\ 
		5 & (35, 75, 35, 5, 5, 5) & 1 & 0 & 100 \\ \hline
		
	\end{tabular}
	\label{tab:sim1}
\end{table} 

\begin{table}[t] 
	\centering
	\caption{Simulated data II with $J=20$. $rep(a,b)$ is a vector of length $b$ with repeated elements $a$. The  $\mbox{Var}_{c}^{true}$ and $\rho_{c}^{true}$ columns show the identical diagonal elements, and identical correlations in $\Sigma_{c}^{true}$ respectively.}
	\begin{tabular}{  m{1.8cm} m{5.2cm} m{1.0cm} m{1.0cm} m{0.8cm} } 
		\hline
		Cluster No. & $\mu_{c}^{true}$ &  $\mbox{Var}_{c}^{true}$ & $\rho_{c}^{true}$ & $n_{c}$ \\ \hline
		1 & (rep(3,5),rep(32,5),rep(35,5),rep(72,5)) & 5& 0.2 & 200 \\ 
		2 & (rep(32,5),rep(5,5),rep(5,5),rep(35,5)) & 5& 0.5 &200 \\ 
		3 & (rep(25,5),rep(15,5),rep(32,5)rep(3,5)) & 5&0.3 & 200 \\ 
		4 & (rep(15,5),rep(75,5),rep(8,5),rep(75,5)) & 5& 0.1&200 \\ 
		5 & (rep(8,5),rep(6,5),rep(25,5),rep(5,5)) & 5& 0.7 & 200 \\ \hline
		
	\end{tabular}
	\label{tab:sim2}
\end{table} 

\begin{figure}[h!] 	
	\centering
	\includegraphics[width=5cm, height=5cm]{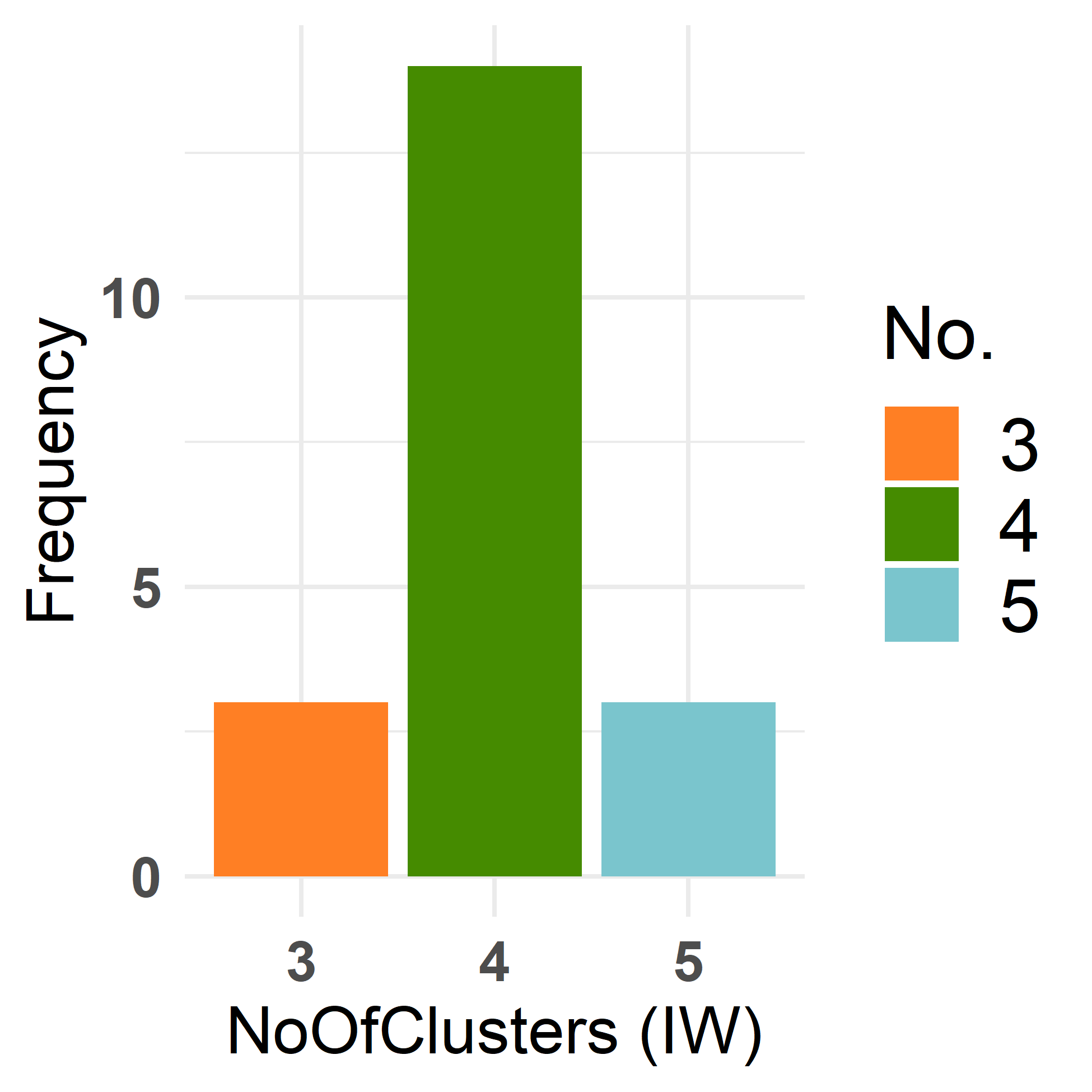} 
	\includegraphics[width=5cm, height=5cm]{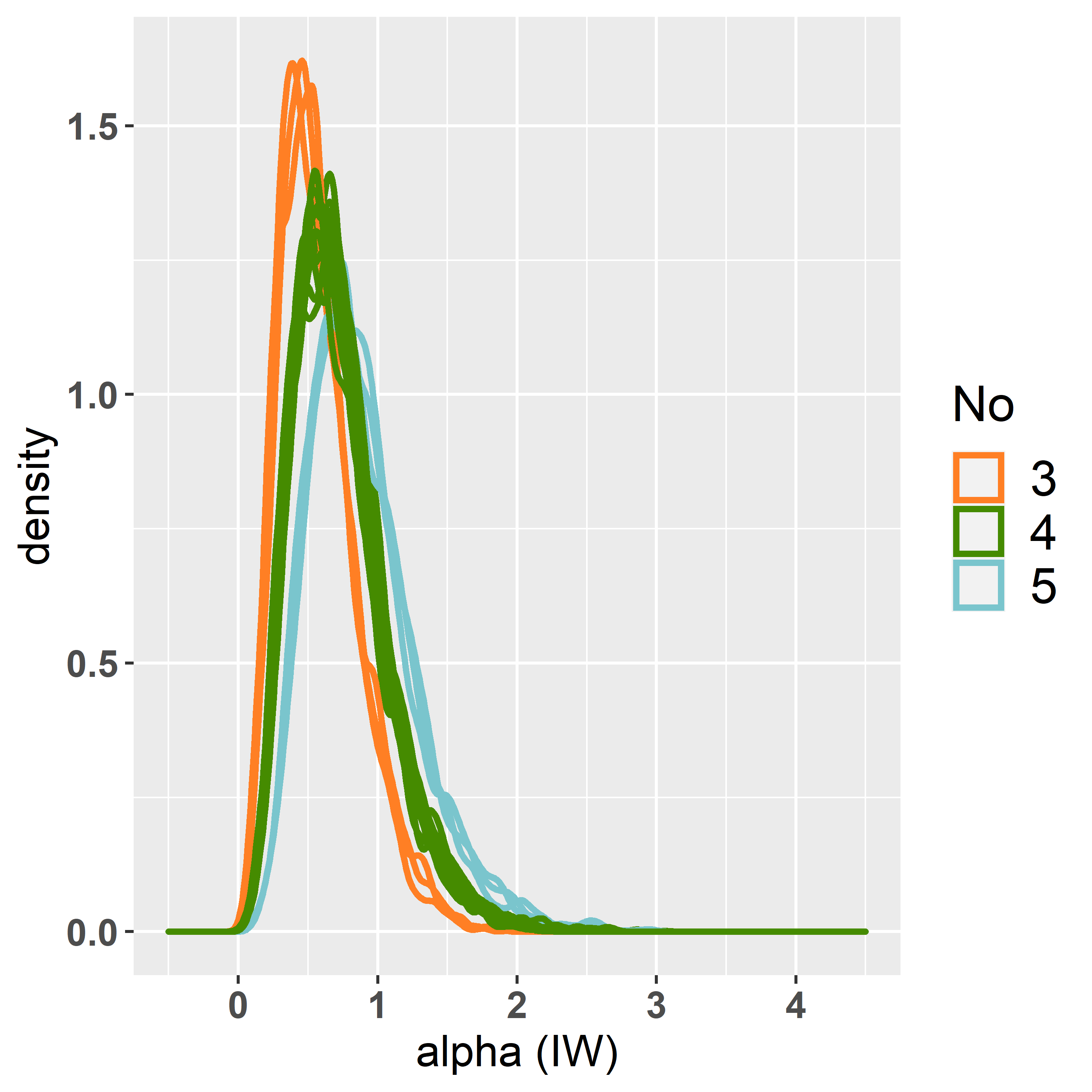} 
	\begin{minipage}{0.8\textwidth}
		\caption{Bar plot of the estimated number of clusters (left) and the density of $\alpha$ (right) for simulated data I. The orange, green and blue colours correspond to 3, 4 or 5 estimated clusters respectively. }
		\label{fig:1}
	\end{minipage}
\end{figure}

\begin{figure}[h!]
	\centering
	\includegraphics[width=5cm, height=5cm]{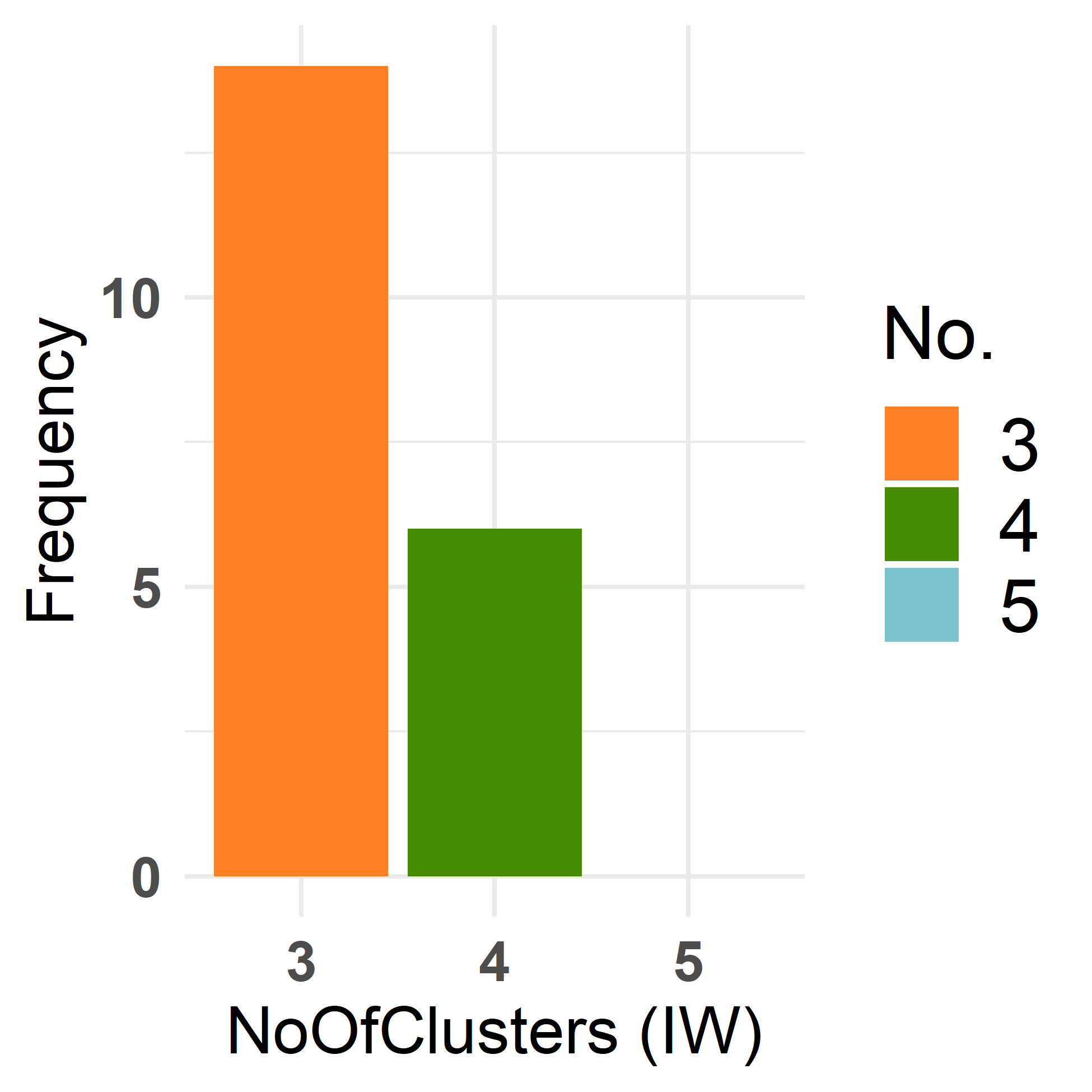} 
	\includegraphics[width=5cm, height=5cm]{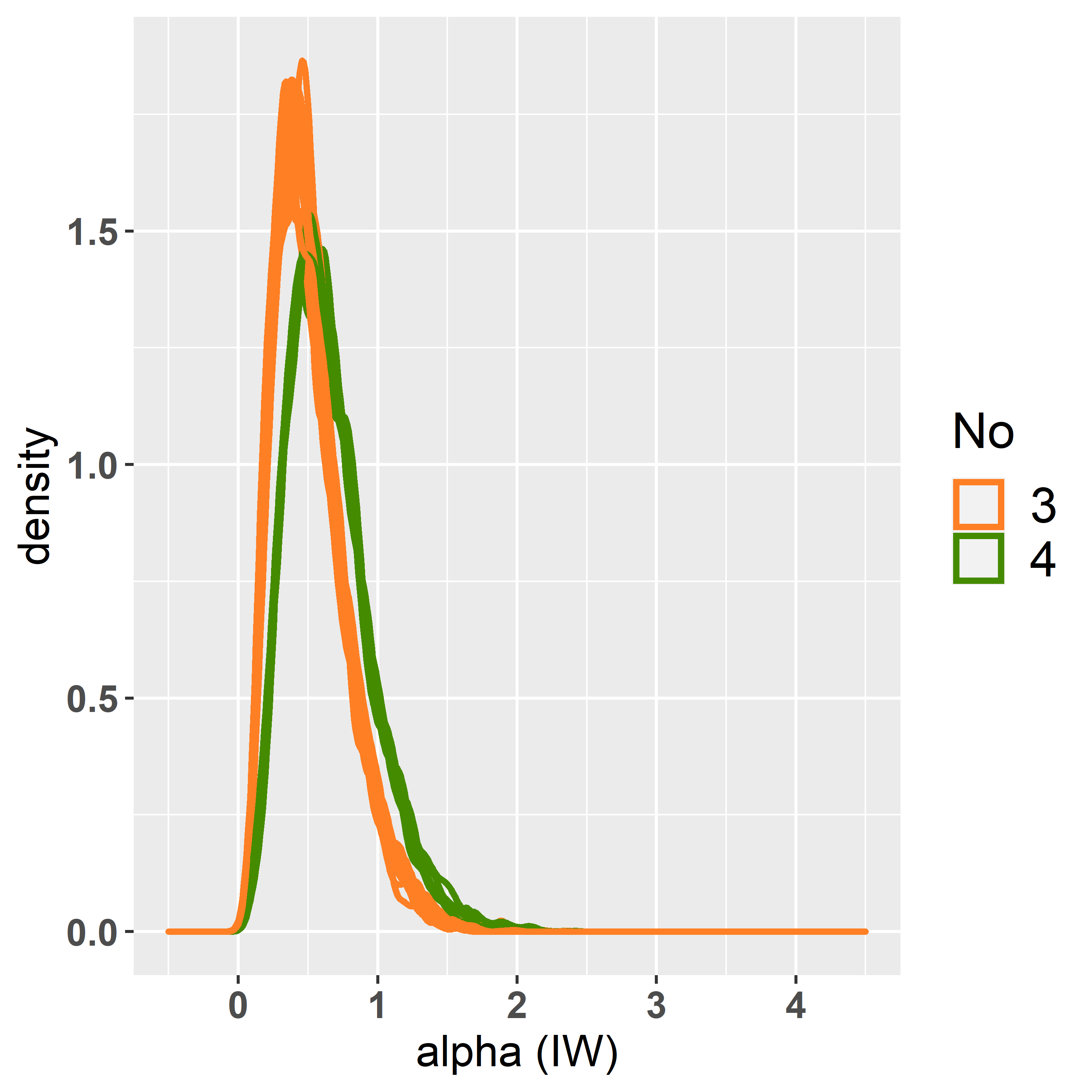} 
	\begin{minipage}{0.8\textwidth}
		\caption{Bar plot of the estimated number of clusters (left) and the density of $\alpha$ (right) for simulated data II. The orange  and green colours corresponds to 3 or 4 final clusters respectively in both plots. }
		\label{fig:2}
	\end{minipage}
\end{figure}

We generate $20$ datasets for each scenario, and adopt the IW prior using the R package PReMiuM \citep{liverani2015premium}. We choose a burn-in period of $10,000$ iterations followed by $6,000$ iterations. Posterior samples for the allocation vector $\bm{Z}$ show that the induced partition becomes stable after this burn-in period. Under both scenarios true clusters are merged, with notably different clustering results for datasets with the same underlying structure. Figure \ref{fig:1} (for simulated data I) shows the bar plot of the estimated number of clusters for the 20 datasets, generated from  post-processing the MCMC output \citep{molitor2010bayesian} for each dataset. It also shows posterior density plots of $\alpha$ for each of the 20 datasets. Figure \ref{fig:2} shows corresponding results for simulation data II. Note that similar results are also obtained by the R package DPackage \citep{jara2011dppackage}.

When true clusters are combined, posterior variances within some $\Sigma_{c}$ are larger than the truth. This can be explored in more detail when the IW is utilised; see the Supplemental material, Section S1 for details. One way to tackle the problem is to set $R_{0}$ in (\ref{eq:7}) to be small, for example, the identity matrix. Although this is effective for Simulation I, it is not for Simulation II or real datasets we have analysed, as the IW is sensitive to the specification of the hyperparameter values. \cite{fruhwirth2018handbook} proposed a strategy for choosing the value of $R_{0}$ in the context of sparse finite mixture model \citep{malsiner2016model}. The setting worked well most times, with $3$ true clusters and diagonal within-cluster covariance matrices. However, the authors also mentioned that when the number of subjects is small, true clusters can be combined. As in (\ref{eq:7}) $\kappa_{0}$ and $R_{0}$ are fixed, one can instead choose to place hyperpriors on those hyperparameters. This can improve the performance for some analyses, but does not provide with a satisfactory solution in general. This is discussed in detail in Section 4. 

Finally, the MCMC sampler initialisation plays a role in the performance of the DPMM, since typically there is a discrepancy between the overall covariance matrix and the within-cluster covariance matrices; see Supplemental material, Section S2. A more detailed discussion about initialization is provided in Section 5. 


The challenges to the DPMM caused by the estimation of $\Sigma_{c}$ are not obvious when one simulates or analyses only a small number of continuous variables (say $J<4$) and the number of target clusters are only, say, $2$ or $3$, as is usually the case for simulation studies that assess DPMM models. As $J$ and the number of target clusters increases, the problem becomes more pronounced since the number of parameters in each $\Sigma_{c}$ increases quadratically. 

In the next section, we review alternative prior specifications for the variance matrix and discuss their implementation. In Section 3, we discuss the prior specification of the mean vector in the Gaussian kernel for the DPMM. In Section 4, we compare the performance of different priors with simulation examples. In Section 5, we consider specifications that influence the performance of the DPMM not discussed before. Section 6 shows a real data comparative analysis. Section 7 summarises our findings on the best performing prior specification, and provides future research directions.

\section{Alternative priors to the Inverse Wishart}

In this Section, we introduce a variety of priors we incorporated into the DPMM framework, with details on their implementation (e.g. full conditional distributions) provided in the Supplemental material, Section S5. This was done within the R package PReMiuM \citep{liverani2015premium} which  accommodates continuous and categorical observations; see also \cite{molitor2010bayesian}. It employs the conditional samplers based on the stick-breaking construction in (\ref{eq:3}). In particular, we modified a version of the slice sampler by \cite{walker2007sampling}, used as the default sampler in PReMiuM. For details of this default sampler see the Supplemental material Section S3 and \cite{liverani2015premium}. 

\subsection{The hierarchical Inverse Wishart prior (HIW)}
This is an extension to the IW, with hyperpriors on $R_{0}$ and $\kappa_{0}$ so that they can be estimated from the data. The simplest hyperprior for $R_{0}$ is the Wishart prior, namely,
\begin{equation}
\begin{aligned} \label{eq:13}
&\Sigma | R_{0}, \kappa_{0} \sim InvWishart_{J}(R_{0},\kappa_{0}), \\
&R_{0}|R_{1}^{-1}, \kappa_{1} \sim Wishart_{J} (R_{0}; R_{1}^{-1}, \kappa_{1} ).
\end{aligned}
\end{equation}
The formulation in (\ref{eq:13}) has already been applied in the DPMM literature; see \cite{gorur2010dirichlet} for instance. It is computationally fast because no Metropolis-Hastings steps are involved and $R_{0}$ is updated all at once. Within the context of the DPMM, the prior of $\Sigma_{c}$ in (\ref{eq:13}) can be written as
\begin{equation}
\begin{aligned} \label{eq:25}
&\Sigma_{c}|R_{0},\kappa_{0} \sim InvWishart_{J}(\Sigma_{c}; R_{0},\kappa_{0}), \\
&R_{0}|R_{1}^{-1}, \kappa_{1} \sim Wishart_{J} (R_{0}; R_{1}^{-1}, \kappa_{1} ),\\
&\kappa_{0}-J |\alpha_{\kappa_{0}}, \beta_{\kappa_{0}} \sim InvGamma(\kappa_{0}-J; \alpha_{\kappa_{0}}, \beta_{\kappa_{0}}),
\end{aligned}
\end{equation}
where $InvGamma$ denotes the Inverse Gamma distribution. Note that $\kappa_{0} > J$. We refer to the prior in (\ref{eq:25}) as HIW1. 
Hyperparameter values are set to be,
\begin{align*}
&R_{1}=I_{J}, \  \ \alpha_{\kappa_{0}}=1/2, \ \ \beta_{\kappa_{0}}=\frac{J}{2}, \ \ \kappa_{1}=J+2,
\end{align*}
which results in relatively small mean within-cluster variances a priori. 
\cite{huang2013simple} suggested another conditional conjugate prior for $\Sigma$ as follows,
\begin{equation}
\begin{aligned} \label{eq:14}
\Sigma_c|\epsilon_{0}, \delta_{1},...,\delta_{J} \sim & InvWishart_{J} (\Sigma_{c}; 2\epsilon_{0}diag( 1/\delta_{1},...,1/\delta_{J}); \epsilon_{0}+J-1 ), \\
\delta_{j} | g_{j} \sim & InvGamma(\delta_{j};1/2,1/g_{j}^{2}).
\end{aligned} 
\end{equation}
\cite{huang2013simple} proved that the standard deviation in the covariance matrix follows a Half-t distribution, which is recommended by Gelman [11].

For $j=1,\hdots,J$ the prior of $\Sigma_{c}$ in (\ref{eq:14}) can be written as 
\begin{equation}
\begin{aligned} \label{eq:27}
& \Sigma_{c} | \epsilon_{0}, \delta_{j, j=1,...,J }\sim InvWishart_{J}\bigg(\Sigma_{c}; 2\epsilon_{0}diag\bigg( \frac{1}{\delta_{1}},...,\frac{1}{\delta_{J}}\bigg), \epsilon_{0}+J-1 \bigg),\\
& \delta_{j} | \alpha_{\delta}, g_{j} \sim InvGamma(\delta_{j};\alpha_{\delta},g_{j}), \\
& (\epsilon_{0}-1) | \alpha_{\epsilon_{0}}, \beta_{\epsilon_{0}} \sim InvGamma(\epsilon_{0}-1; \alpha_{\epsilon_{0}}, \beta_{\epsilon_{0}}). \\
\end{aligned}
\end{equation}
We refer to the prior in (\ref{eq:27}) as HIW2. 
The hyperparameter values are set to be, 
\begin{align*}
&g_{j}=\frac{J*20.0}{range(\bm{X}_{j})*range(\bm{X}_{j})}, \  \ \alpha_{\delta}=0.2, \ \ \beta_{\epsilon_{0}}=\frac{J}{2}, \ \ \alpha_{\epsilon_{0}}=0.5,
\end{align*}
which also results in relatively small mean within-cluster variances a priori.


\subsection{The separation prior}
This construction relies on the idea of decomposing the covariance matrix to variance and correlation matrices \citep{barnard2000modeling}. Specifically, 
\begin{equation} \label{eq:15}
\Sigma=SRS,
\end{equation}
where $S=\{s_{j,j}\}$ is a diagonal matrix with the standard deviation of each variable as diagonal elements and $R=\{r_{i,j}\}$ is a correlation matrix. Prior distributions can be specified for $S$ and $R$, so that they are independent or dependent a priori. We refer to priors whose specification is based on (\ref{eq:15}) as separation priors. 

Setting a prior for $R$ that allows for efficient  sampling from its full conditional distribution is not straightforward. \cite{barnard2000modeling} suggested two priors. Firstly,
\begin{equation} \label{eq:16}
p_{R} \propto |R|^{\frac{J^{2}-J-2}{2}}(\prod_{j=1}^{J} |R_{jj} |)^{-\frac{J+1}{2}},
\end{equation}
where $|A|$ denotes the determinant of matrix $A$, and $R_{jj}$ represents the $j$th principle submatrix of $R$. The prior in (\ref{eq:16}) results in a uniform prior on each correlation element $r_{i,j}$. The second prior is a joint uniform prior on $R^{J}$, namely,
\begin{equation} \label{eq:17}
p_{R} \propto 1.
\end{equation}
Updating $R$ element by element is suggested in \cite{barnard2000modeling}. Keeping $R$ positive definite is computationally expensive as it involves solving a quadratic equation before updating each $r_{i,j}$. 
\cite{o2008domain} proposed an Inverse Wishart prior for $R$, which is not restricted to be a correlation matrix any more. The advantage of this setting is the sampling convenience gained by using the conditional conjugate Inverse Wishart distribution. However,  variances and correlations in $\Sigma_{c}$ can no longer be independent a priori. 
We adopt a similar specification to \cite{o2008domain} as shown below. 
\begin{equation}
\begin{aligned}  \label{eq:28}
&R_{c} |R_{R}, \kappa_{R} \sim InvWishart_{J}(R_{c};R_{R}^{-1}, \kappa_{R}), \\
&(\kappa_{R}-J) | \alpha_{\kappa_{R}}, \beta_{\kappa_{R}} \sim InvGamma(\kappa_{R}-J; \alpha_{\kappa_{R}}, \beta_{\kappa_{R}}), \\
&s_{c,j} | \alpha_{s}, \beta_{sj} \sim InvGamma(s_{c,j}; \alpha_{s}, \beta_{sj}), \\
&\beta_{sj}| \alpha_{0}, \beta_{0j}  \sim Gamma(\beta_{sj}; \alpha_{0}, \beta_{0j}), \\
&\Sigma_{c}=S_{c}R_{c}S_{c},
\end{aligned}
\end{equation}
where $S_{c}=\{s_{c,j}\}$ is a diagonal matrix for cluster $c$ and $R_{c}=\{r_{c, i,j}\}$ is a dense matrix for cluster $c$. 
Note that if no constraint is set on  $R_{c}$, matrices $S_{c}$ and $R_{c}$ are not identifiable. However, when the focus is on $\Sigma_{c}$, this may not be of concern. Also, notice that a hyperprior is placed on $\beta_{sj}$ which is similar to the setting in \cite{richardson1997bayesian}. Hyperparameter values are listed below. 
\begin{align*}
&R_{R}=I_{J}, \  \ \alpha_{\kappa_{R}}=1/2, \ \ \beta_{\kappa_{R}}=\frac{J}{2}, \\
&\alpha_{s}=2, \ \  \alpha_{0}=0.2, \ \ \beta_{0,j}=\frac{10}{range(X_{j})^{2}}.
\end{align*}
This results in relatively small mean within-cluster variances a priori. 
Note that under the framework in (\ref{eq:15}), another prior available for matrix $R$ is the LKJ (Lewandowski-Kurowicka-Joe) prior  \citep{lewandowski2009generating}, with Hamiltonian Markov Chain Monte Carlo (HMCMC) typically  used to sample from the posterior. We did not adopt this specification for reasons of computational efficiency. 

\subsection{The log prior}
Another prior specification for the  covariance matrix $\Sigma$ relies on the log transformation $A=\log(\Sigma)$, 
and the spectral decomposition $\Sigma=EDE^{T}$. Here, $D$ is a diagonal matrix with the eigenvalues of $\Sigma$ as diagonal elements and $E$ is an orthonormal matrix. The columns of $E$ are normalized eigenvectors, and correspond to the eigenvalues of $D$. Based on the above, 
\begin{align*}
A=E\log(D)E^{T},
\end{align*} 
where $\log(D)$ is a diagonal matrix with the logarithm of the eigenvalues of $\Sigma$ as its diagonal elements. When $A$ is sampled as a symmetric matrix, $\exp(A)=\Sigma$ is positive definite. This specification (referred to as the log prior) was first considered by \cite{leonard1992bayesian}, who represented $A$ as a vector,
\begin{align*}
\bm{a}=(a_{1,1}, a_{2,2},...,a_{J,J}; a_{1,2},...,a_{J-1,J};...;a_{1,J-1}, a_{2,J}; a_{1,J}).
\end{align*} 
A uniform prior, $\bm{a} \propto 1$, or a multivariate normal prior can be placed on $\bm{a}$. 
We adopt the multivariate normal prior as it is relatively flexible and the elements of $\log(\Sigma)$ can be sampled all at once. In the context of the DPMM, using the subscript $c$ to denote cluster labels,
\begin{equation} \label{eq:43}
\bm{a}_{c} | \mu_{\bm{a}}, \Sigma_{\bm{a}} \sim N_{q}(\bm{a}_{c}; \mu_{\bm{a}}, \Sigma_{\bm{a}}).
\end{equation}
Leonard and Hsu \cite{leonard1992bayesian} utilize the Taylor series approximation for the Normal likelihood,  
\begin{equation} \label{eq:37}
\prod_{i=1}^{n_{c}}f(X_{i}|\bm{Z},\mu_{c}, \bm{a}_{c}) \simeq
(2\pi)^{-\frac{n_{c}}{2}} e^{-Jn_{c}}
|S_{c}^{*}|^{-\frac{n_{c}}{2}}\exp\bigg\{-\frac{1}{2}(\bm{a}_{c}-\bm{\lambda}_{c})^{T} Q_{c} (\bm{a}_{c}-\bm{\lambda}_{c})\bigg\},
\end{equation}
where $\bm{\lambda}_{c}$ is the mean vector, $Q_{c}$ is the precision matrix and  $S_{c}^{*}=\frac{1}{n_{c}}\sum(X_{i}-\mu_{c})(X_{i}-\mu_{c})^{T}$. The details of how to compute $\bm{\lambda}_{c}$ and $Q_{c}$ are given in the Supplemental material, Section S5.

In this conjugate setting, the posterior for $\bm{a}_{c}$ is  approximately Normal. 
In the context of DPMM, we use Metropolis-Hastings MCMC to sample from the full conditional distribution of $\Sigma_{c}$. To better explore the sample space of $\Sigma_{c}$, we set the multivariate $t$ distribution as the proposal, which can also serve as the importance function for the importance sampling algorithm in \cite{leonard1992bayesian}. 

Because the multivariate normal prior for $\bm{a}_{c}$ is already quite flexible, we assigned fixed values to the hyperparameters $\mu_{\bm{a}}$ and $\Sigma_{\bm{a}}$. We set $\mu_{\bm{a}}=(-1,\hdots ,-1,\\ 0,\hdots ,0)$, where the first $J$ elements are non-zero. The specification of $-1$ makes the variances in $\Sigma_{c}$ relatively small. We also set $\Sigma_{\bm{a}}=diag(3,\hdots ,3,1,\hdots ,1)$, with the first $J$ elements in the diagonal equal to 3. 

The advantage of using the log transformation is that it is relatively easy to implement and the sampling method allows to update the vector $\bm{a}_{c}$ all at once. However, whether the sampler for $\bm{a}_{c}$ can mix well depends on the quality of the approximation in (\ref{eq:37}). In addition, the calculation of matrix $Q_{c}$ is non-trivial especially for large $J$.

\subsection{The sparse prior}
For the estimation of a sparse precision or covariance matrix, the graphical and adaptive graphical lasso within the frequentist framework [ \cite{yuan2007model, friedman2008sparse, fan2009network}] are equivalent to placing the Laplace prior on the off-diagonal elements and the exponential prior on the diagonal elements of the precision matrix [\cite{wang2012bayesian, khondker2013bayesian}]. Specifically,
\begin{equation}
\begin{aligned} \label{eq:24}
\boldsymbol{X}_{i}|T & \sim N_{J}(\boldsymbol{0}, T^{-1}), \\
p(T|M_{0}) & =C_{0}^{-1}\prod_{i<j}
\Big\{Laplace(T_{ij}|0,1/m_{0,ij})\Big\} \prod_{i}\Big\{Exp(T_{ii}|m_{0,ii}/2)\Big\}, \  T \in P^{+},
\end{aligned}
\end{equation}
where $T$ denotes the precision matrix and $P^{+}$ is the set of all positive definite matrices. $C_{0}^{-1}$ is a normalizing constant. The hyperparameters $m_{0,ij}$ control the amount of shrinkage of $T_{ij}$ towards zero and can be arranged into matrix $M_{0}=\{m_{0,ij}\}$. 
MCMC algorithms for sampling from the resulting posterior are given by \cite{wang2012bayesian} and \cite{khondker2013bayesian}. 

The construction in (\ref{eq:24}) can not result in exact zero posterior estimates of $T_{ij}$. To avoid a thresholding approach, a spike-and-slab [\cite{mitchell1988bayesian, george1993variable, george1997approaches}] type prior puts a point mass at $T_{ij}=0$ and a continuous distribution when $T_{ij} \neq 0$, as in \cite{banerjee2013bayesian}. However, posterior computation becomes cumbersome and \cite{banerjee2013bayesian} utilized the Laplace approximation. For this reason, we focus on the prior in (\ref{eq:24}). We implement the sampler proposed by \cite{wang2012bayesian} as it does not require Metropolis-Hastings steps. 

The Laplace distribution can be represented as a scale mixture of Normals. This suggests that the prior for $T$ can be written as,
\begin{equation}  \label{eq:47}
p(T|M_{0}, M_{1})=C_{1}^{-1}\prod_{i<j}
\Big\{N(T_{ij}|0,m_{1,ij}) \Big\} \prod_{i}\Big\{Exp(T_{ii}|m_{0,ii}/2)\Big\}, \ T \in P^{+},
\end{equation}
where $m_{1,ij}$ are auxiliary variables that can be arranged into a matrix $M_{1}=\{m_{1,ij}\}$ with diagonal elements zero. $C_{1}^{-1}$ is another normalizing constant. \cite{wang2012bayesian} constructed the prior of $M_{1}$ as, 
\begin{equation} \label{eq:n-sp-1}
P(M_{1}|M_{0}) \propto C_{1} \prod_{i<j} Exp(m_{1,ij}|m_{0,ij}^{2}/2).
\end{equation}
After integrating out $m_{1,ij}$, $T$ follows the Laplace distribution given in (\ref{eq:24}). Therefore, the specification of the sparse prior in the context of the DPMM can be provided as,
\begin{equation}
\begin{aligned} \label{eq:49}
p(T_{c}|M_{0}, M_{1,c}) & =C_{1,c}^{-1}\prod_{i<j}
\Big\{N(T_{c,ij}|0,m_{1,c,ij}) \Big\} \prod_{i}\Big\{Exp(T_{c,ii}|m_{0,ii}/2)\Big\}, \ T_{c} \in P^{+} , \\
P(M_{1,c}|M_{0}) & \propto C_{1,c} \prod_{i<j} Exp(m_{1,c,ij}|m_{0,ij}^{2}/2).
\end{aligned}
\end{equation}
Here, $M_{1,c}$ is the cluster specific matrix that corresponds to $M_1$. 
The prior contains the intractable normalizing constant $C_{1,c}$ $\{c \in P \}$, but this cancels out when sampling from the joint prior, 
\begin{align*}
& P(T_{c}, M_{1,c} | M_{0}) \\
\propto &
\prod_{i<j}
\Big\{N(T_{c,ij}|0,m_{1,c,ij}) Exp(m_{1,c,ij}|m_{0,ij}^{2}/2) \Big\} \prod_{i}\Big\{Exp(T_{c,ii}|m_{0,ii}/2)\Big\}, \ T_{c} \in P^{+}.
\end{align*}

It is required to sample from this prior when sampling for the parameters of empty clusters. (For non-empty clusters, in the full conditional distribution of $M_{1,c}$, the normalizing constant $C_{1,c}$ also cancels out.) 
To increase the efficiency of the MCMC sampler, we set what we argue are preferred initial values for $T_{c}$ and $M_{1,c}$. Additional details are provided in the Supplementary material, Section S5.6.

Figure \ref{fig:4} shows empirical prior distributions for the correlations, and empirical bivariate density plots for the variances for different $M_{0}$. These and other empirical results not shown here indicate that smaller values of diagonal elements $m_{0,ii}$ correspond to smaller variances in $\Sigma_{c}$ and a prior density for the correlations that is more concentrated around zero. This is also the effect of larger off-diagonal values of $m_{0,ij}$. 


Within the context of the DPMM, we observed that values of $M_{0}$ that imply correlations very close to zero 
are restrictive and clustering results are unsatisfactory. $M_{0}$ values that lead to a less spiky density for the correlations result in much better performance. Note that the prior distribution of the correlations should not be too flat. It should still imply a higher probability of generating small correlations, but should allow for larger correlations in absolute value in the tail. The prior range of the within-cluster variances is also important. When variances are too small a priori, the final clustering can contain many small clusters; whereas, when variances are too large, the final clustering can combine true clusters. We suggest tuning the values in $M_{0}$ as follows: plot the dataset using some dimension reduction technique (as discussed in Section 4 and the supplementary material, Section S7) and calculate the sample variances from the the whole dataset. If there is no clear separation of groups of data points (which is normally the case for real datasets), then set the value of $M_{0}$ such that the expected within-cluster variances are concentrated just below the largest overall variances. If there is clear separation of data clusters, set the value of $M_{0}$ so that the within-cluster expected variances are concentrated at approximately half of the largest overall variance. In both cases, set the value of $M_{0}$ such that the prior distributions for the correlations are relatively uninformative. For this, we have found that a reasonable setting can be $\frac{m_{0,ij}}{m_{0,ii}}=3$. 
For the simulated and real datasets analysed in this manuscript, we set 
$m_{0,ii}=10$ and $ m_{0,ij}=30$, for $i \neq j$.

In the setting given in (\ref{eq:24}), all of the elements in $M_{0}$ are fixed. In the context of the DPMM, one can also place a hyperprior on $M_{0}$. The difficulty in doing so lies in the derivation of the posterior of $M_{0}$. \cite{wang2012bayesian} proves that only when $m_{0,ij}$ and $m_{0,ii}$ are all equal, $C_{0}^{-1}$ is not a function of $M_{0}$. Therefore, to set hyperpriors for $M_{0}$, we let $M_{0}$ be component specific, namely

\begin{equation} \label{eq:n_2}
\begin{aligned}
&p(T_{c}|M_{0, c}) =C_{0,c}^{-1}\prod_{i<j}
\Big\{Laplace(T_{c,ij}|0,1/m_{0,c,ij})\Big\} \prod_{i}\Big\{Exp(T_{c,ii}|m_{0,c,ii}/2)\Big\}, \\
&p(M_{0, c}) \propto C_{0,c}
\prod_{i < j}\Big\{Gamma(m_{0, c,ij}|\alpha_{m_{0}},\beta_{m_{0}})\Big\}\prod_{i  }\Big\{Gamma(m_{0,c,ii}|\alpha_{m},\beta_{m})\Big\}, 
\end{aligned}
\end{equation}

so that the posterior conditional distributions of $M_{0,c}$ do not involve $C_{0,c}$. However, empirical results showed that the fixed $M_{0}$ setting generates better clustering results. 

For different choices of $\{ \alpha_{m}, \beta_{m},\alpha_{m_{0}},\beta_{m_{0}} \}$ we observed that, as the hyperprior placed on $M_{0,c}$ is cluster specific, the prior for $T_{c}$ after integrating $M_{0,c}$ out is less flexible than the prior of $T_{c}$ based on $M_{0}$. In addition, the model includes a large number of parameters after making $M_{0}$ cluster specific, which slows down computational speed. Therefore, we choose the specification with a fixed $M_{0}$ and refer to it as the sparse prior in the rest of this manuscript.

\section{The base distribution of \texorpdfstring{$\mu_{c}$}{} }
Besides the base distribution of $\Sigma_{c}$, it is also beneficial to place hyperpriors on the parameters $\mu_{0}$ and $\Sigma_{0}$ of the base distribution for $\mu_{c}$ in (\ref{eq:7}). Since $\mu_{c}$ and $\Sigma_{c}$ are independent a priori, adding hyperpriors on $\mu_{0}$ and $\Sigma_{0}$ does not affect the algebraic form of the conditional posterior distributions for $\Sigma_{c}$ given in Section 2. One choice of hyperpriors for $\mu_{0}$ and $\Sigma_{0}$ is given below. 
\begin{equation}
\begin{aligned} \label{eq:57}
& \mu_{c} | \mu_{0}, \Sigma_{0} \sim N_{J}(\mu_{c}; \mu_{0}, \Sigma_{0}), \\
& \mu_{0} | \mu_{00}, \Sigma_{00} \sim N_{J}(\mu_{0}; \mu_{00}, \Sigma_{00}), \\
& \Sigma_{0} | R_{00}^{-1}, \kappa_{00} \sim InvWishart_{J}(\Sigma_{0}; R_{00}^{-1}, \kappa_{00}). 
\end{aligned}
\end{equation}

The values of the hyperparameters $\{\mu_{00}, \Sigma_{00}, R_{00}^{-1}, \kappa_{00}\}$ are set to be 
\begin{align*}
&\mu_{00}= \frac{1}{n} \sum_{i=1}^{n} X_{i} \ , \ \Sigma_{00}=Diag(range(X_{1})^{2}, range(X_{2})^{2},...,range(X_{J})^{2}), \\
&R_{00}=\Sigma_{00}^{-1}  \ , \
\kappa_{00}=J+2,
\end{align*}
which results in reasonably vague priors for $\mu_{0}$ and $\Sigma_{0}$.

The specification in (\ref{eq:57}) is conditional conjugate, as the conditional posterior distributions of $\mu_{0}$ and $\Sigma_{0}$ are of standard form; see Supplemental material, Section S6.

Empirical results from implementing the priors in (\ref{eq:57}) showed that placing a hyperprior on $\mu_{0}$ can improve the performance of the DPMM. Simulation studies did not show that adding a hyperprior on $\Sigma_{0}$ improves clustering results. Therefore, we only implement the hyperprior for $\mu_{0}$ in (\ref{eq:57}) in all subsequent analyses. 

\section{Comparison of different priors using simulated data}

In this Section, we compare the performance of the priors discussed in Section 2. We add to the comparisons the setting where within-cluster independence is assumed for the observed variables (see Supplemental material Section S5.1). An Inverse Gamma prior is specified for the variance parameters. We refer to this settings as the independent prior. The adjusted Rand index is used to compare the different priors. Assuming the cluster allocation vector $\bm{Z}$ is of length $n$, the Rand index is defined as,
\begin{equation} \label{eq:58}
Rand \ index =\frac{TP+TN}{ \binom{n}{2} },
\end{equation}
where $TP$ denotes the number of pairs of observations correctly allocated to the same cluster. $TN$ denotes the number of pairs correctly  allocated to different clusters. \cite{hubert1985comparing} proposed the adjusted Rand index, which is defined as follows
\begin{equation} \label{eq:6.2}
\frac{index - expected \ index}{maximum \ index - expected \ index}.
\end{equation}  

 
We adopt the adjusted Rand index for our comparative study, recommended by \cite{milligan1986study} 
in their review. 

Observations are generated using multivariate normal distributions with cluster specific means and covariance matrices. Two simulation settings for the within-cluster correlation matrices are considered. In the first dense setting, all correlations are non-zero. Within a matrix, all $\rho_{c}^{true}$ correlations are identical. In the second sparse setting,  within-cluster correlation matrices are block diagonal. Variables are highly correlated within each block and independent between blocks. Non-zero correlations within a matrix are identical. Two simulation scenarios are considered within each setting. One with larger within-cluster variances where the clusters are not well separated, and another with smaller within-cluster variances where, although the clusters are more separated, the simulations are still challenging. 


$20$ datasets are generated for each scenario and setting combination. Burn-in consists of $10,000$ iterations, with $6,000$ additional iterations. Posterior allocation samples were stable after the burn-in period. We compare the performance of the  priors in terms of: (a) the number and profile of clusters in the `best-representation' clustering obtained from the post-processing of the MCMC output \citep{molitor2010bayesian}; (b) the adjusted Rand index; (c) computing time. 

\subsection{Simulations with dense within-cluster covariance matrices}

\begin{figure}[h!]
	\centering
	\includegraphics[width=5cm, height=5cm]{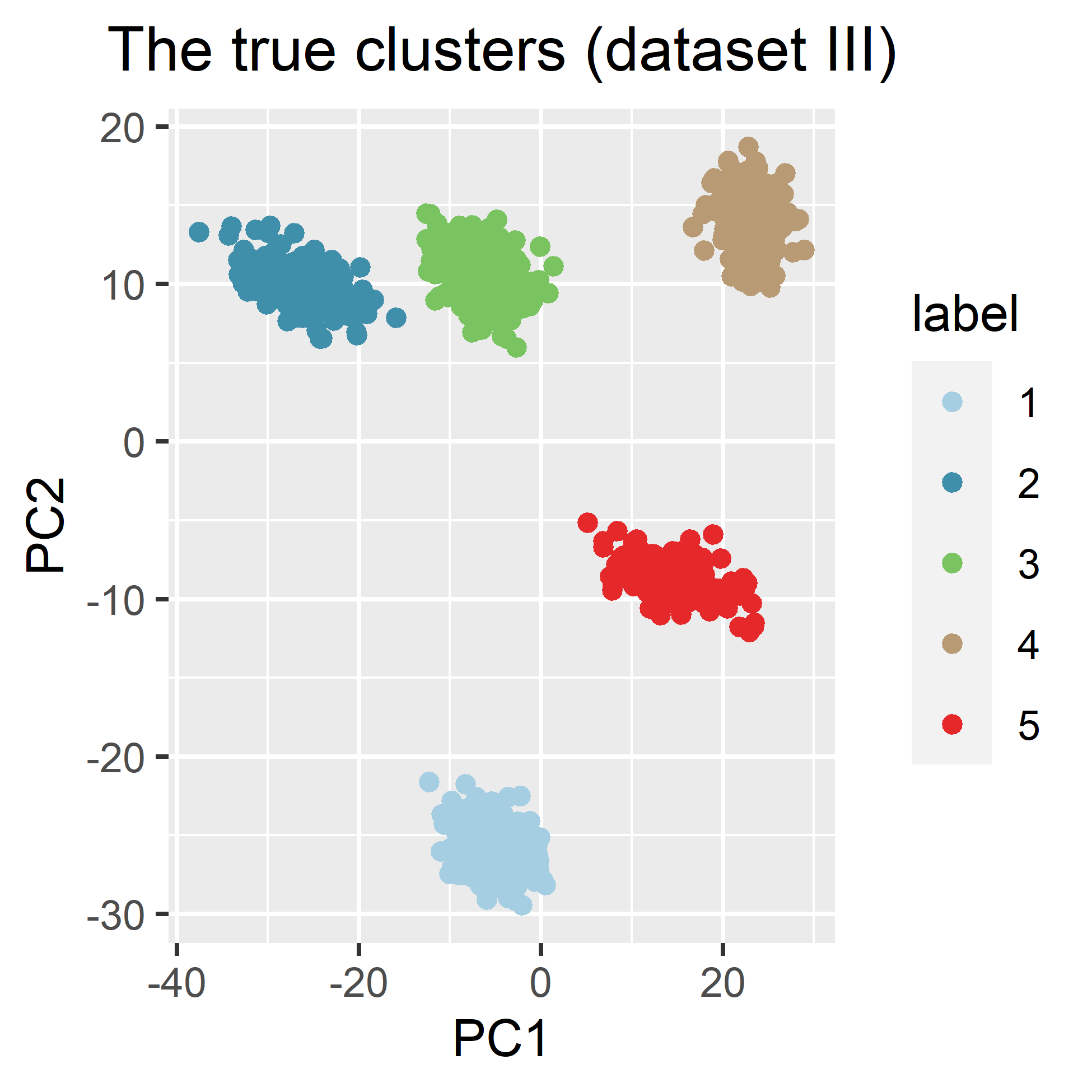} 
	\includegraphics[width=5cm, height=5cm]{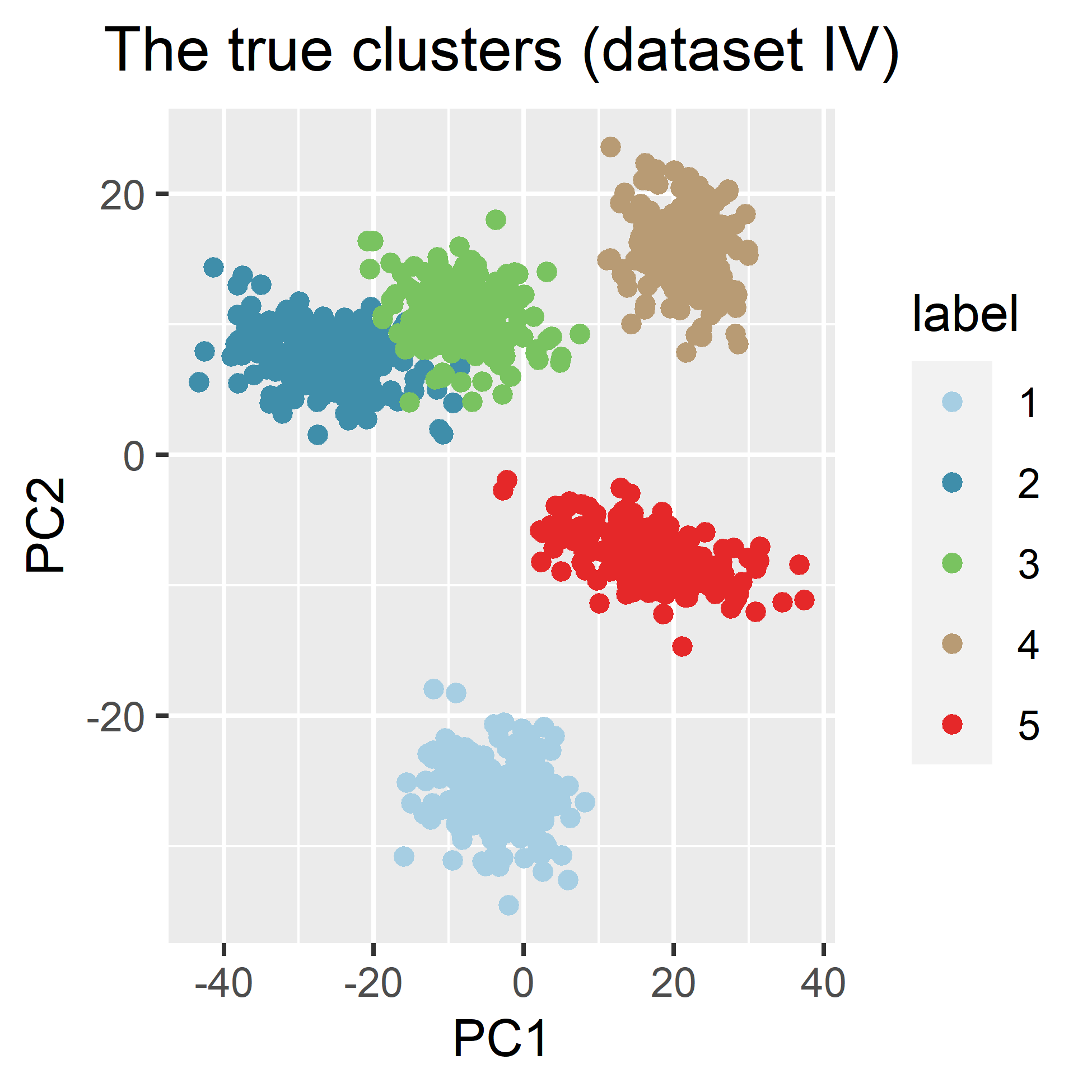} \\
	\includegraphics[width=5cm, height=5cm]{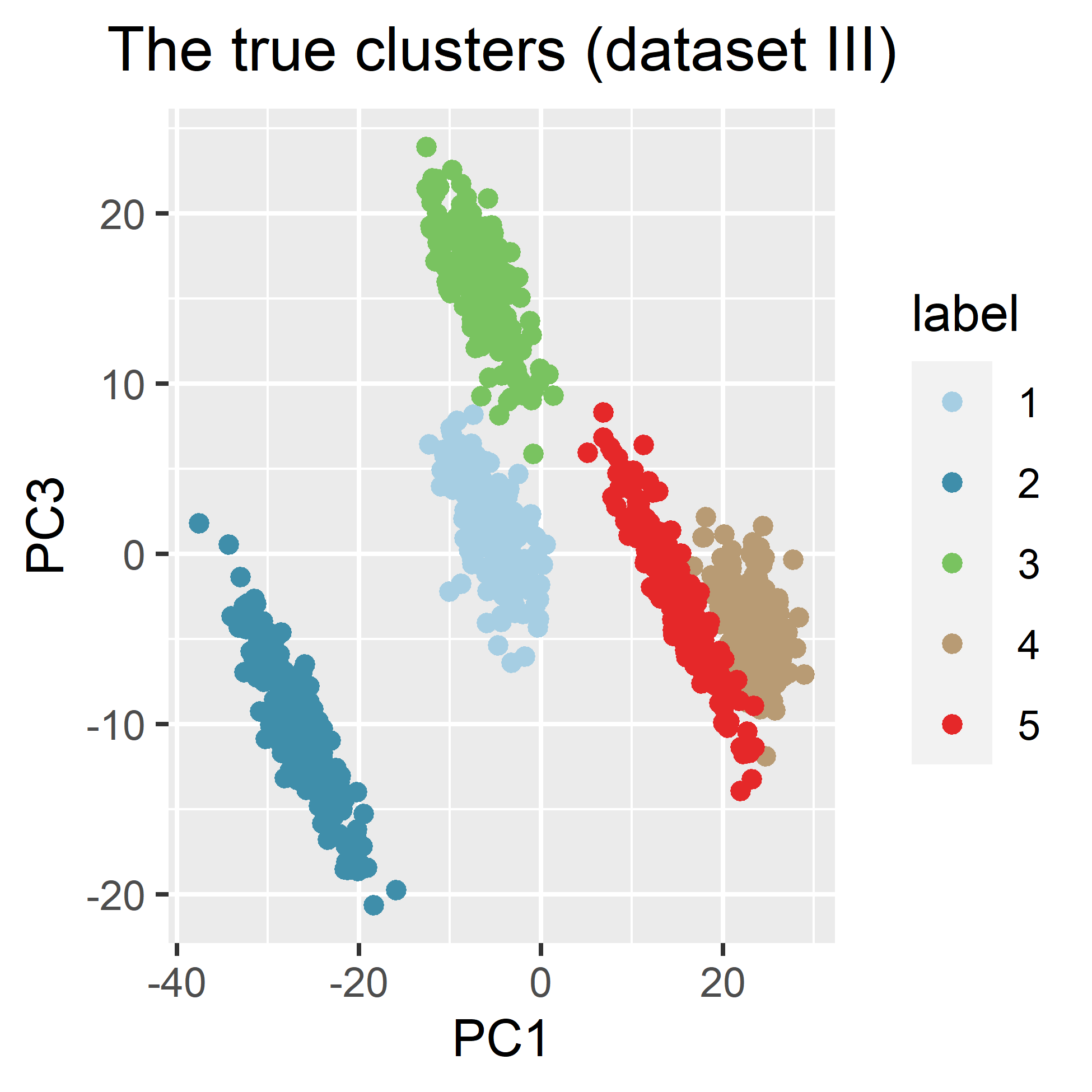} 
	\includegraphics[width=5cm, height=5cm]{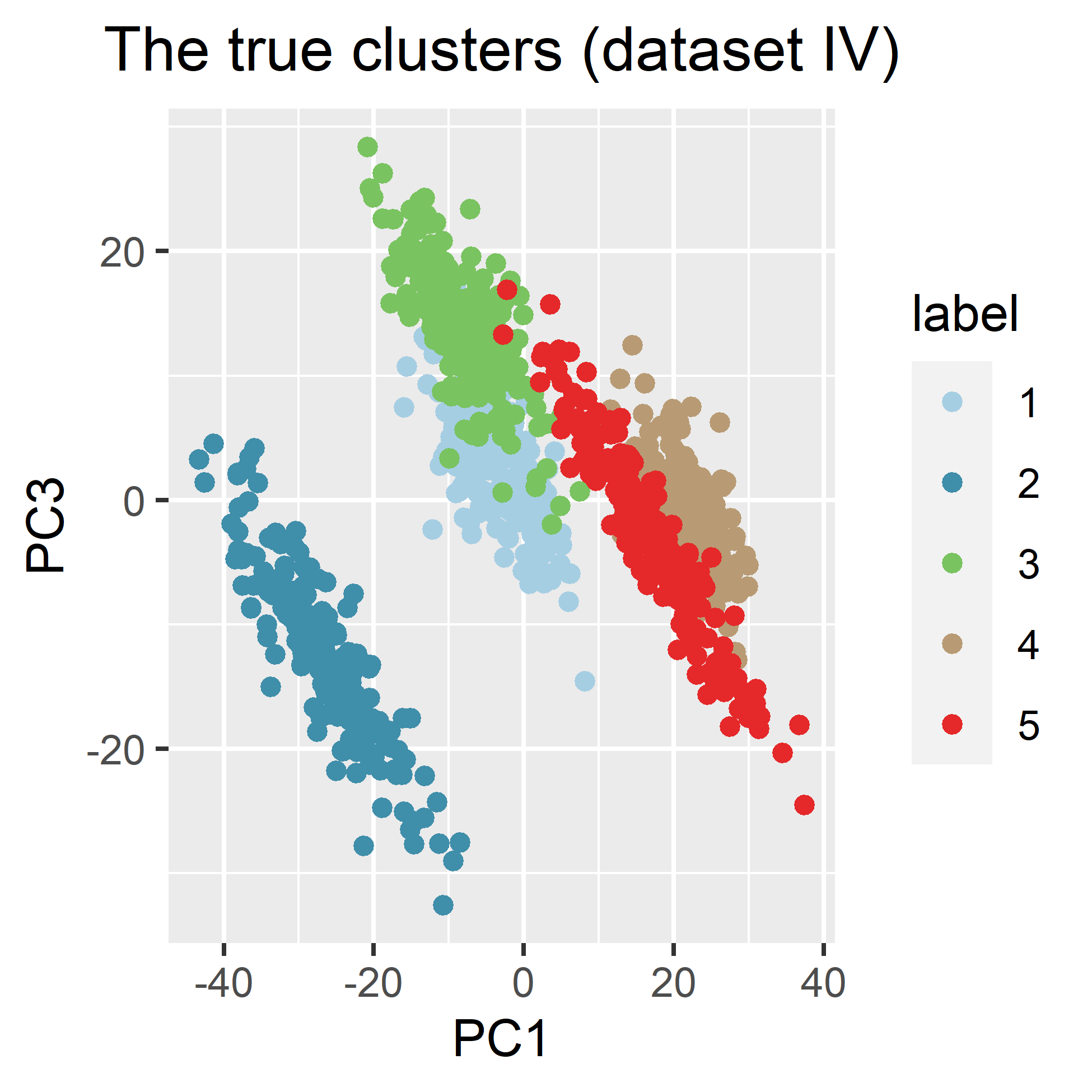} \\
	\includegraphics[width=5cm, height=5cm]{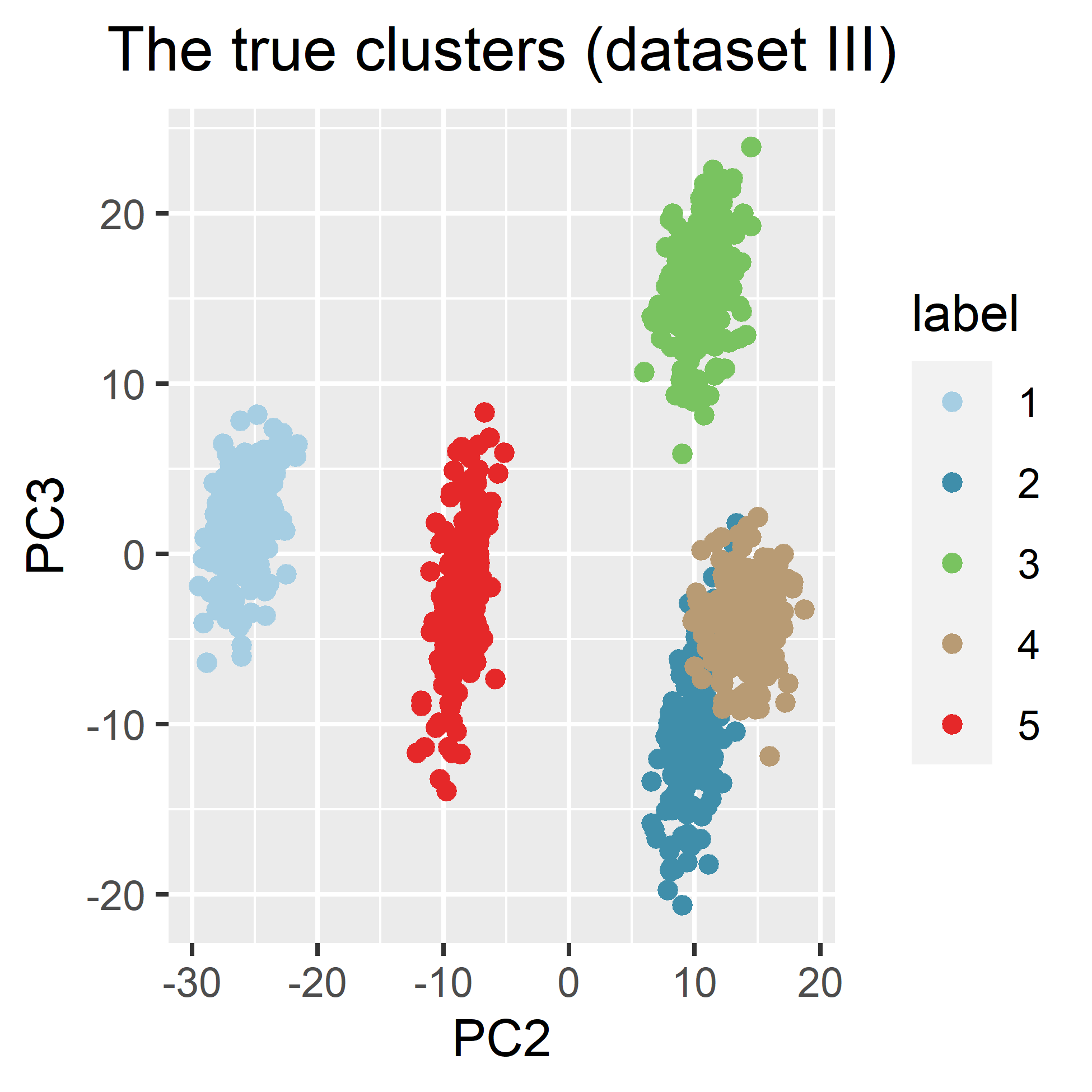} 
	\includegraphics[width=5cm, height=5cm]{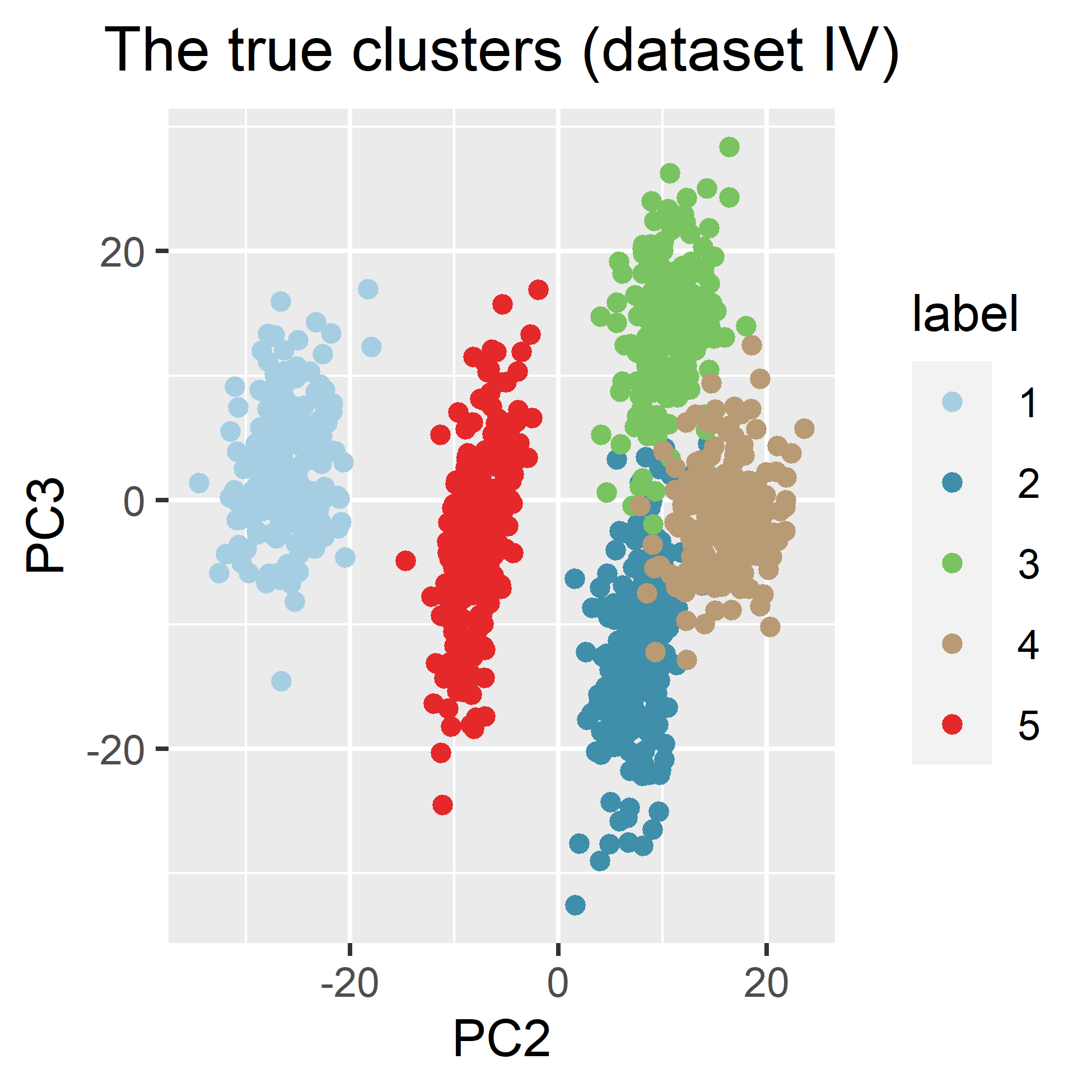} \\
	\vspace{1cm}
	\begin{minipage}{0.8\textwidth}
		\caption{Reduced space plots of the clusters for simulated data III (Table \ref{tab:6.2.1}) and IV (Table \ref{tab:6.2.2}). A representative dataset is shown for each setting III and IV. PC1, PC3 and PC3 stand for the first, second and third principle component respectively. }
		\label{fig:6.2.1}
	\end{minipage} 
\end{figure}

The two dense simulation settings are given in Tables \ref{tab:6.2.1} and \ref{tab:6.2.2}. Five clusters are simulated. 
The sample size for each cluster ($n_c=200$) is relatively small compared to the number of cluster-specific parameters for $J=20$. More than 80\% of the variability is explained by the first three Principle Components (PCs); see Supplemental material, Section S8. Those PCs are used to visualize the clusters. Reduced space plots  for one
 representative simulated dataset III and another dataset IV are given in Figure \ref{fig:6.2.1}. Simulated data III (Table \ref{tab:6.2.1}) have smaller variances, which translates to a stronger clustering signal. Clusters in simulated data IV (Table \ref{tab:6.2.2}) are closer and the signal is less strong. Plots of the overall covariance matrices are given in the Supplemental material, Section S8.  

\begin{table}[H]
	\centering
	\caption{Simulated data III specifications. $J=20$. $rep(a,b)$ represents a vector of length $b$ with repeated elements $a$.}
	\begin{tabular}{  m{1.6cm} m{5.2cm} m{1.0cm} m{1.0cm} m{0.8cm} } 
		\hline
		Cluster No. & $\mu_{c}^{true}$ & $\Var_{c}^{true}$ & $\rho_{c}^{true}$& $n_{c}$ \\ \hline
		1 & (rep(3,5),rep(12,5),rep(18,5),rep(12,5)) & 3& 0.2 & 200 \\ 
		2 & (rep(12,5),rep(18,5),rep(3,5),rep(18,5)) & 3 & 0.5 & 200 \\ 
		3 & (rep(18,5),rep(18,5),rep(12,5),rep(8,5)) & 3& 0.3 & 200 \\ 
		4 & (rep(18,5),rep(3,5),rep(8,5),rep(3,5)) & 3 & 0.1 & 200 \\ 
		5 & (rep(8,5),rep(8,5),rep(12,5),rep(3,5)) & 3& 0.7 & 200 \\ \hline	
	\end{tabular}
    \label{tab:6.2.1}
\end{table} 

\begin{table}[H]
	\centering
	\caption{Simulated data IV specifications. $J=20$.}
	\begin{tabular}{  m{1.6cm} m{5.2cm} m{1.0cm} m{1.0cm} m{0.8cm} } 
		\hline
		Cluster No. & $\mu_{c}^{true}$ & $\Var_{c}^{true}$  & $\rho_{c}^{true}$ & $n_{c}$ \\ \hline
		1 & (rep(3,5),rep(12,5),rep(18,5),rep(12,5)) & 9& 0.2 & 200 \\ 
		2 & (rep(12,5),rep(18,5),rep(3,5),rep(18,5)) & 9& 0.5 &200 \\ 
		3 & (rep(18,5),rep(18,5),rep(12,5)rep(8,5)) & 9&0.3 & 200 \\ 
		4 & (rep(18,5),rep(3,5),rep(8,5),rep(3,5)) & 9& 0.1&200 \\ 
		5 & (rep(8,5),rep(8,5),rep(12,5),rep(3,5)) & 9& 0.7 & 200 \\ \hline		
	\end{tabular}
    \label{tab:6.2.2}
\end{table}

\begin{figure}[h!]
	\centering
	\includegraphics[width=5.0cm, height=5cm]{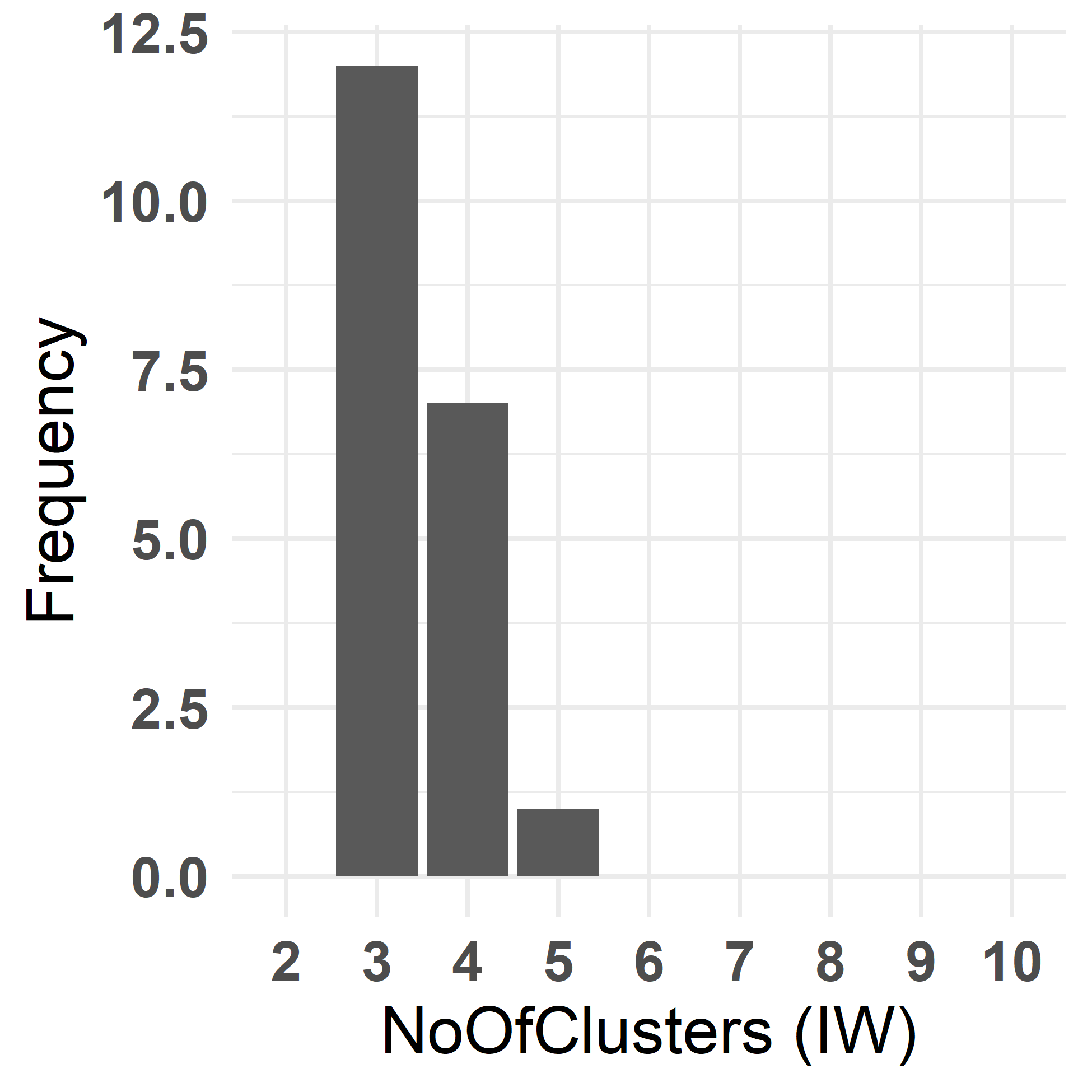} 
	\includegraphics[width=5.0cm, height=5cm]{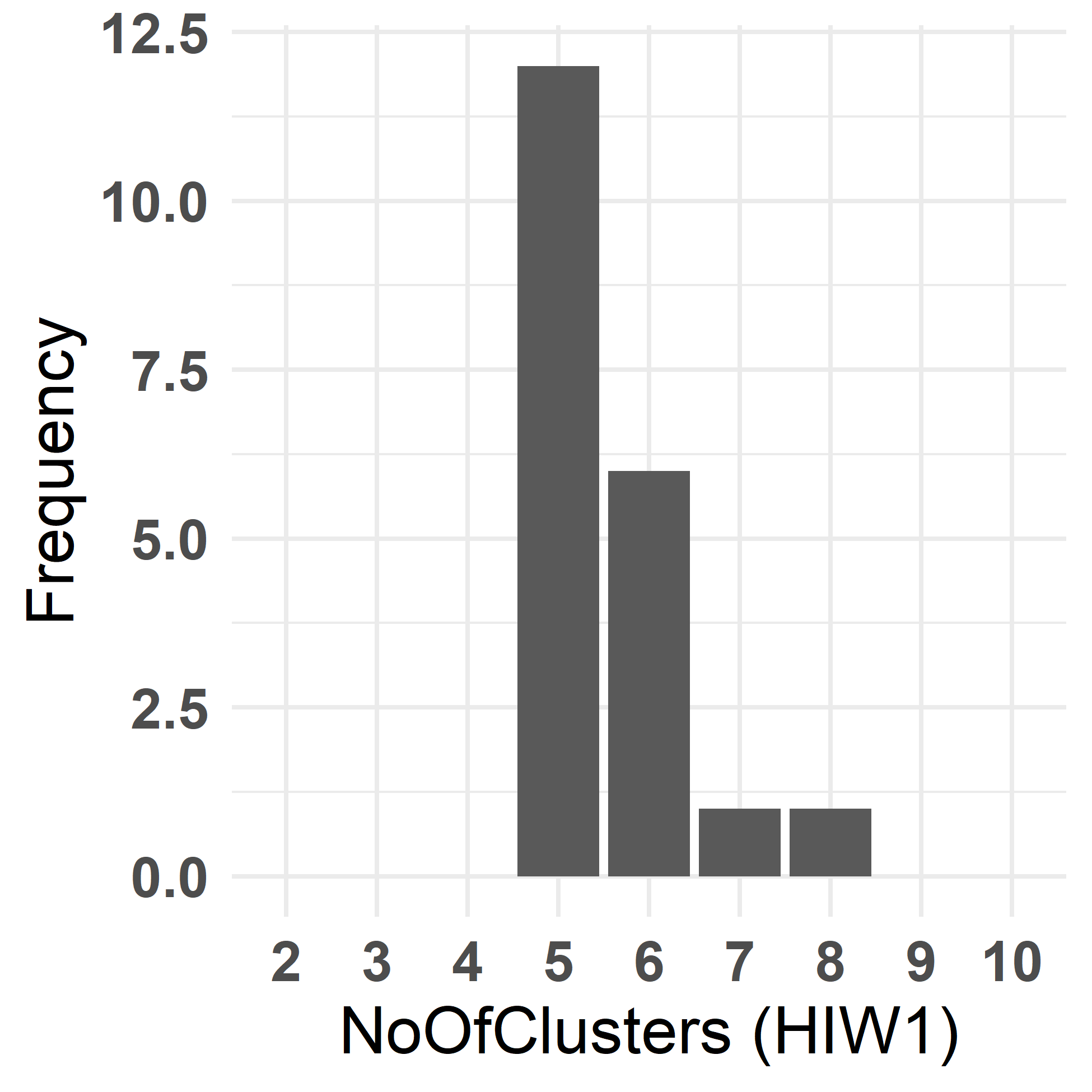} \\
	\includegraphics[width=5.0cm, height=5cm]{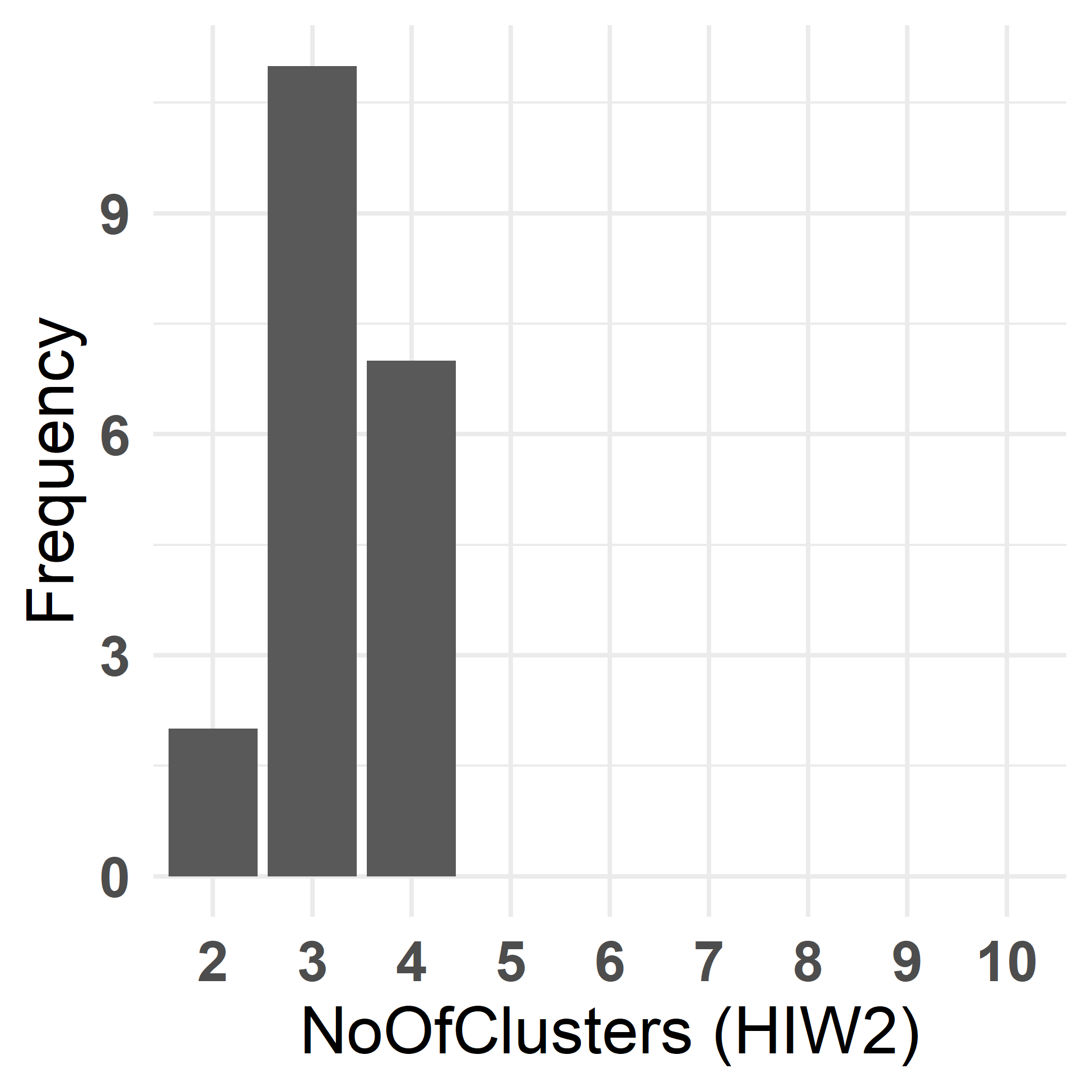} 
	\includegraphics[width=5.0cm, height=5cm]{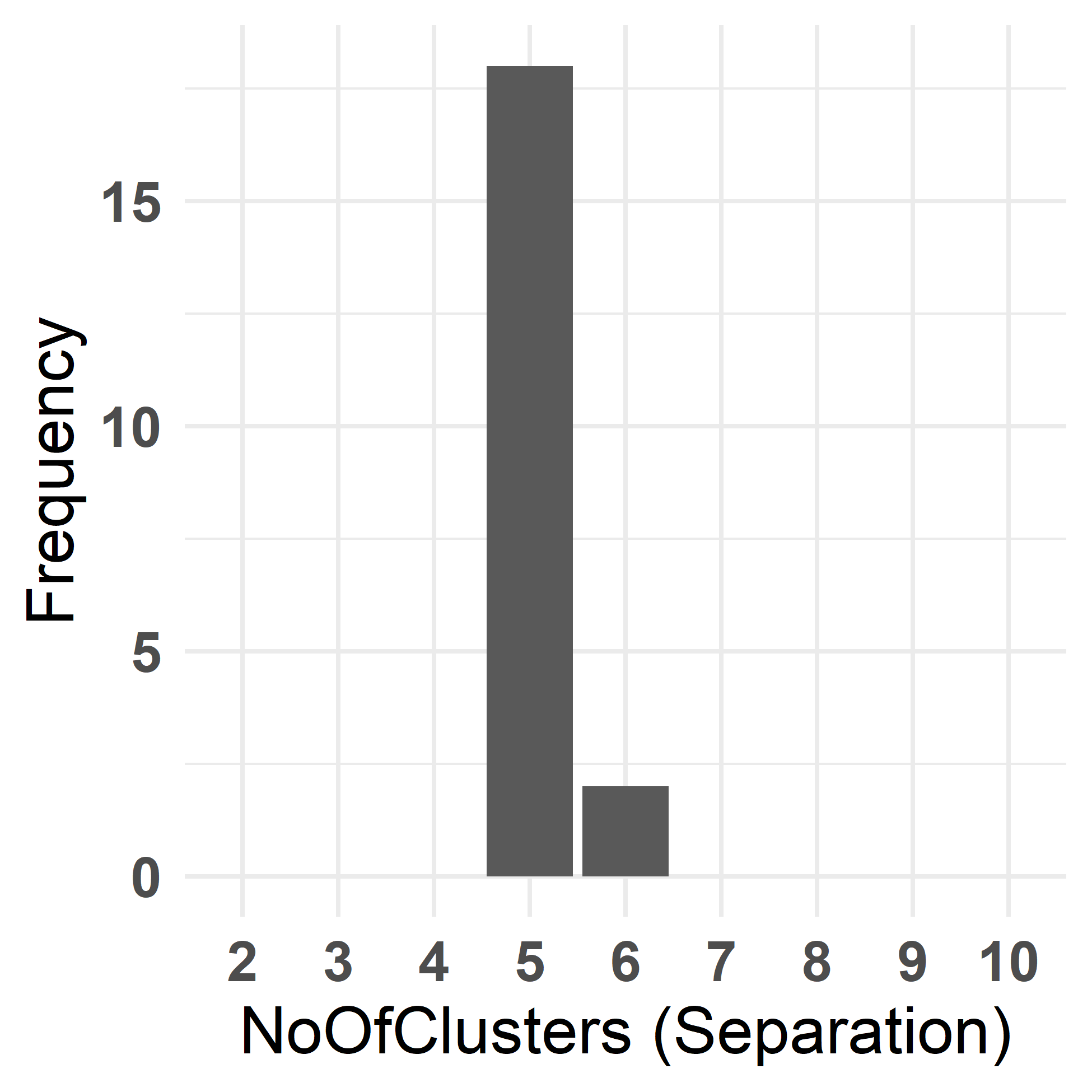} \\ 
	\includegraphics[width=5.0cm, height=5cm]{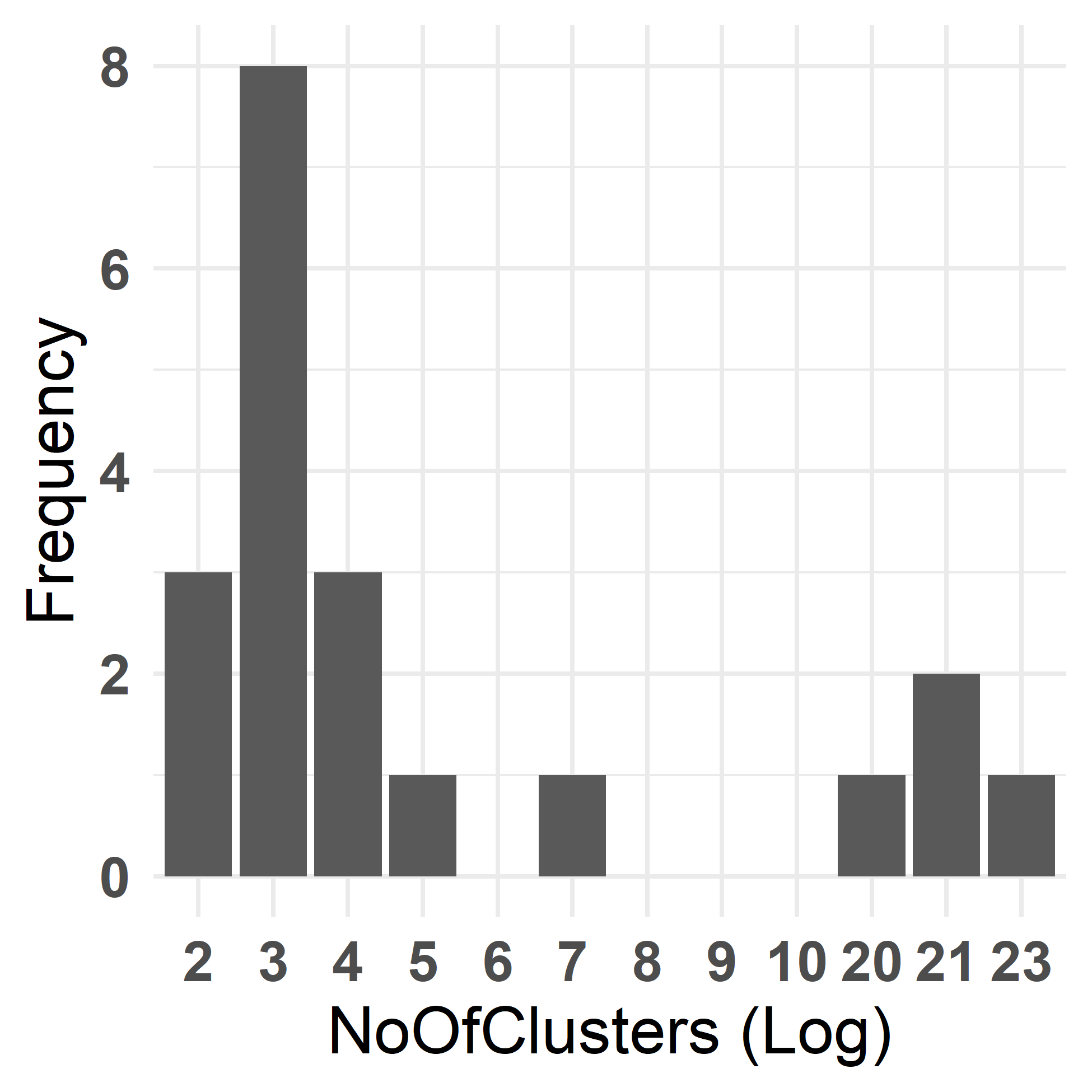} 
	\includegraphics[width=5.0cm, height=5cm]{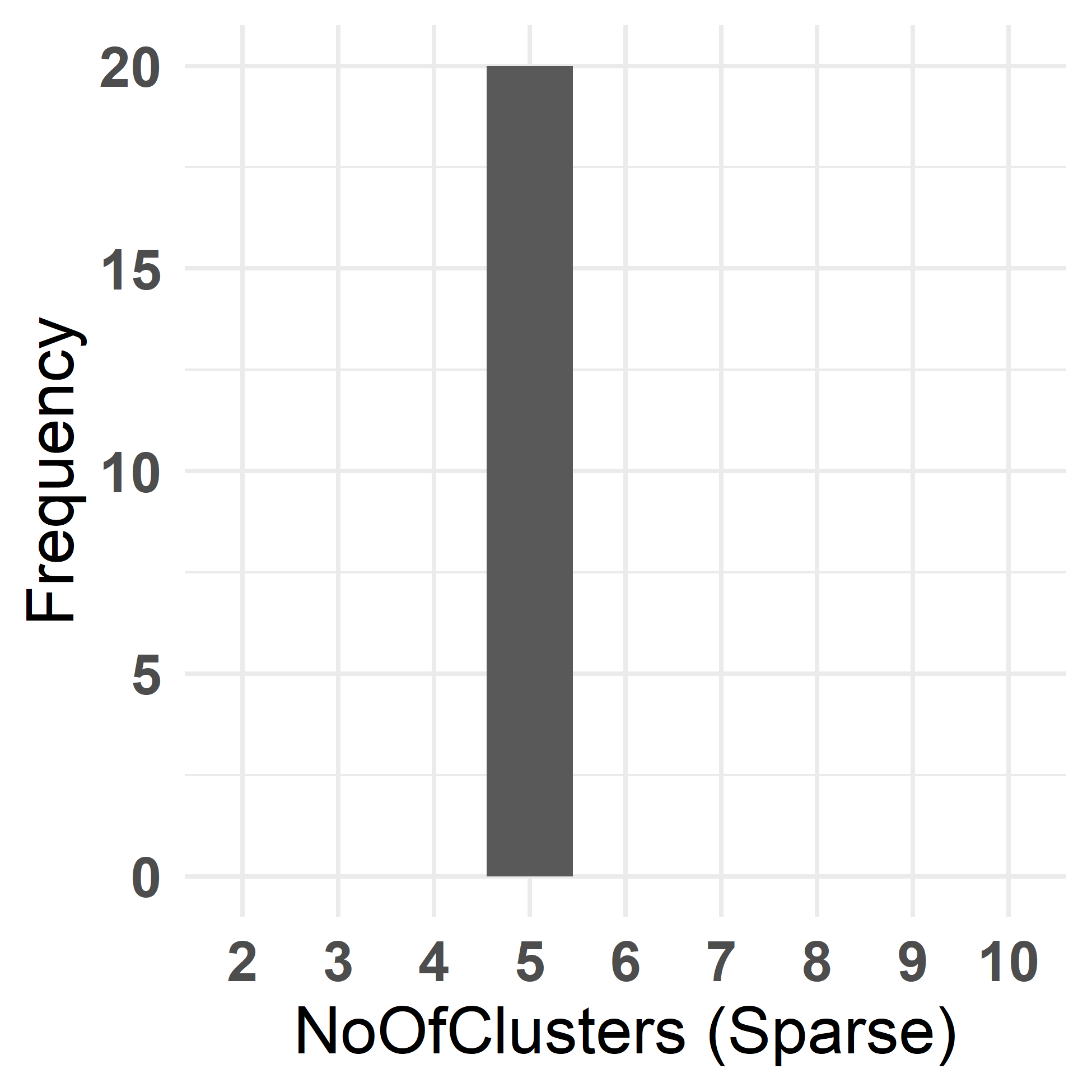}\\
	\includegraphics[width=5.0cm, height=5cm]{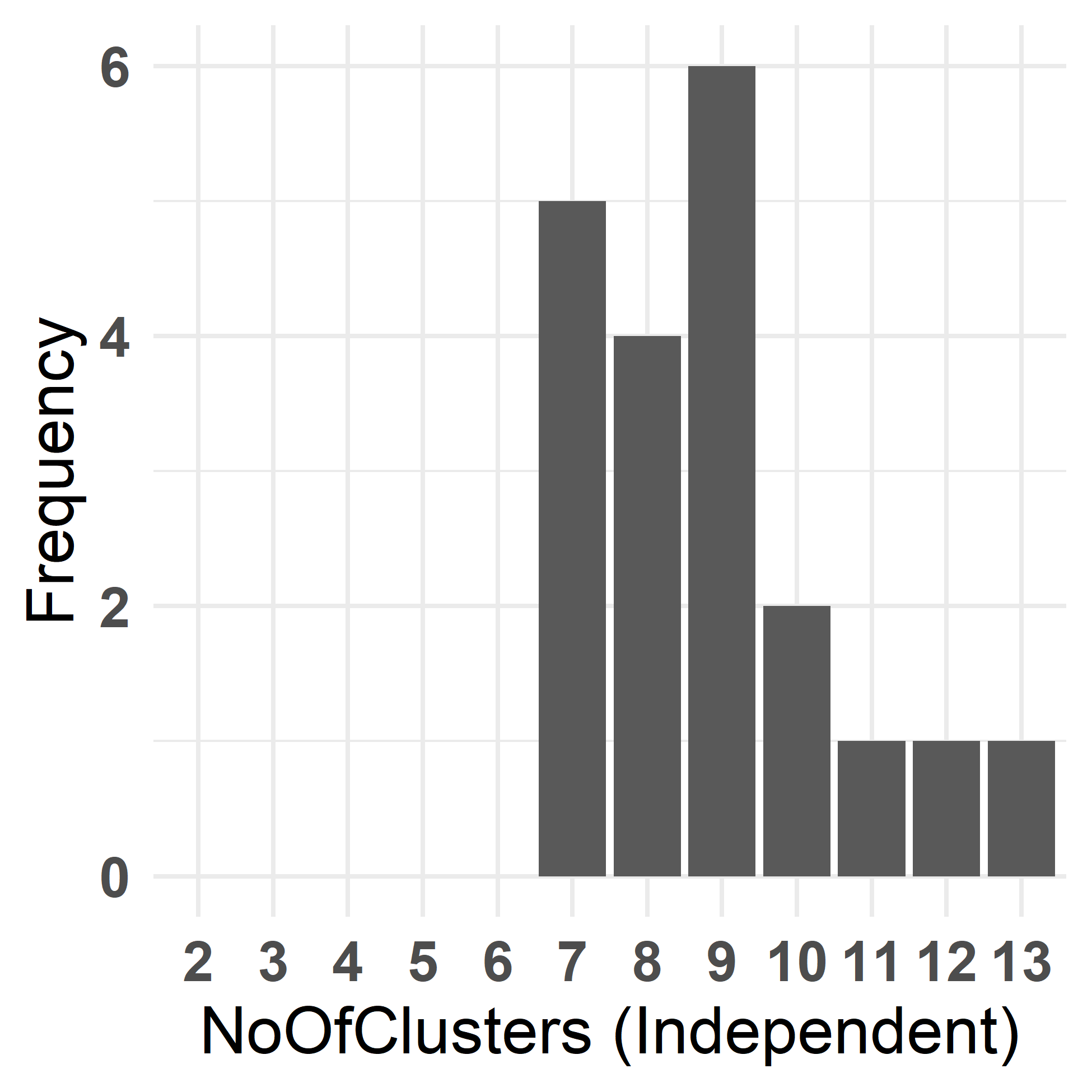} \\
	\vspace{0.5cm}
	\begin{minipage}{0.8\textwidth}
		\caption{Bar plots of the number of clusters for the 20 datasets under set-up III (Table \ref{tab:6.2.1}).}
	\label{fig:6.2.3}
	\end{minipage}
\end{figure}

\begin{figure}[h!]
	\centering
	\includegraphics[width=5.0cm, height=5cm]{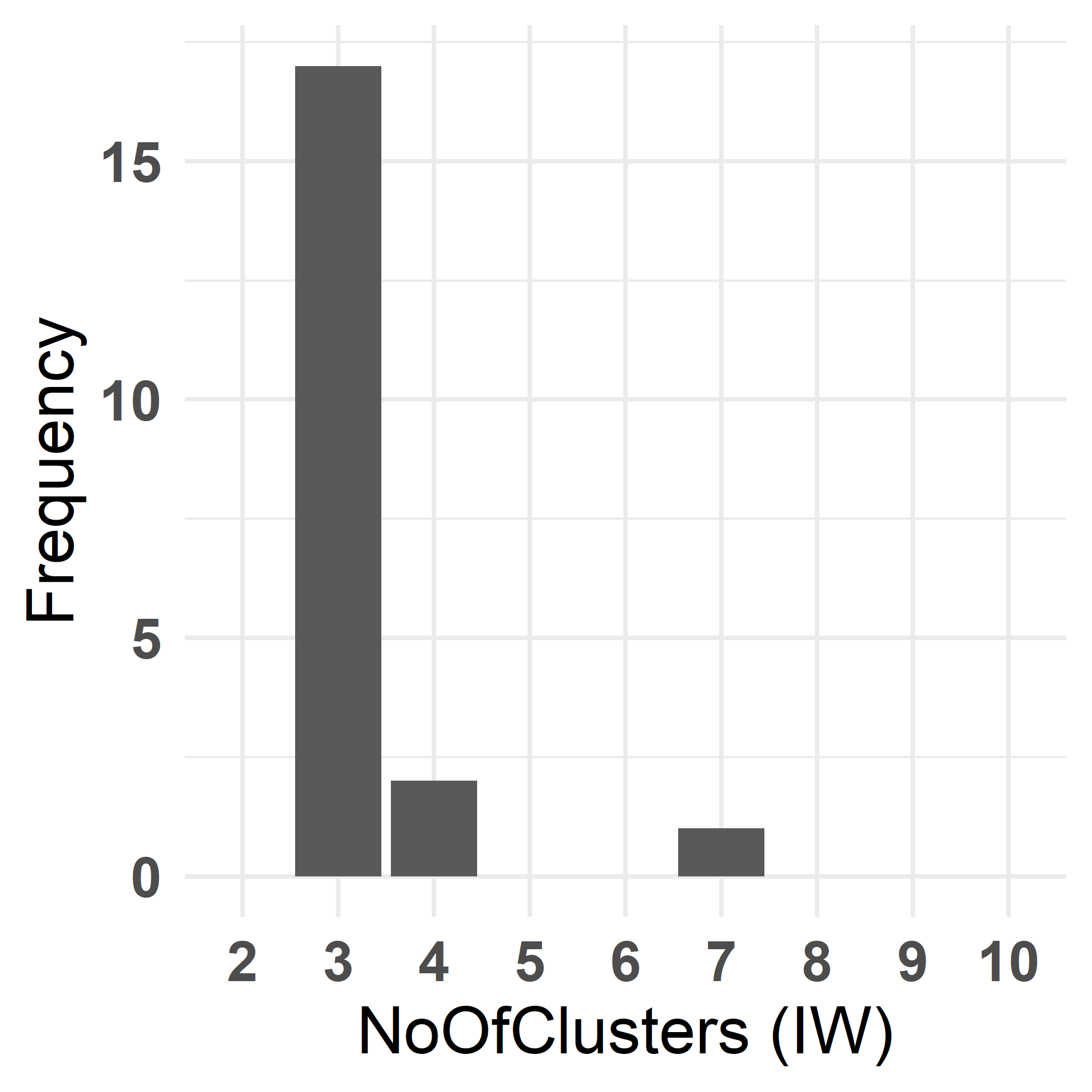} 
	\includegraphics[width=5.0cm, height=5cm]{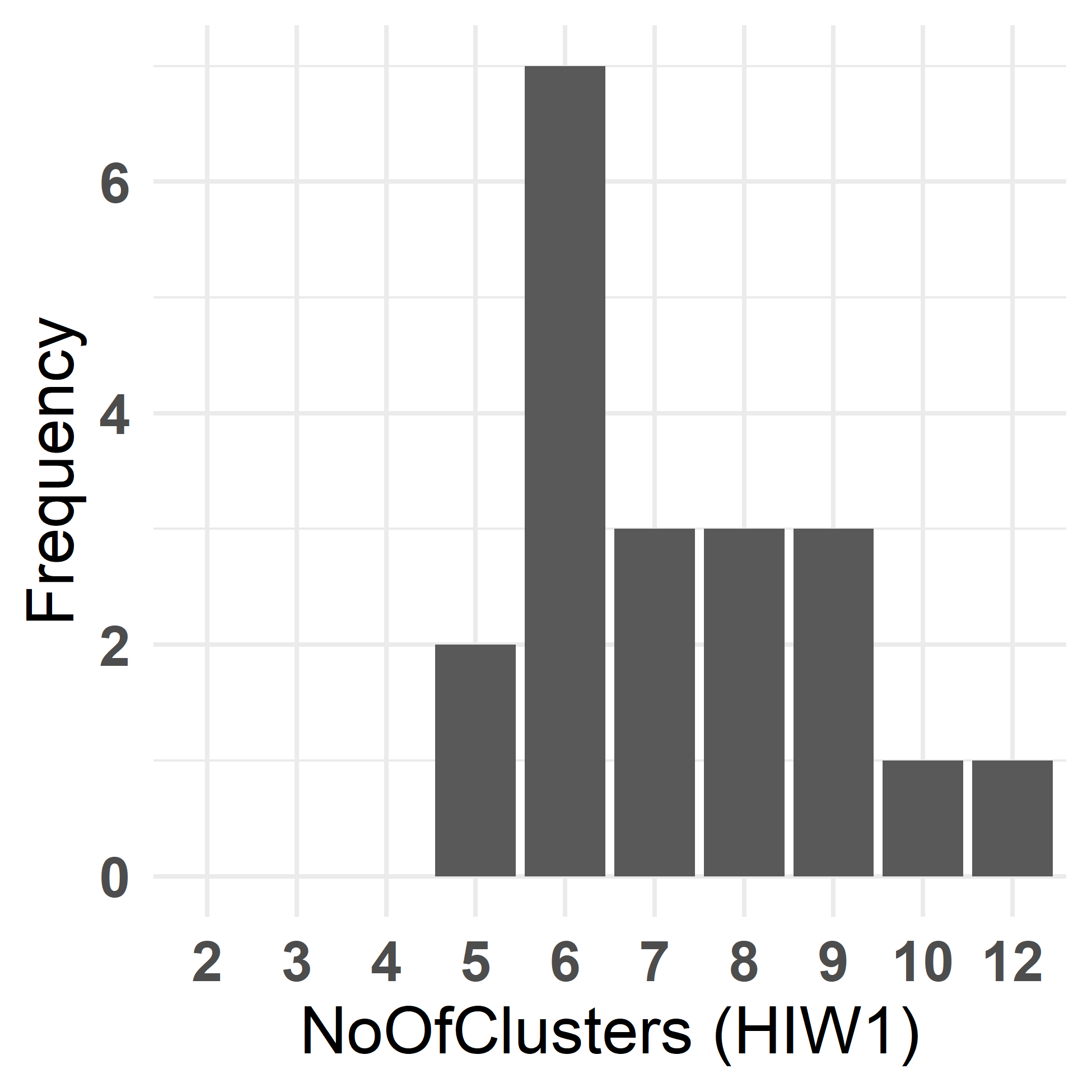} \\
	\includegraphics[width=5.0cm, height=5cm]{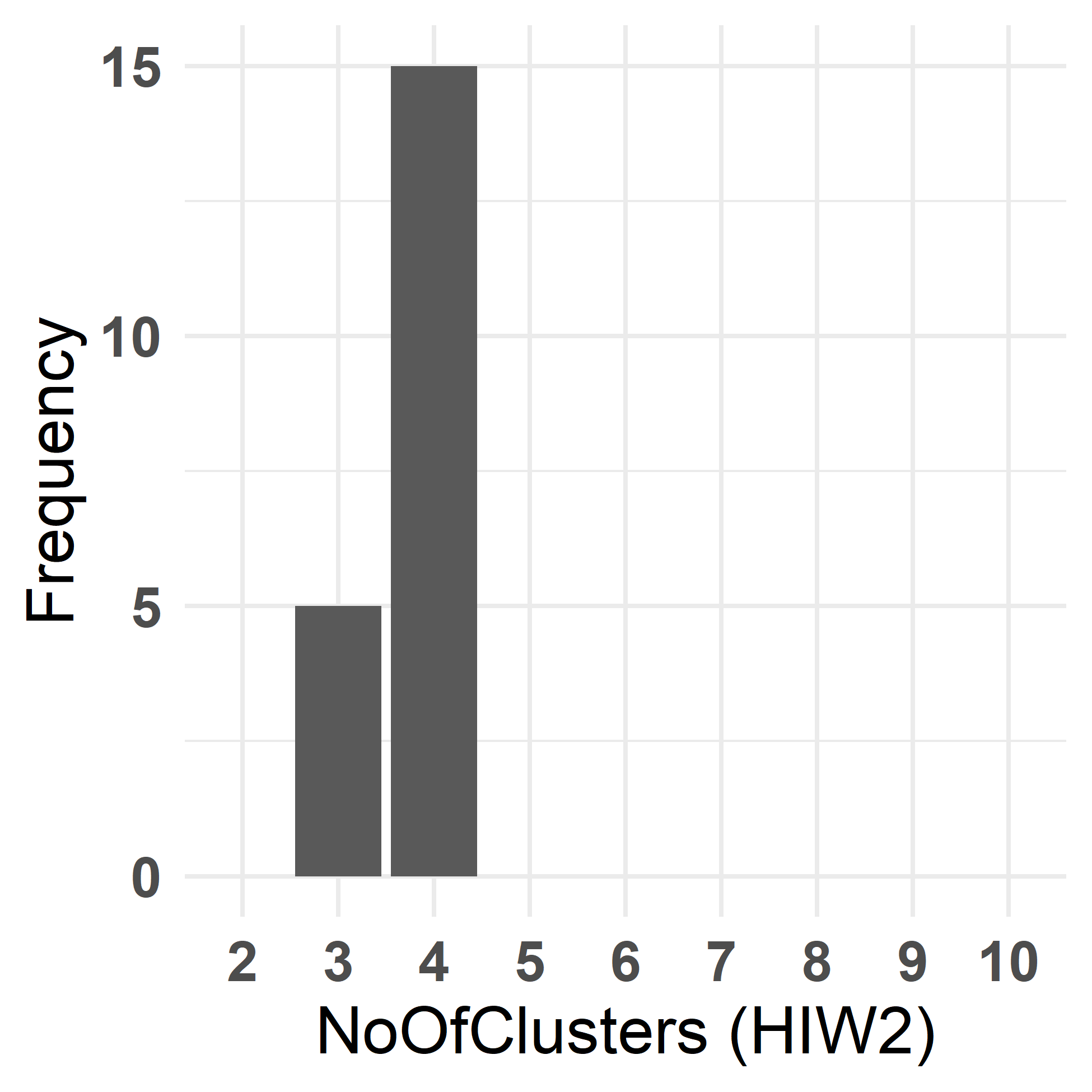} 
	\includegraphics[width=5.0cm, height=5cm]{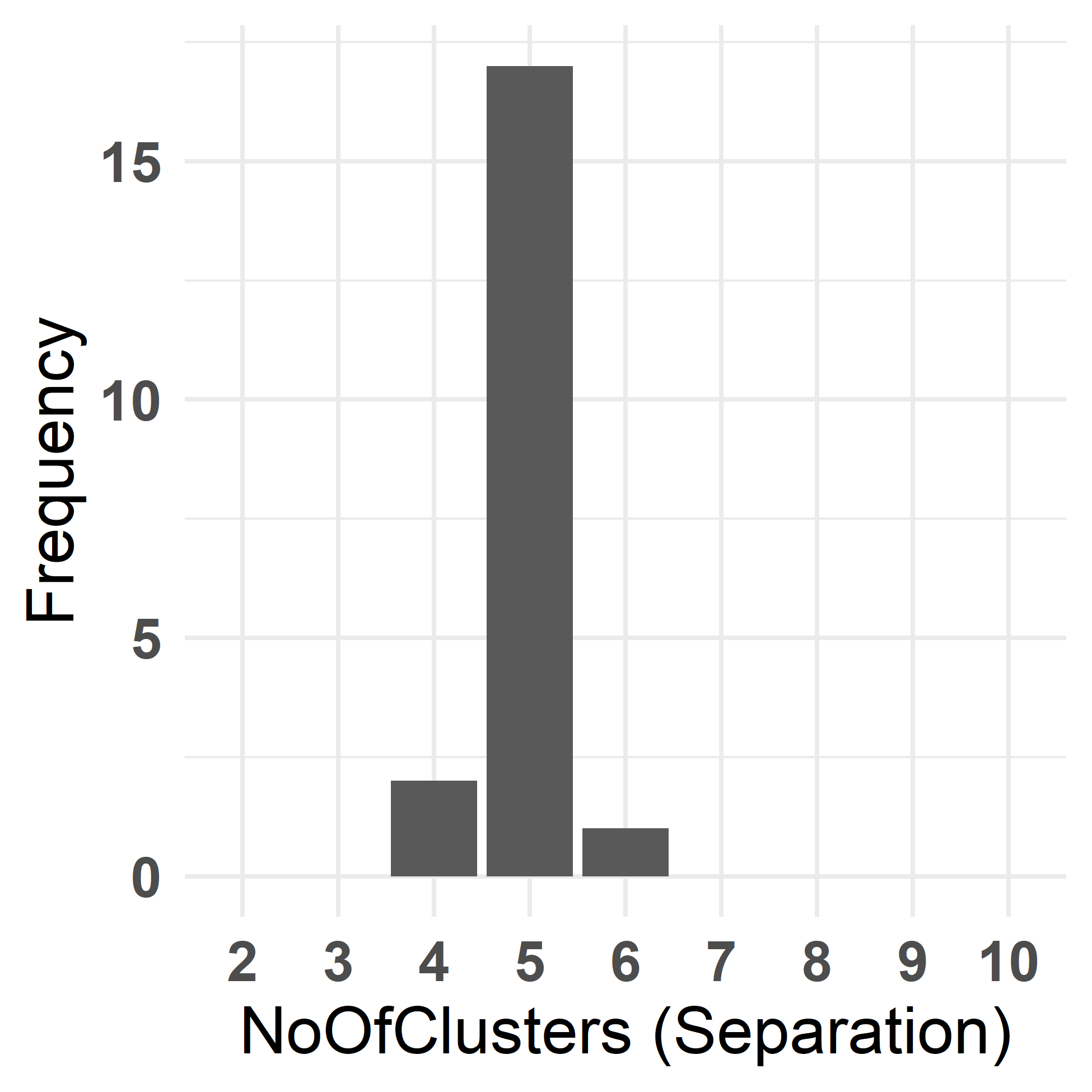} \\ 
	\includegraphics[width=5.0cm, height=5cm]{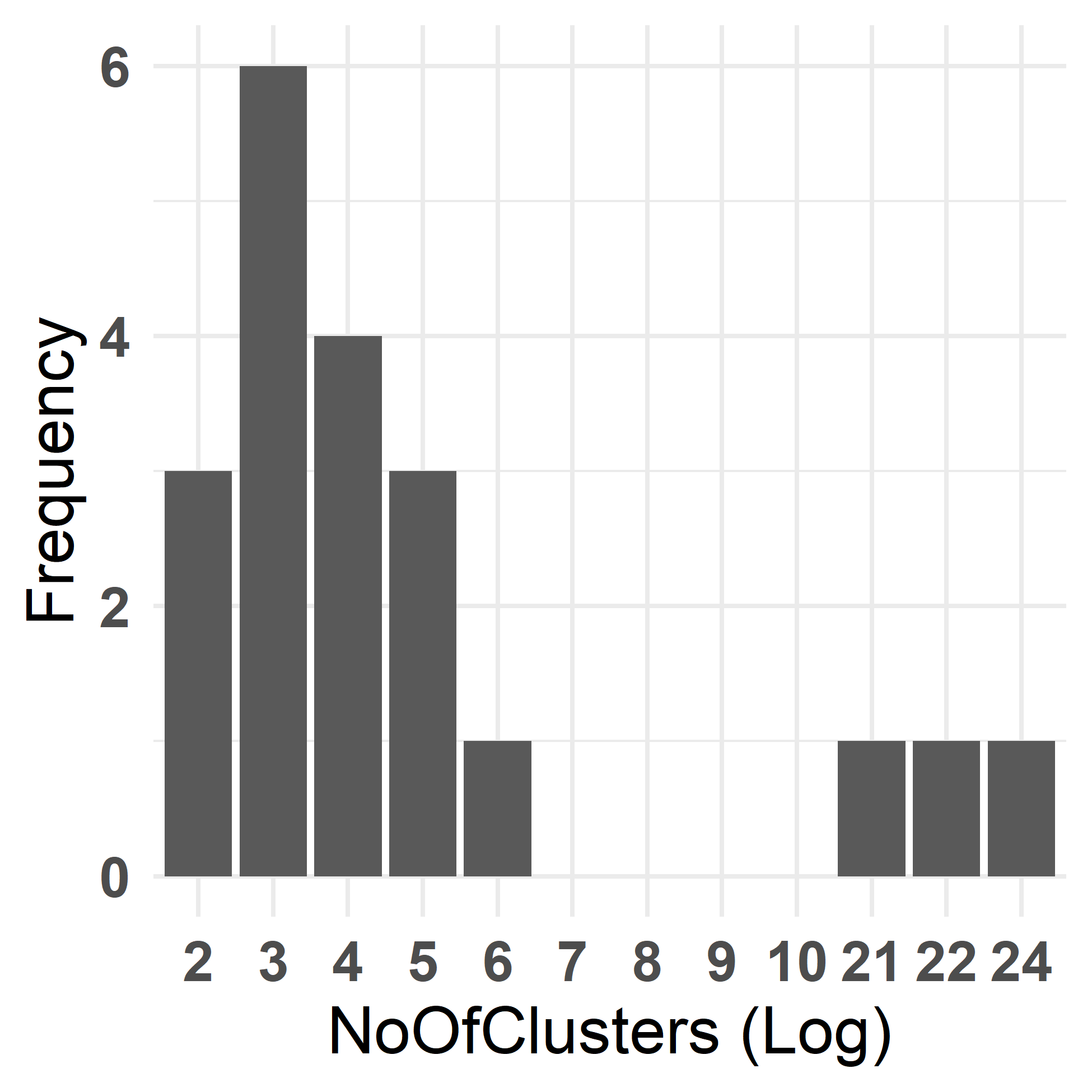} 
	\includegraphics[width=5.0cm, height=5cm]{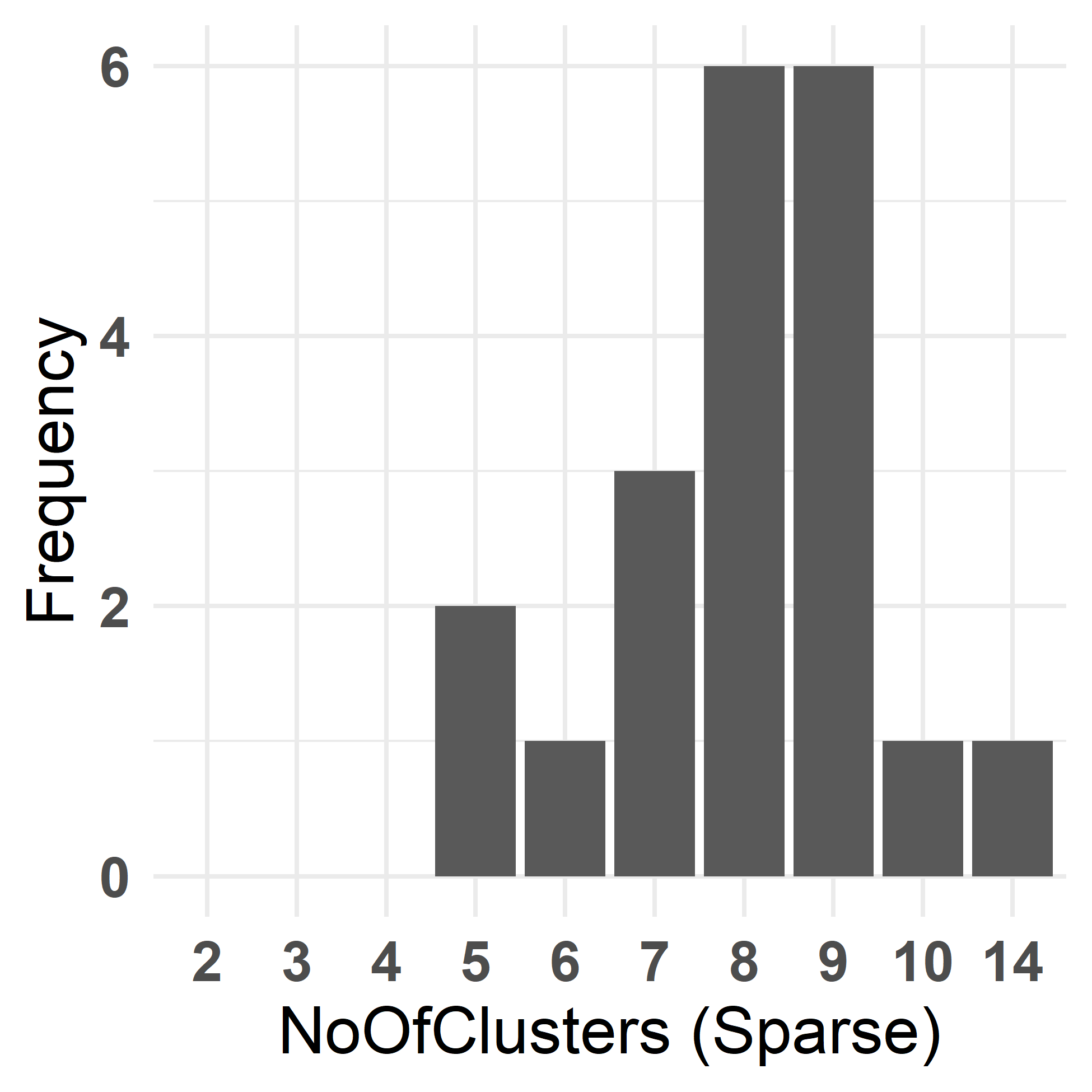}\\
	\includegraphics[width=5.0cm, height=5cm]{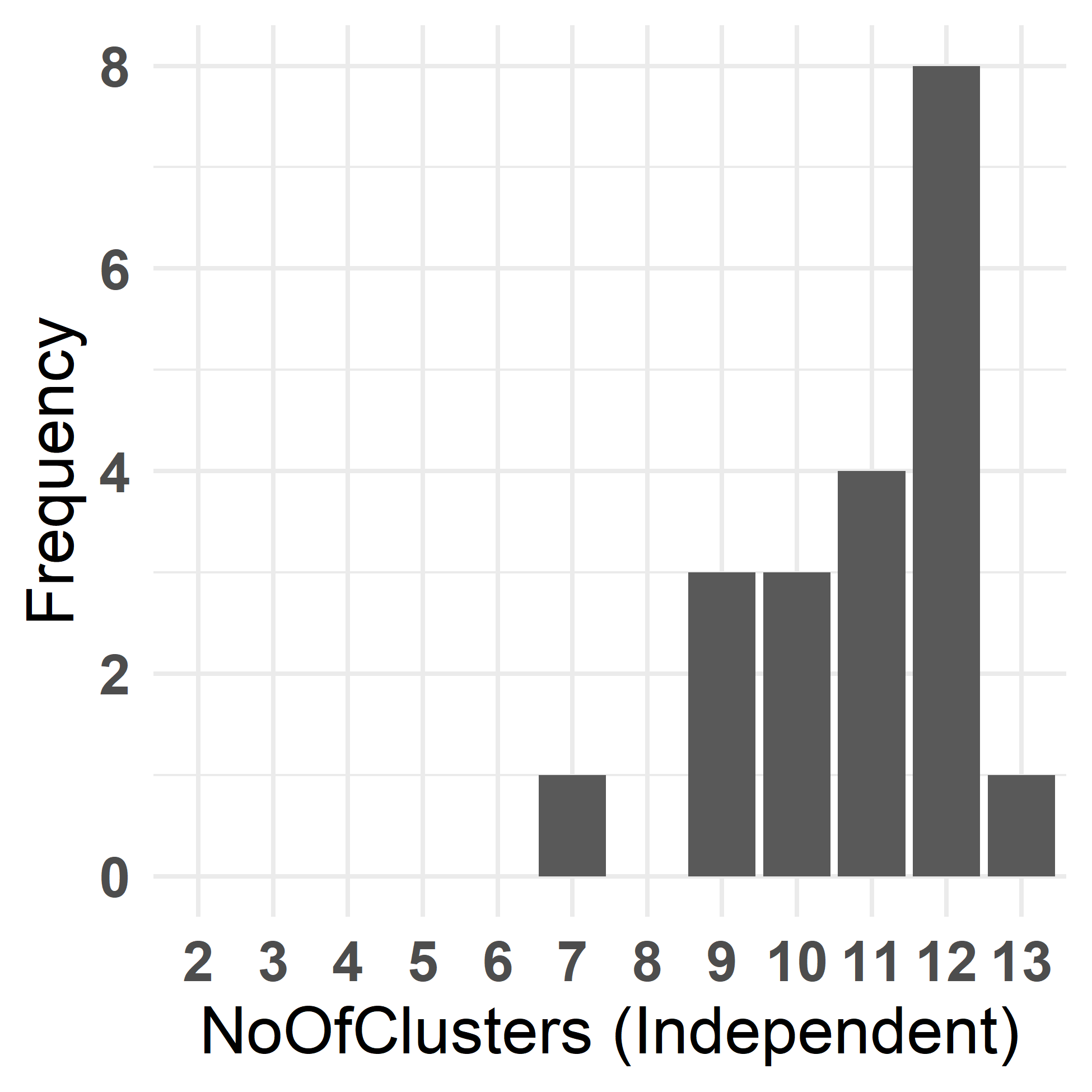} \\
	\vspace{0.5cm}
	\begin{minipage}{0.8\textwidth}
		\caption{Bar plots of the number of clusters for the 20 datasets under set-up IV (Table \ref{tab:6.2.2}).}
	 \label{fig:6.2.4}
	\end{minipage}
\end{figure}

\begin{figure}[h!]
	\centering
	\includegraphics[width=5cm, height=5cm]{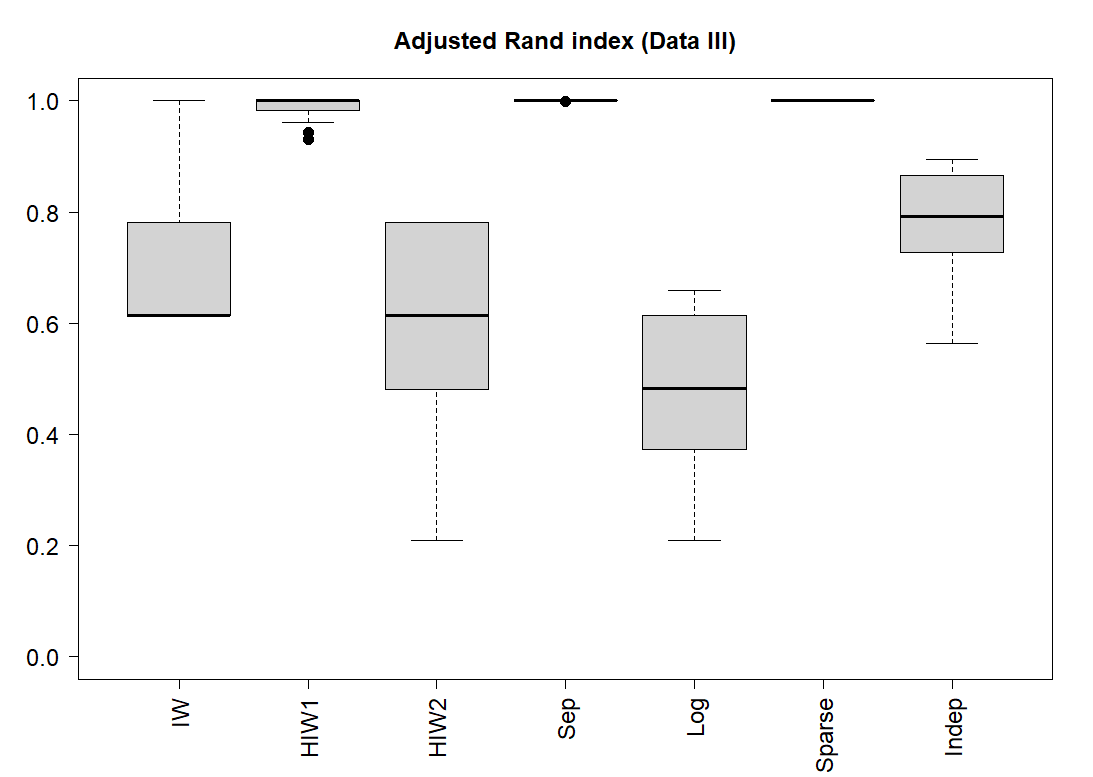} 
	\includegraphics[width=5cm, height=5cm]{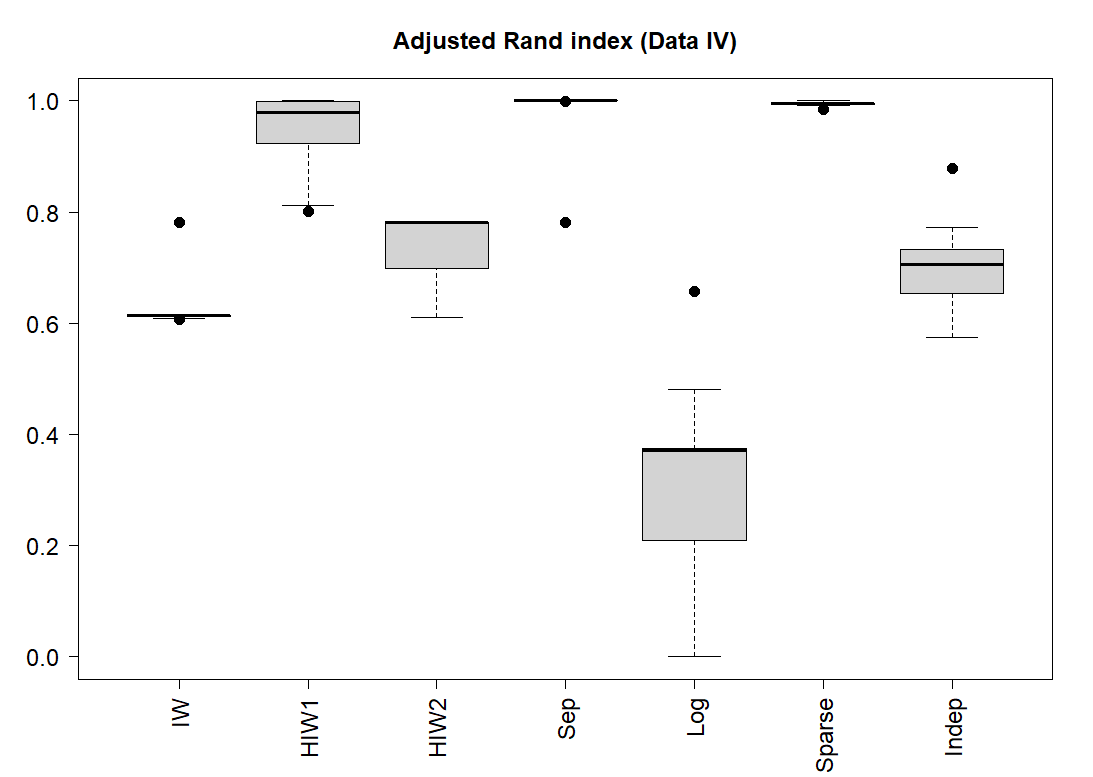} \\
	\vspace{0.5cm}
	\begin{minipage}{0.8\textwidth}
		\caption{The boxplots of the adjusted Rand indices for different priors. The plot on the left corresponds to datasets simulated according to Table \ref{tab:6.2.1} (Data III). The plot on the right corresponds to datasets simulated according to Table \ref{tab:6.2.2} (Data IV).}
	\label{fig:6.2.5}
	\end{minipage}
\end{figure}

Bar plots of the number of clusters recognised for each of the $7$ priors are shown in Figures \ref{fig:6.2.3} and  \ref{fig:6.2.4}. More important for assessing performance is the profile of the clusters, in terms of containing observations that were truly generated by the same mixture component. Figure \ref{fig:6.2.5} shows  boxplots of the adjusted Rand indices for different priors. 

\begin{figure}[h!]
	\centering
	\includegraphics[width=5cm, height=5cm]{pic_main/s1-9-1} 
	\includegraphics[width=5cm, height=5cm]{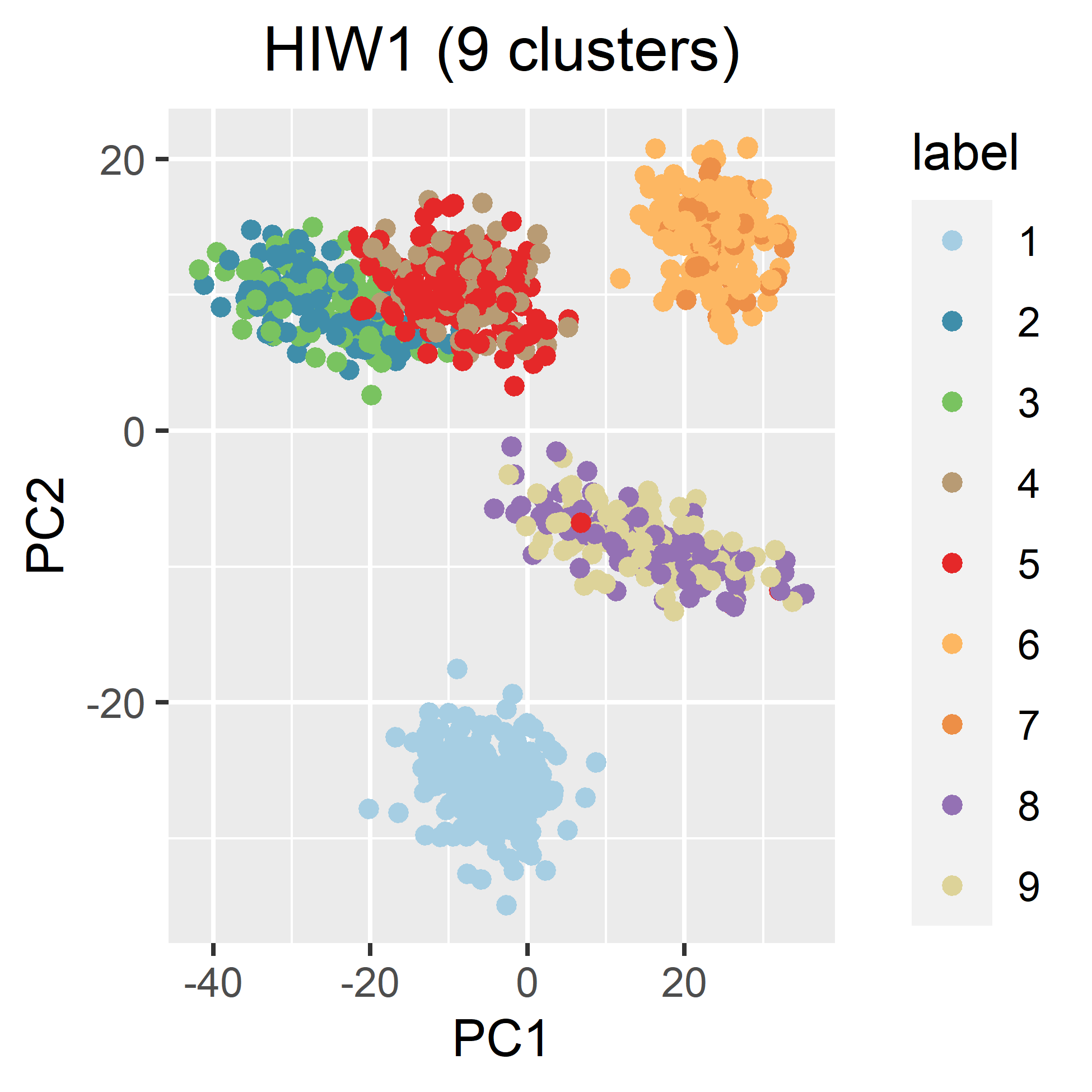} \\
	\includegraphics[width=5cm, height=5cm]{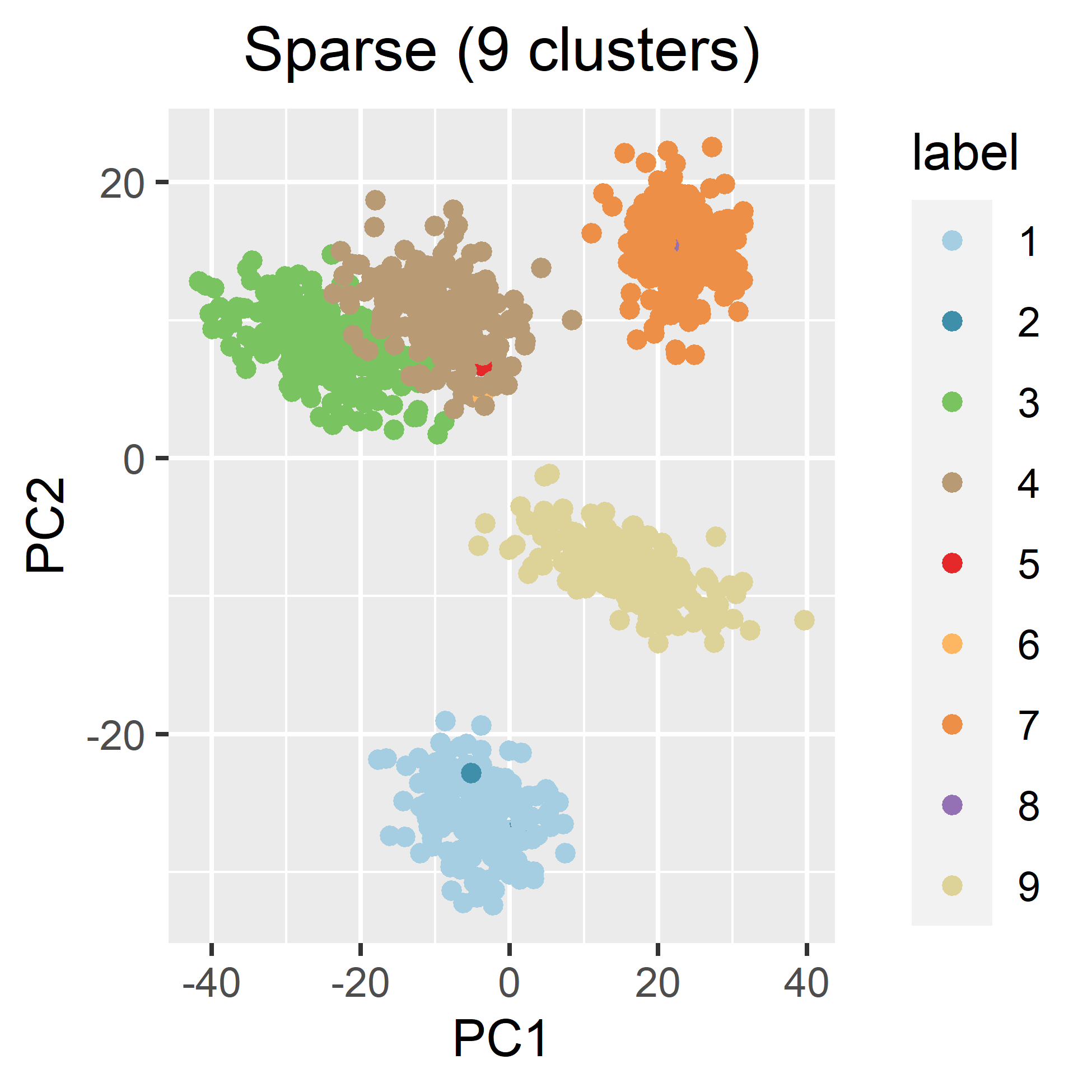} 
	\includegraphics[width=5cm, height=5cm]{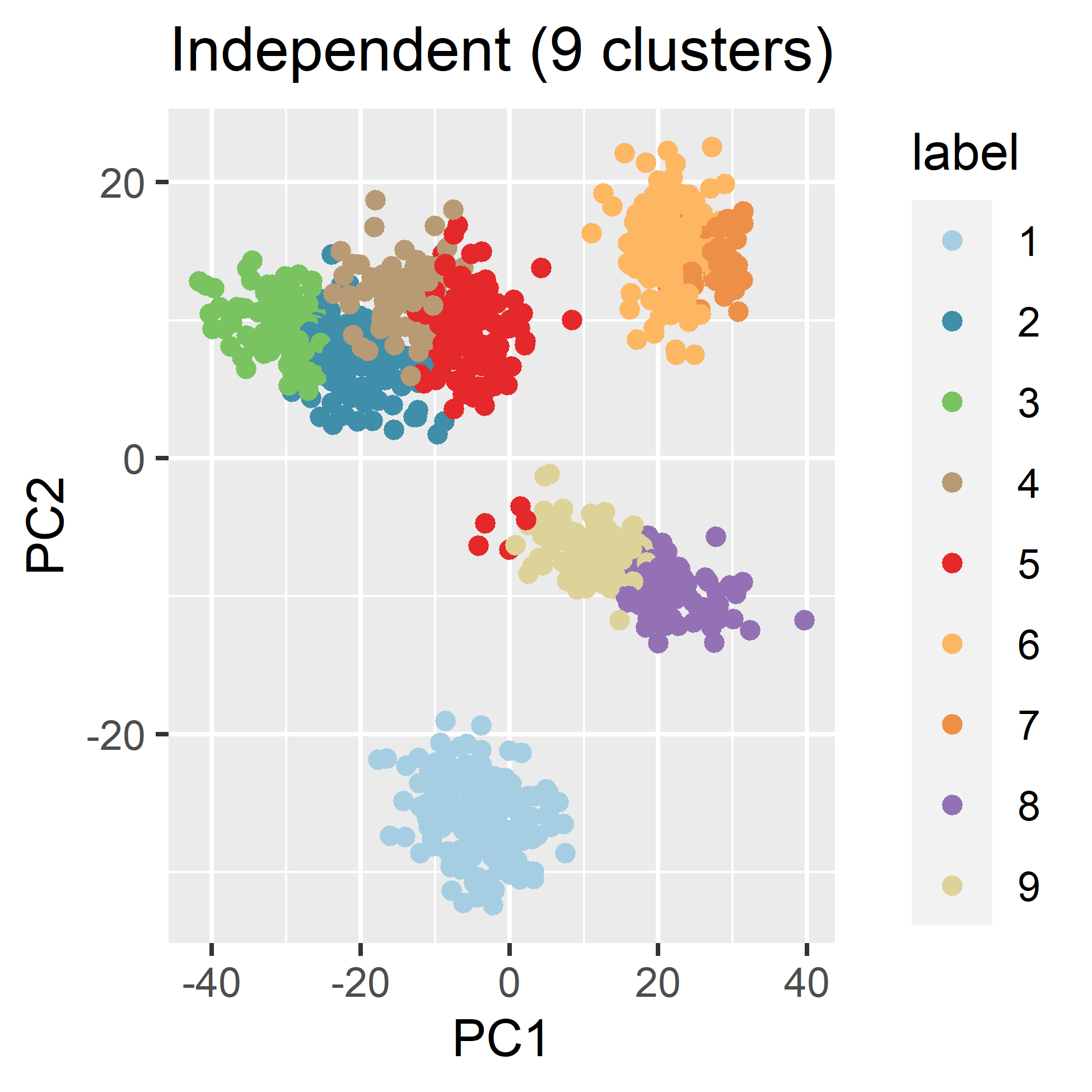} \\
	\vspace{0.5cm}
	\begin{minipage}{0.8\textwidth}
		\caption{Reduced space plots of the true partition and one representative partition for the HIW1, sparse and  independent priors for simulated data IV.}
		\label{fig:6.2.6}
	\end{minipage}
\end{figure}

The number of identified clusters when the IW and HIW2 priors are employed is usually lower than the true number of clusters since true clusters are combined together. This leads to unsatisfactory performance in terms of adjusted Rand index. The independent prior consistently recognizes more clusters than the truth, as it splits true clusters into clusters of substantial size. This is illustrated in Figure \ref{fig:6.2.6}, where representative partitions corresponding to the HIW1,  sparse and independent priors are shown, when all three priors identify $9$ clusters for simulated data IV. 

The log prior is the worst performer. The number of identified clusters can either be small or very large (beyond 20 clusters). The adjusted Rand indices for both data III and IV are volatile and low in value. The bad performance is partly due to the poor approximation of the likelihood in (\ref{eq:37}). The approximation assumes that $\mu_{c}$ is known and, furthermore, only uses up to the quadratic term in the Taylor expansion. This leads to notably bad performance when the dimensionality increases in the context of the DPMM. 

Overall, the sparse prior is the best performer in terms of the adjusted Rand index and the profile of the generated clusters. For simulated data III, its Rand indices always equal one and for simulated data IV they are always very close to one. The Rand indices for the separation prior are usually close to one, except for two outliers for simulated data IV. Such low outliers in the box-plot typically occur when true clusters are combined together. There is more variability in the performance of the HIW1 prior compared to the sparse and separation priors. The superfluous clusters generated by the HIW1 are not clearly defined; see Figure \ref{fig:6.2.6} for instance. The sparse prior also identifies superfluous clusters, but its behaviour is different. For example, when the sparse prior identifies $9$ clusters, superfluous clusters  only contain one or two subjects per cluster; see Figure \ref{fig:6.2.6}. Thus, the posterior partition is similar to the true partition with adjusted Rand index close to one. 

It is worth noting that the HIW2 performs worse than the HIW1 possibly because when $R_{0}$ is restricted to diagonal, some information from the data can not be incorporated into $R_{0}$. This can be seen by comparing the conditional posteriors (Supplemental material, Section S5), where $R_{0}$ is updated using all the information in $T_{c}$ (for $c \in A$) but $\delta_{j}$ is updated using only the information in the diagonal of $T_{c}$, for $j = 1,...,J$. 

\begin{figure}[h!]
	\centering
	\includegraphics[width=5cm, height=5cm]{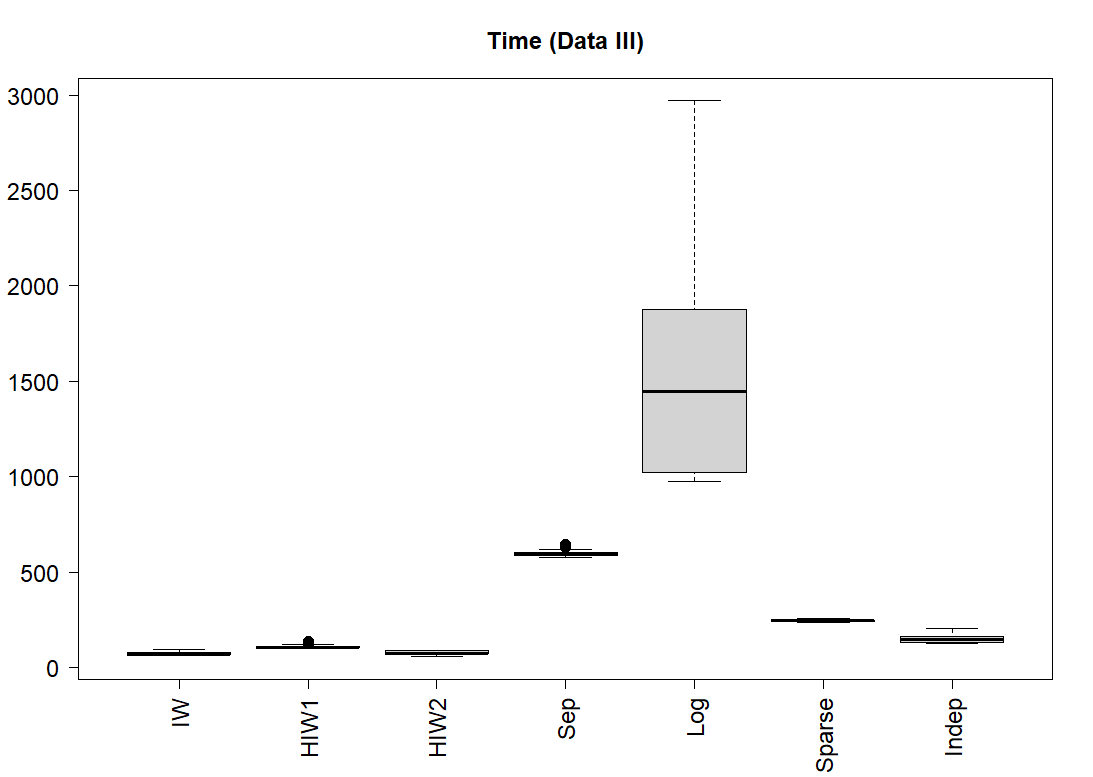} 
	\includegraphics[width=5cm, height=5cm]{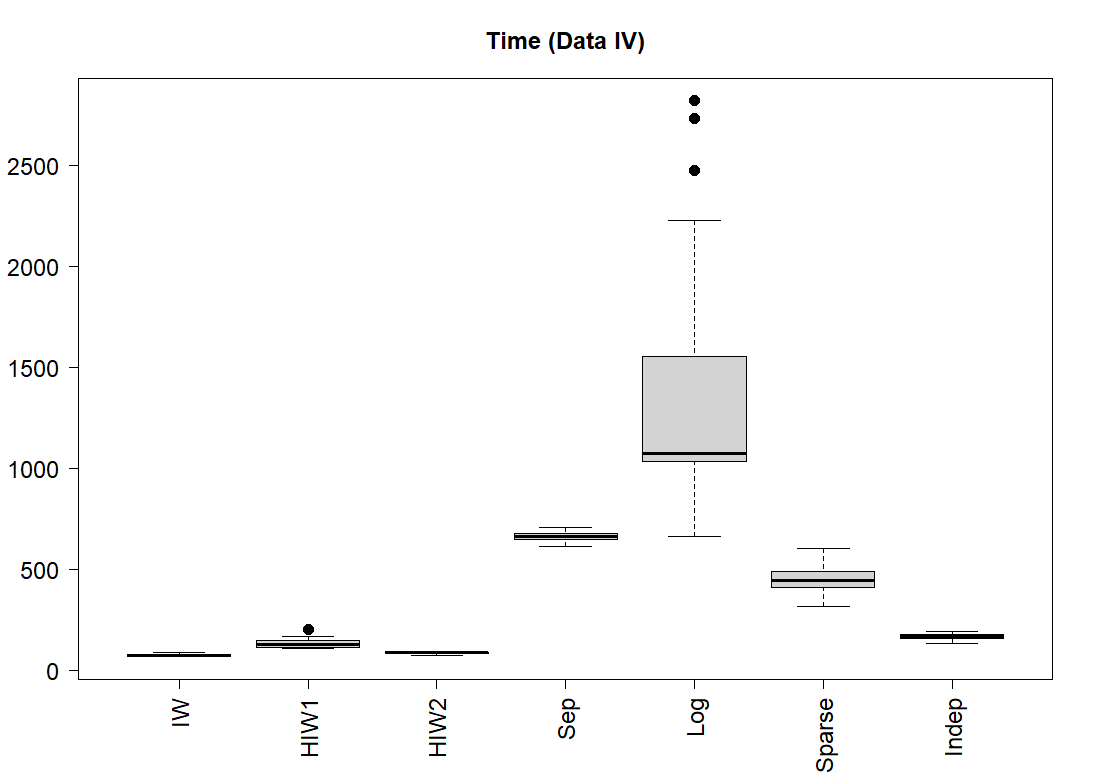} \\
	\vspace{0.5cm}
	\begin{minipage}{0.8\textwidth}
		\caption{Computational time for the 20 datasets simulated according to Table \ref{tab:6.2.1} (Data III) on the left and Table \ref{tab:6.2.2} (Data IV) on the right in seconds.}
	\label{fig:6.2.7}
	\end{minipage}
\end{figure}

Figure \ref{fig:6.2.7} displays the box plot of run-times for each simulated dataset. The log prior is much slower than the others because of the complexity in  calculating the precision matrix $Q_{c}$ of $a_{c}$, which is of order $O(J^4)$. Therefore, as $J$ increases, it becomes impractical to use the log prior, not to mention its poor performance. The separation prior is the second slowest because the elements in $T_{S{c}}$ are updated element by element using Metropolis-Hastings. Although the time required by the sparse prior is about $2$ to $3$ times the time required by HIW1, this still allows for the implementation of the sparse prior in practice. The IW is usually the quickest since its sampler is the simplest. The HIW1 and HIW2 priors are only slightly slower than the IW.  

In the Supplemental material, Section S8, plots of the posterior distributions of $\alpha$, for the different set-ups and datasets are shown. As the value of $\alpha$ is closely related to the number of clusters identified, the bandwidth of the posteriors is closely associated with the results in Figures \ref{fig:6.2.3} and \ref{fig:6.2.4}.

\subsection{Simulations with sparse within-cluster covariance matrices}

The specifications for the two sparse  simulation scenarios are given in Tables \ref{tab:6.3.1} (data V) and \ref{tab:6.3.2} (data VI). Correlation matrices are block diagonal with $5$ blocks. Simulated data V are generated with smaller variances compared to simulated data VI, and thus carry a stronger clustering signal.

\begin{table}[b!]
	\centering
	\caption{Information of the simulated data V with $J=20$. $rep(a,b)$ represents a vector of length $b$ with repeated elements $a$.}
	\begin{tabular}{  m{1.6cm} m{5.2cm} m{1.0cm} m{1.0cm} m{0.8cm} } 
		\hline
		Cluster & $\mu_{c}^{true}$ & $Var_{c}^{true}$ & $\rho_{c}^{true}$& $n_{c}$ \\ \hline
		1 & (rep(3,5),rep(12,5),rep(18,5),rep(12,5)) & 3& 0.7 & 200 \\ 
		2 & (rep(12,5),rep(18,5),rep(3,5),rep(18,5)) & 3 & 0.7 & 200 \\ 
		3 & (rep(18,5),rep(18,5),rep(12,5),rep(8,5)) & 3& 0.7 & 200 \\ 
		4 & (rep(18,5),rep(3,5),rep(8,5),rep(3,5)) & 3 & 0.7 & 200 \\ 
		5 & (rep(8,5),rep(8,5),rep(12,5),rep(3,5)) & 3& 0.7 & 200 \\ \hline	
	\end{tabular}
	\label{tab:6.3.1}
\end{table} 

\begin{table}[b!]
	\centering
	\caption{Information of the simulated data VI with $J=20$.}
	\begin{tabular}{  m{1.6cm} m{5.2cm} m{1.0cm} m{1.0cm} m{0.8cm} } 
		\hline
		Cluster No. & $\mu_{c}^{true}$ & $\Var_{c}^{true}$  & $\rho_{c}^{true}$ & $n_{c}$ \\ \hline
		1 & (rep(3,5),rep(12,5),rep(18,5),rep(12,5)) & 9& 0.7 & 200 \\ 
		2 & (rep(12,5),rep(18,5),rep(3,5),rep(18,5)) & 9& 0.7 &200 \\ 
		3 & (rep(18,5),rep(18,5),rep(12,5)rep(8,5)) & 9&0.7 & 200 \\ 
		4 & (rep(18,5),rep(3,5),rep(8,5),rep(3,5)) & 9& 0.7&200 \\ 
		5 & (rep(8,5),rep(8,5),rep(12,5),rep(3,5)) & 9& 0.7 & 200 \\ \hline		
	\end{tabular}
	\label{tab:6.3.2}
\end{table}

In the Supplemental material, Section S9, visualisation plots for the generated clusters are presented. It is shown that both specifications generate challenging data sets, with the clusters being considerably close for datasets VI. In the same section, we also present two Figures that display the number of clusters recognized for each prior for simulated data V and VI.

\begin{figure}[h!]
	\centering
	\includegraphics[width=5cm, height=5cm]{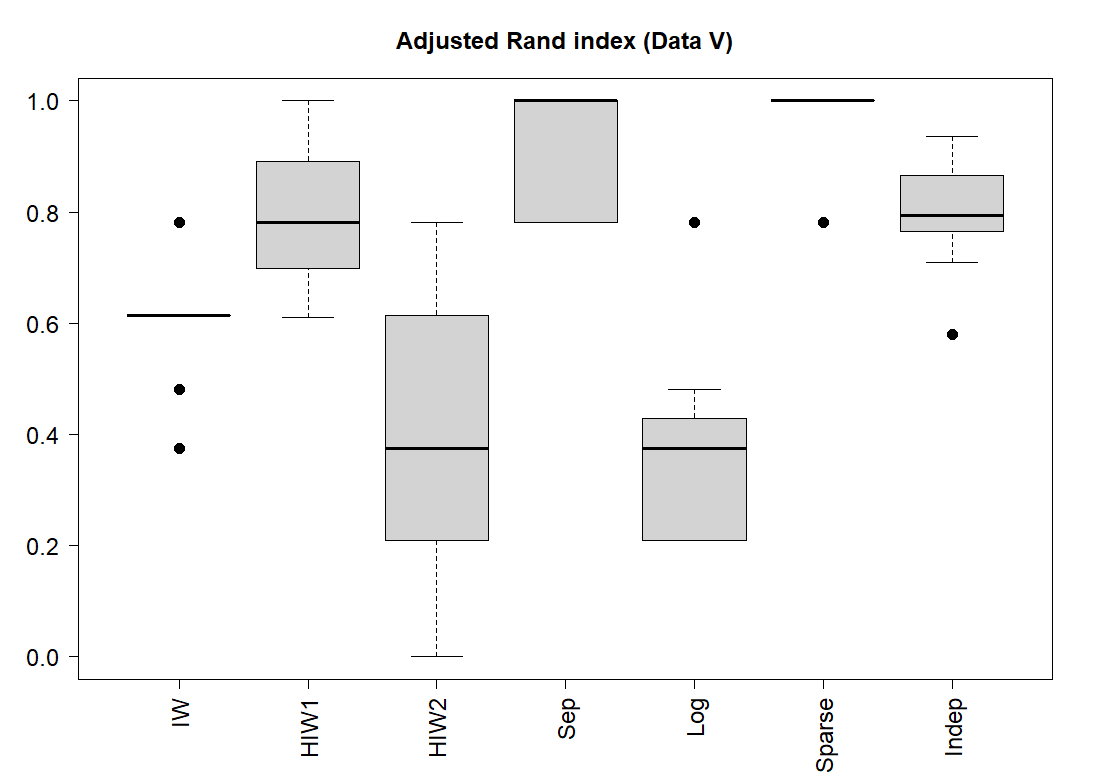} 
	\includegraphics[width=5cm, height=5cm]{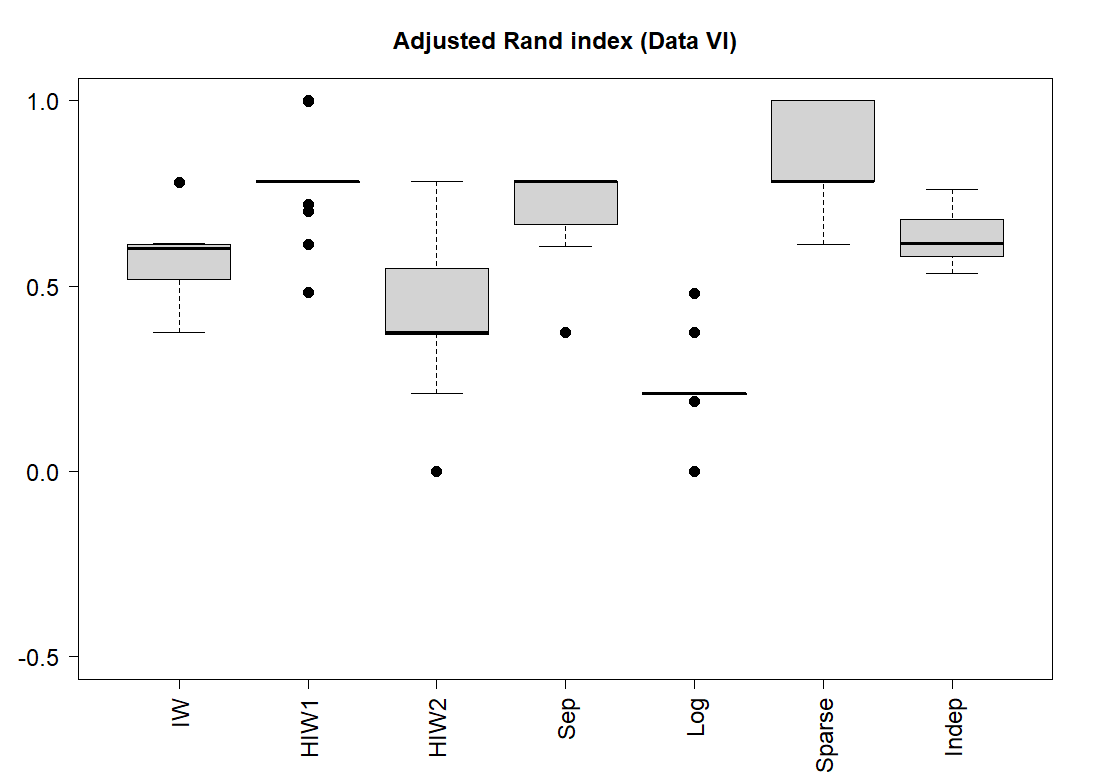} \\
	\vspace{0.5cm}
	\begin{minipage}{0.8\textwidth}
		\caption{Boxplots of the adjusted Rand indices for different priors. The plot on the left corresponds to datasets simulated according to Table \ref{tab:6.3.1} (data V). The plot on the right corresponds to datasets simulated according to Table \ref{tab:6.3.2} (data VI).}
		\label{fig:6.3.4}
	\end{minipage}
\end{figure}

Boxplots of the adjusted Rand indices for each prior are shown in Figure \ref{fig:6.3.4}. The log and HIW2 priors are the worst performers, as they only identify two clusters for both data V and VI most of the time.

The sparse prior is the best performer overall. It recognizes $5$ groups more often than all other priors for both settings. Adjusted Rand index values are consistently 1 for datasets V, as the model identifies true clusters perfectly, with three exceptions when it combined clusters. For the simulated data VI, adjusted Rand indices become more volatile, however, values are higher than those from the other priors on average.

When the clustering signal is strong, the separation prior also identifies $5$ clusters quite often. However, its performance deteriorates greatly for a weaker signal. Adjusted Rand indices exhibit more variability than the sparse prior, as it combines true clusters more often. 
The HIW1 prior is not performing as well as the separation prior when the signal is relatively strong. When the signal becomes weak, however, it outperforms the separation prior. Finally, the independent prior again splits true clusters, which results in more groups identified in the final clustering. 


Inferences on computational speed and posterior distributions for $\alpha$ are shown in the Supplemental material, Section S9, as they are similar to those in the previous subsection. Note that stability regarding the posteriors of $\alpha$ is not sufficient to indicate whether the DPMM generates sensible and stable partition results. 

Further checks on the convergence of the MCMC chains were made. Two simulated datasets were randomly selected, one from data III and one from data V. We obtained trace plots for the adjusted Rand index, number of clusters and $\alpha$, after running 6 MCMC chains with different initialisations. We also considered the posterior similarity matrices resulting from the different initialisations. For both data sets, the diagnostic plots indicated very good convergence for the Separation and Sparse priors, noting that the trace plot for the number of clusters under the Separation prior (data V) showed stickiness at 4 or 5 clusters. The diagnostic plots indicated that convergence was a problem under the other priors, with the HIW1 performing better compared to the HIW2, IW, Independent and Log priors. The fact that convergence was best achieved when the Separation and Sparse priors were used demonstrates that the choice of prior may not only improve on the modelling, but also on the convergence of the corresponding MCMC sampler, especially for challenging datasets such as the ones we simulated. Although the choice of prior appears to be crucial for achieving convergence and valid clustering outcomes, a different type of sampler (e.g a marginal sampler with split/merge steps) may be a plausible additional tool for addressing the problems described in Section 1. This investigation is beyond the scope of this manuscript.  Output from the convergence checks for one dataset is included in Section S9 in the Supplemental material. 

\section{The role of sample size and initial values}

Sample size is particularly important for generating sensible clustering results with mixture modelling, especially when the signal in the data is less strong. Sample size can be viewed as large or small after considering the cluster sizes $n_{c}$ in relation to the number of parameters to be estimated. 


Consider the previously specified data VI, but with an increased sample size of $n_{c}=500$. We refer to this new specification as data VII.   
In Section S10 in the Supplemental material we display bar plots of the number of clusters identified by different priors. We also present boxplots of the adjusted Rand indices for different priors. Except for the independent prior, the tendency to combine true clusters is alleviated for all priors. True clusters are identified more frequently, which translates to higher adjusted Rand indices.  For the sparse prior, it can be observed that the adjusted Rand indices are always close to one, which is similar to the case when the signal is strong in simulated data V. Therefore, empirically, with a larger sample size, the resulting partitioning improves significantly. Note that, for the sparse prior, even with a sample size of $500$ per cluster, the clustering results sometimes contain small clusters consisting of one or two subjects. This may be due to the tendency of the DPMM to create extra small clusters; see \cite{miller2014inconsistency} and \cite{miller2017mixture} for more details. As the conditional independence assumption does not hold, the independent prior identifies even more clusters when the sample size increases and adjusted Rand indices are lower than those for the simulated data VI.  

Starting clustering allocation values for the DPMM sampler are also important. (This is true for other numerical approaches too, such as the EM algorithm; see \cite{fraley2007bayesian}.) Typically, there is a discrepancy between the overall covariance matrix and the within-cluster covariance matrices. When the DPMM is randomly initialized in terms of cluster allocation, the starting values of $\Sigma_{c}$ will be similar to the overall covariance matrix. Then, when the signal is not strong, or the sample size is relatively small,  clustering results can be unstable. One remedy is to generate more non-empty clusters than expected when randomly initializing the algorithm, as suggested by \cite{hastie2015sampling}. One could also initialize the DPMM using some distance-based approach. For instance, one may use k-means, one of the proposed approaches in \cite{fruhwirth2018handbook} where the initialization of the sampler for the sparse finite mixture model for large $J$ is discussed.  Such approaches usually require to set a fixed number of clusters. Since the goal is to make the starting values of $\Sigma_{c}$ smaller, one can set the number of clusters to be much larger than the maximum number of clusters expected in the data. Starting values for $\Sigma_{c}$ are then more likely to be reasonable. Based on our experience when using this approach to initialize the DPMM sampler, the clustering results can indeed be more stable most of the time. However, when the distance-based approach is not suitable given the shape of the clusters or the nature of the problem in general, this will reduce its effectiveness. \cite{fruhwirth2018handbook}
observed lack of mixing when initializing using k-means in high-dimensionality, and proposed to set the prior expected within-cluster variances and covariances to be larger as $J$ increases. This suggests that different initialization approaches can be relevant to specifying hyperparameter values for the prior of covariance matrices. 

For the analyses in this manuscript we have not used a distance-based approach to derive starting values, as we wanted to focus purely on the comparative performance of the prior specifications under standard random initialisation. 

\section{Analysis of a real dataset}

To further assess the priors' performance a real data set is analysed where true labels are known. The dataset is taken from the UCI Machine Learning Repository and is part of the RNA-Seq PAN-cancer dataset. The dataset contains 20,531 covariates and 801 subjects. Covariates contain gene expression measurements from patients with five distinct types of tumors, BRCA, KIRC, COAD, LUAD and PRAD. The cancer labels form the benchmark partition for this analysis. 
 Since the number of covariates $J$ is much larger than the number of subjects $n$, it is necessary to pre-select genes. All genes with zero observations were removed, reducing their number to 12,356. Genes with  strong associations with the labels were selected, to effect a relatively strong clustering signals. Welch's t-test was adopted as it is  robust under skewed distributions \citep{fagerland2012t, fagerland2009performance}. Boxplots of resulting p-values are shown in the Supplemental material, Section S11. We selected the $20$ genes with the smallest corresponding p-values. The Shapiro-Wilk test for Normality (p-values are shown in the Supplemental material Section S11) indicated that the observations are not normally distributed. This allows to draw inferences on the behaviour of the different priors when the normality assumption is not satisfied.   

%

\begin{figure}[h!]
	\centering
	\includegraphics[width=5cm, height=5cm]{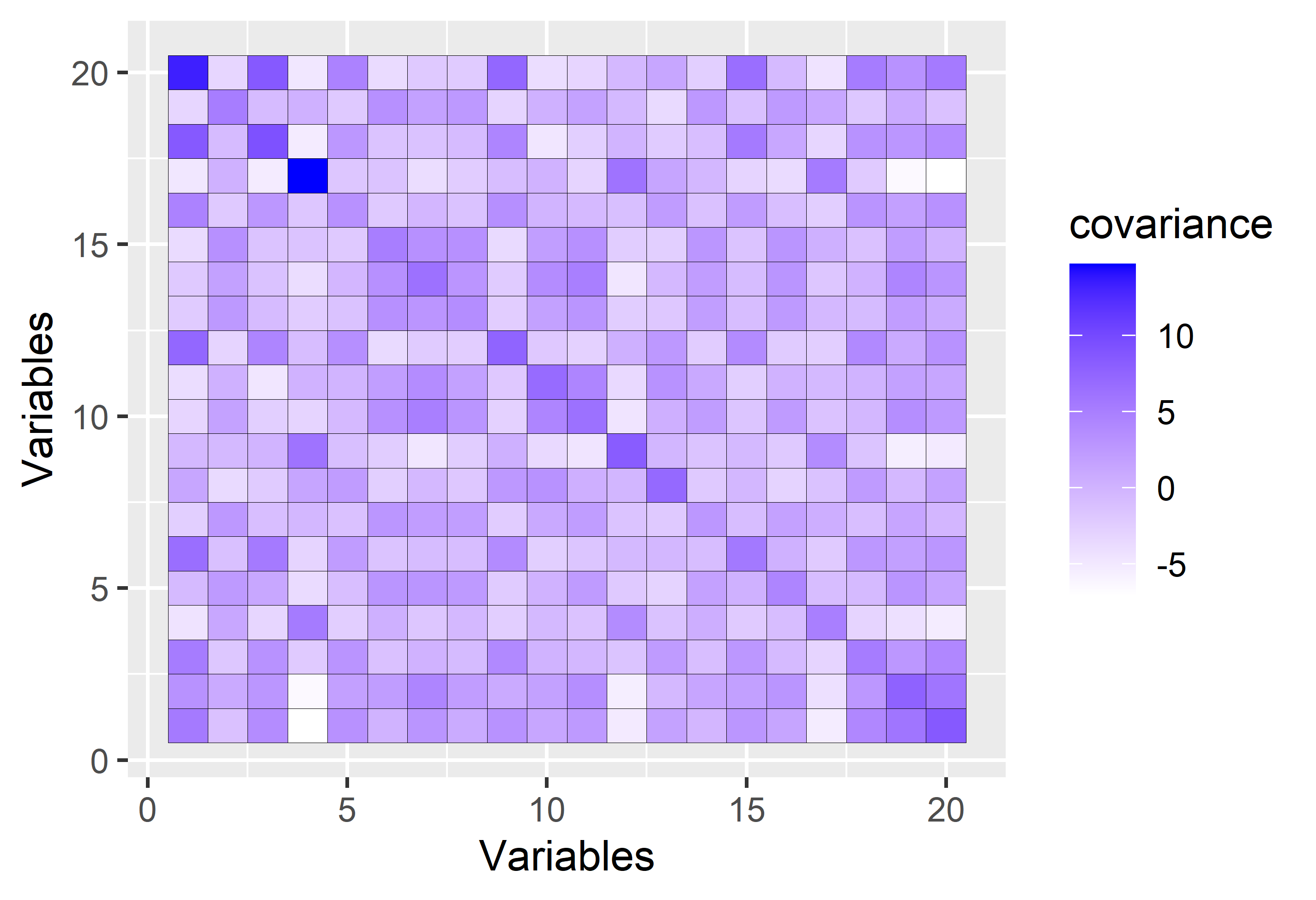} 
	\includegraphics[width=5cm, height=5cm]{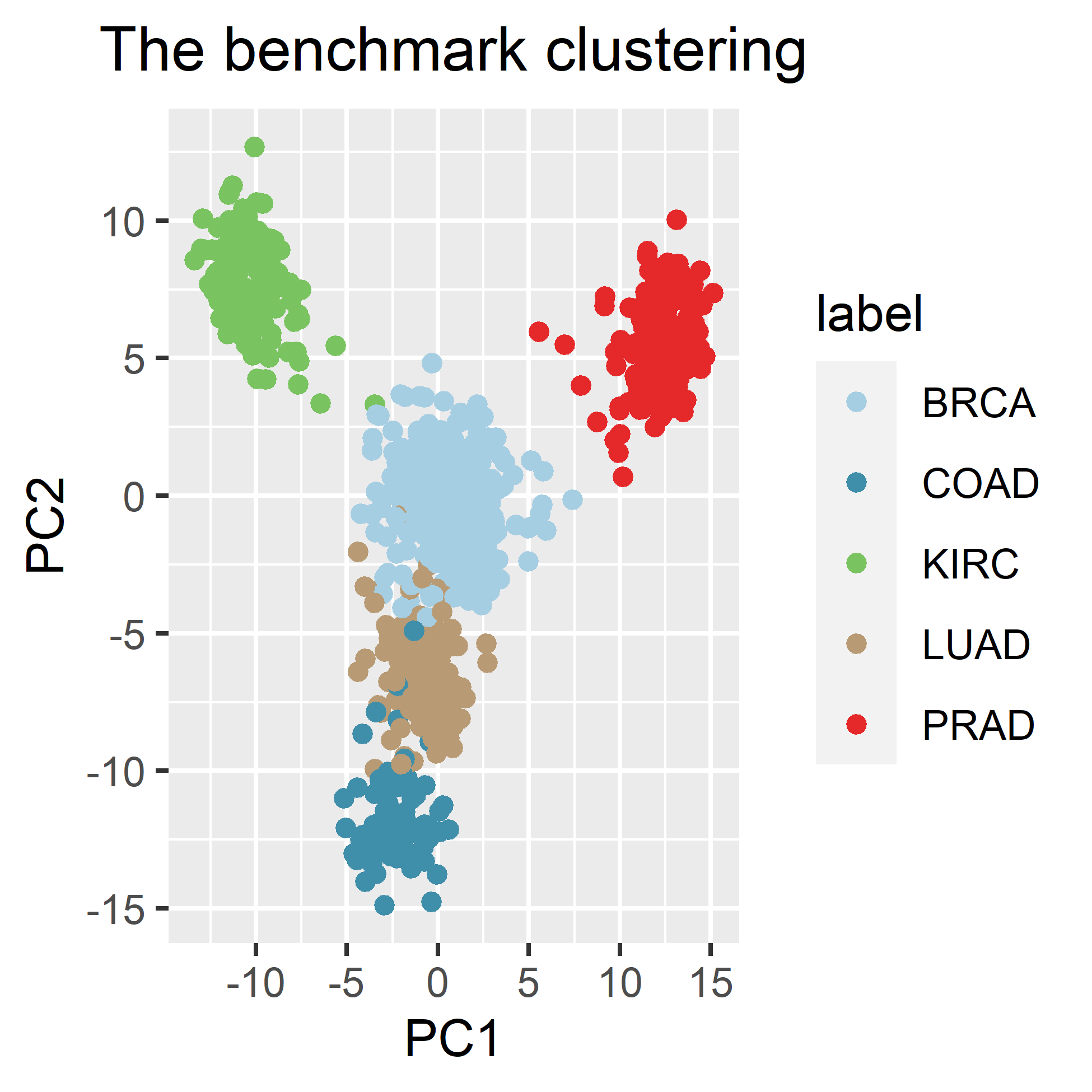} \\
	\includegraphics[width=5cm, height=5cm]{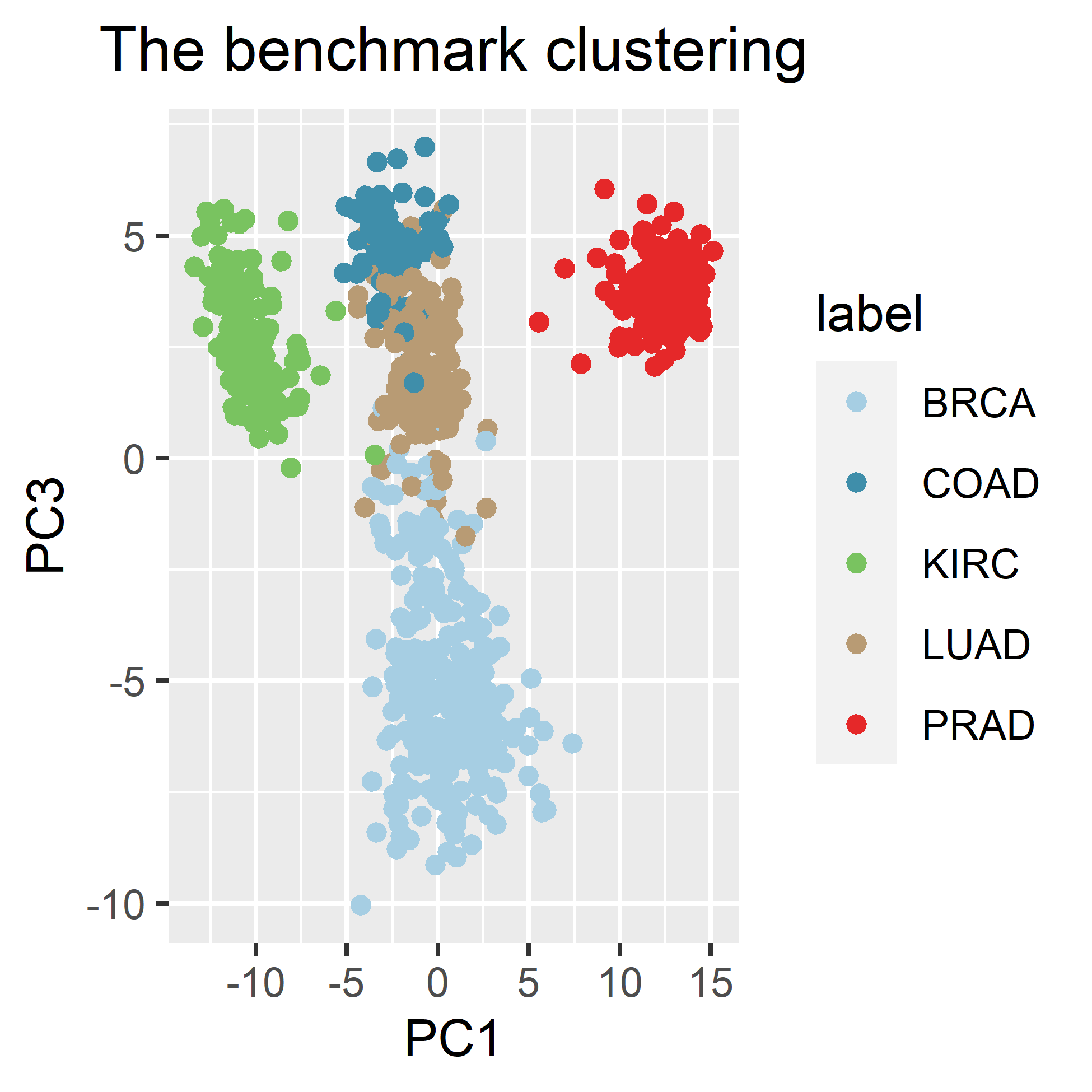} 
	\includegraphics[width=5cm, height=5cm]{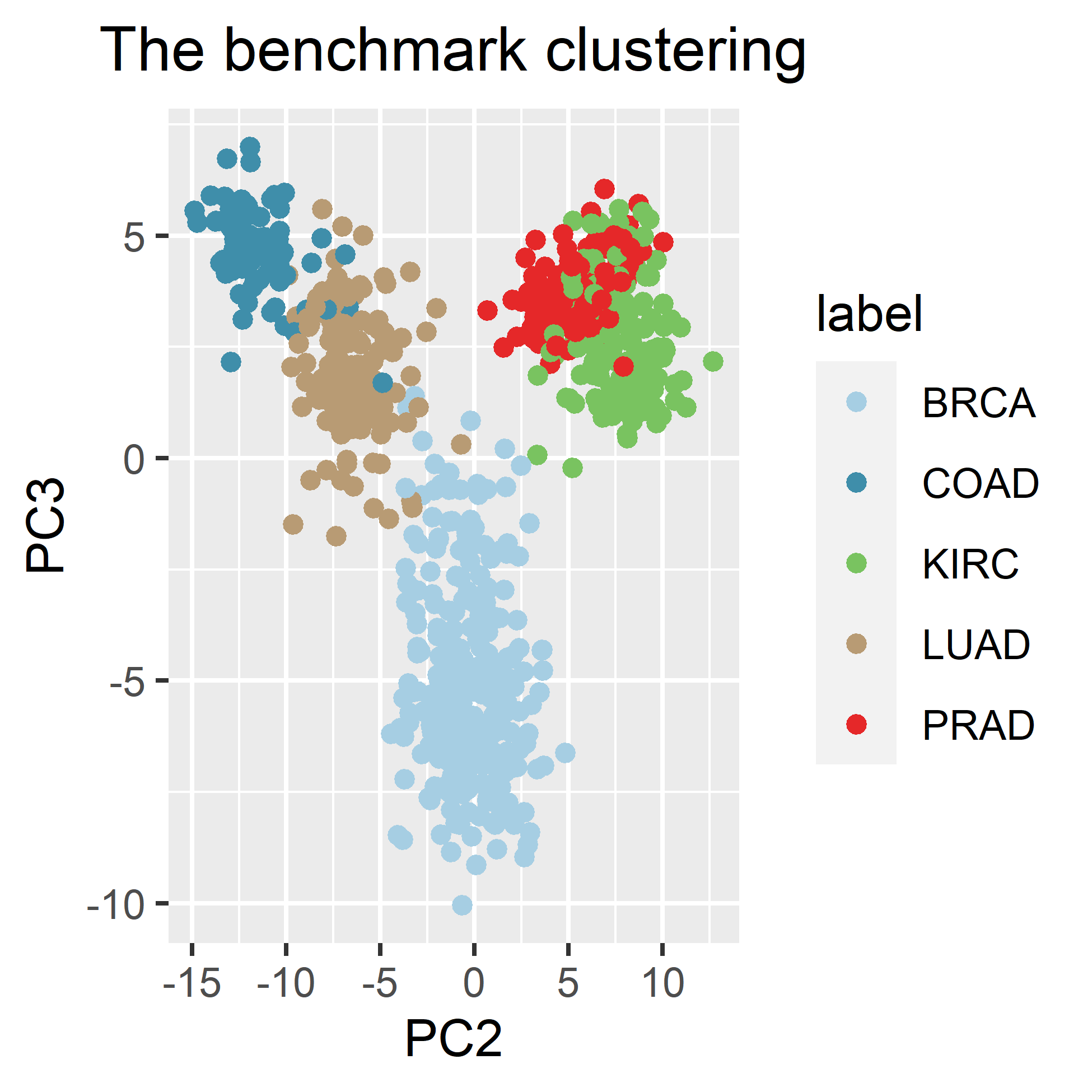} \\
	\begin{minipage}{0.8\textwidth}
		\caption{Overall correlation matrix of the 20 selected variables on the top-left. Remaining illustrations show reduced space plots of the benchmark clustering according to the cancer labels.}
		\label{fig:8.2}
	\end{minipage}
\end{figure}

\begin{figure}[h!]
	\centering
	\includegraphics[width=5.0cm, height=5cm]{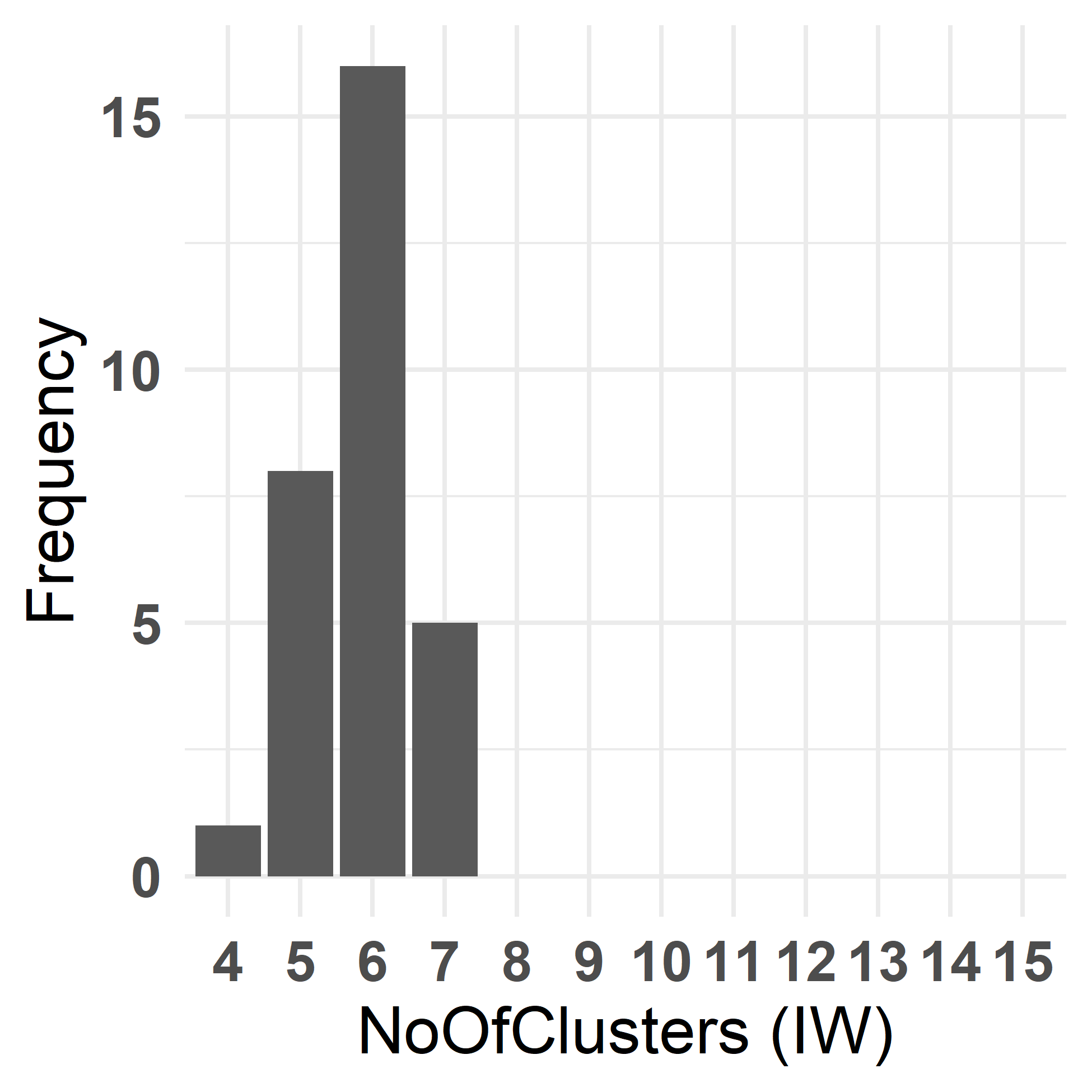} 
	\includegraphics[width=5.0cm, height=5cm]{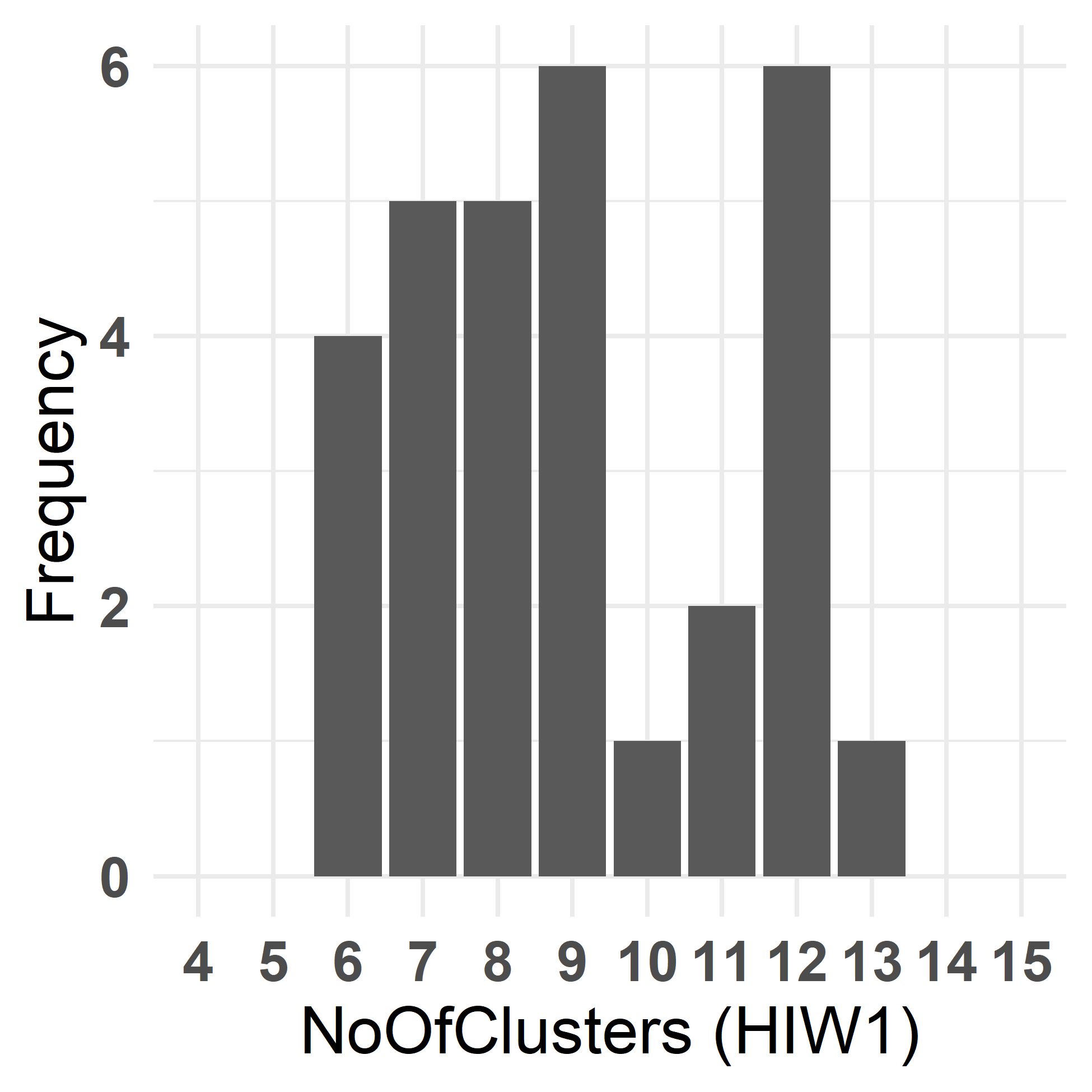} \\
	\includegraphics[width=5.0cm, height=5cm]{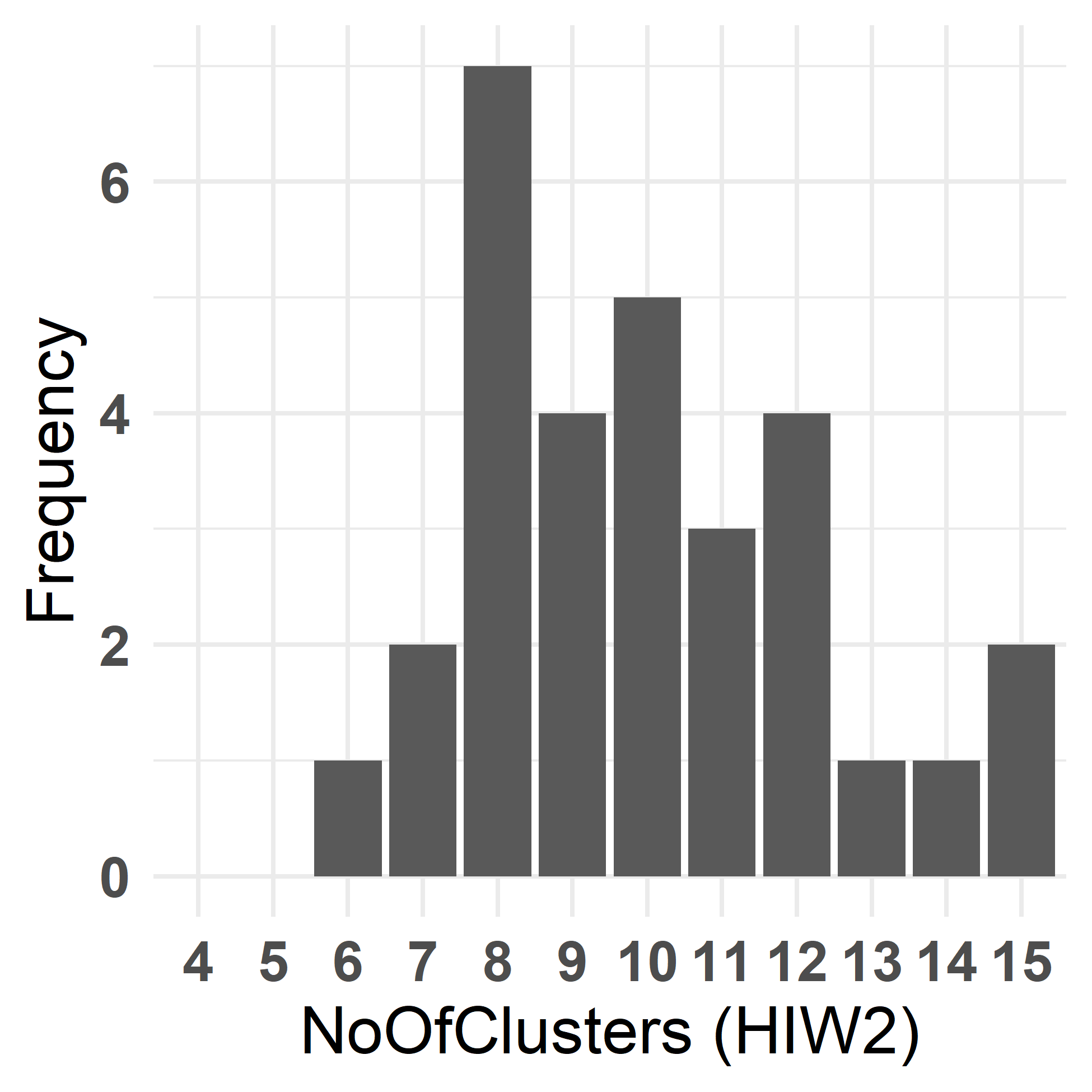} 
	\includegraphics[width=5.0cm, height=5cm]{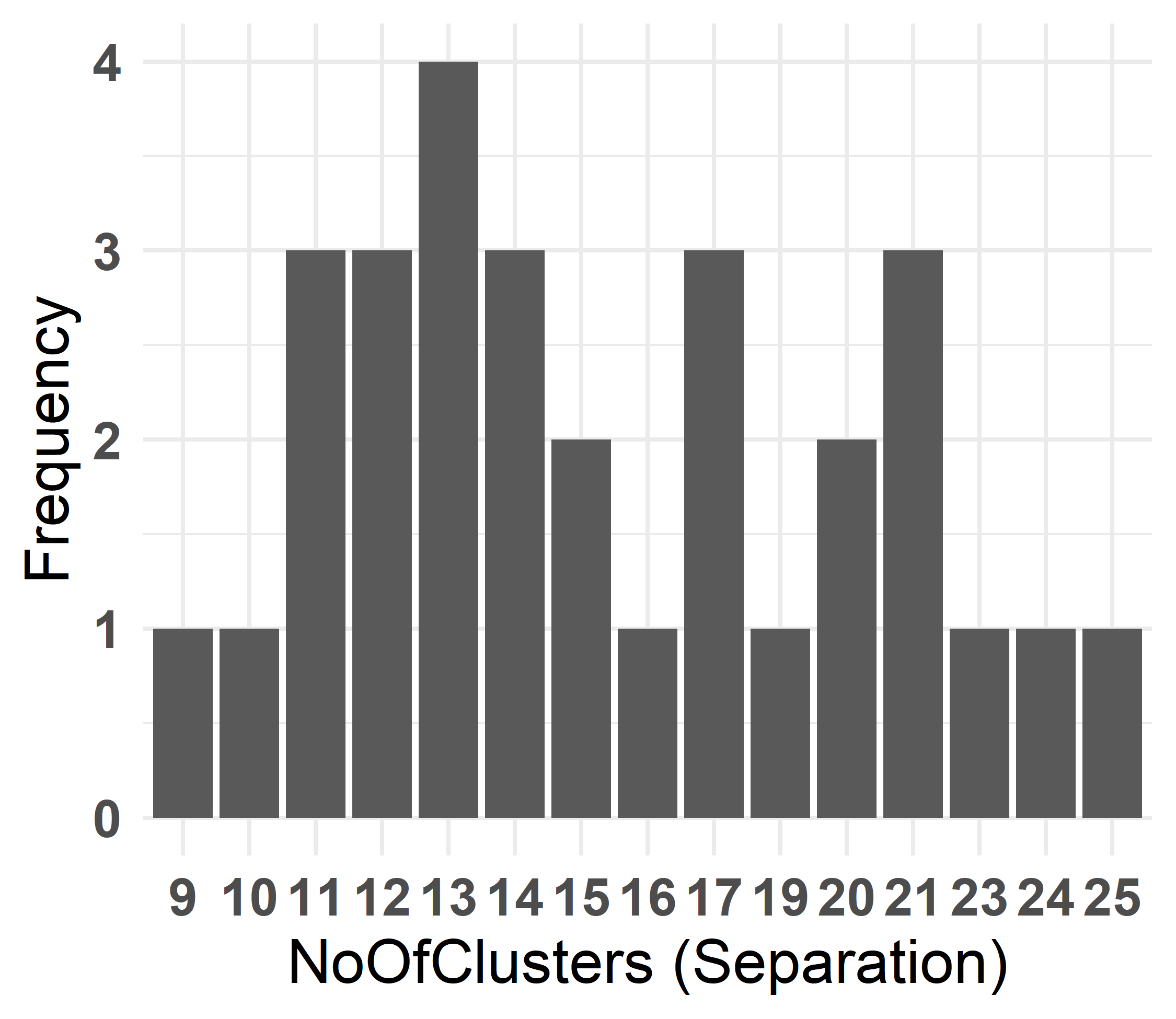} \\
	\includegraphics[width=5.0cm, height=5cm]{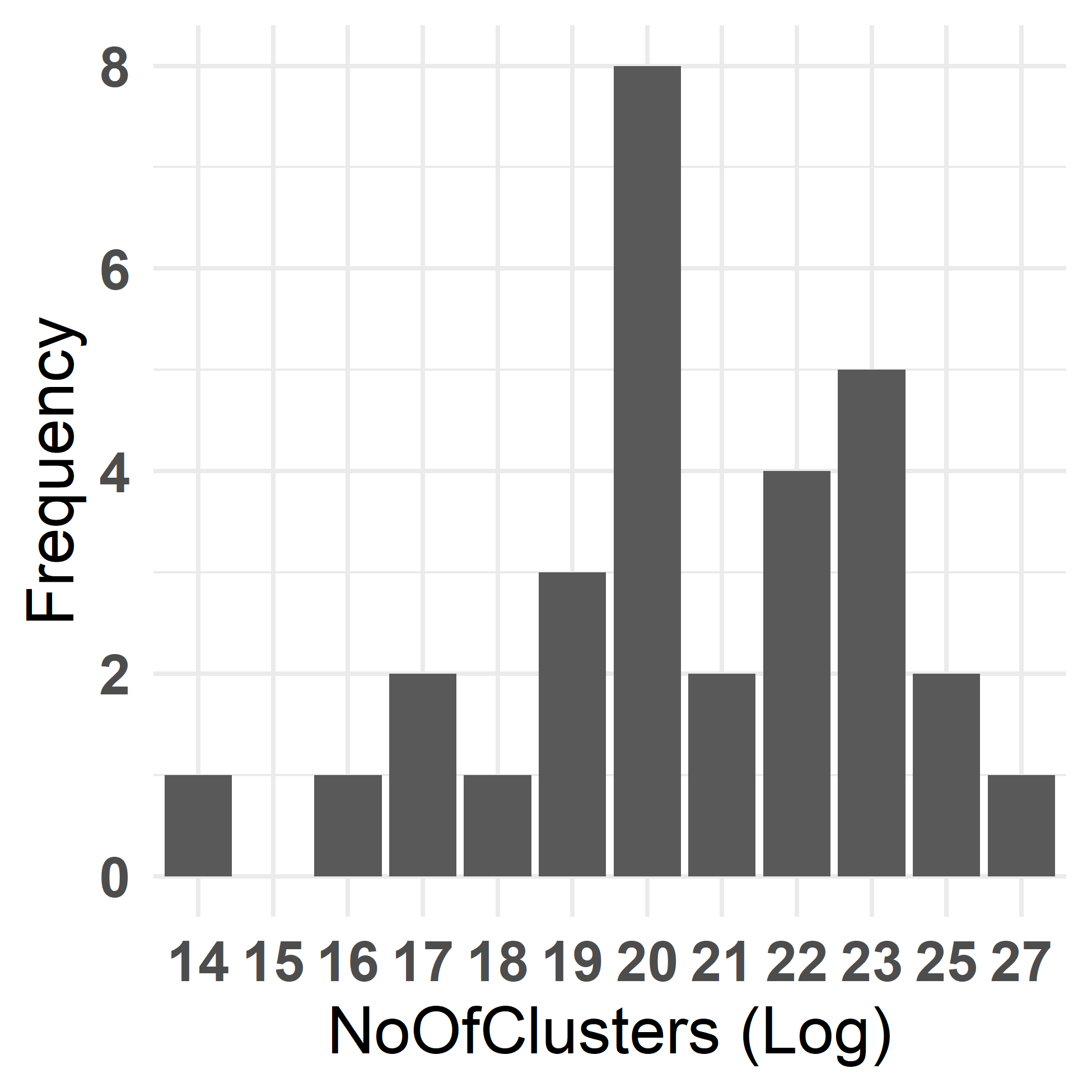} 
	\includegraphics[width=5.0cm, height=5cm]{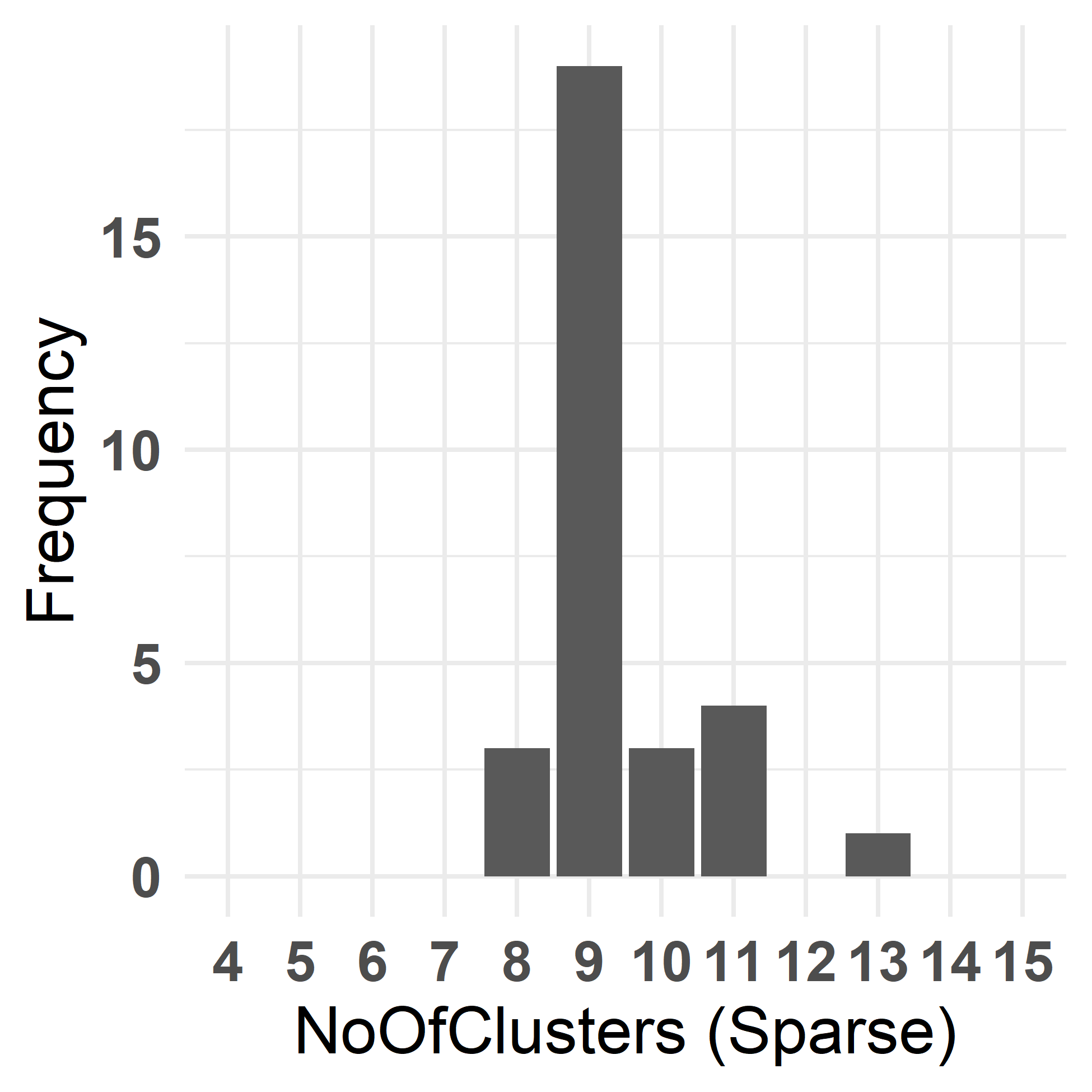} \\
	\includegraphics[width=5.0cm, height=5cm]{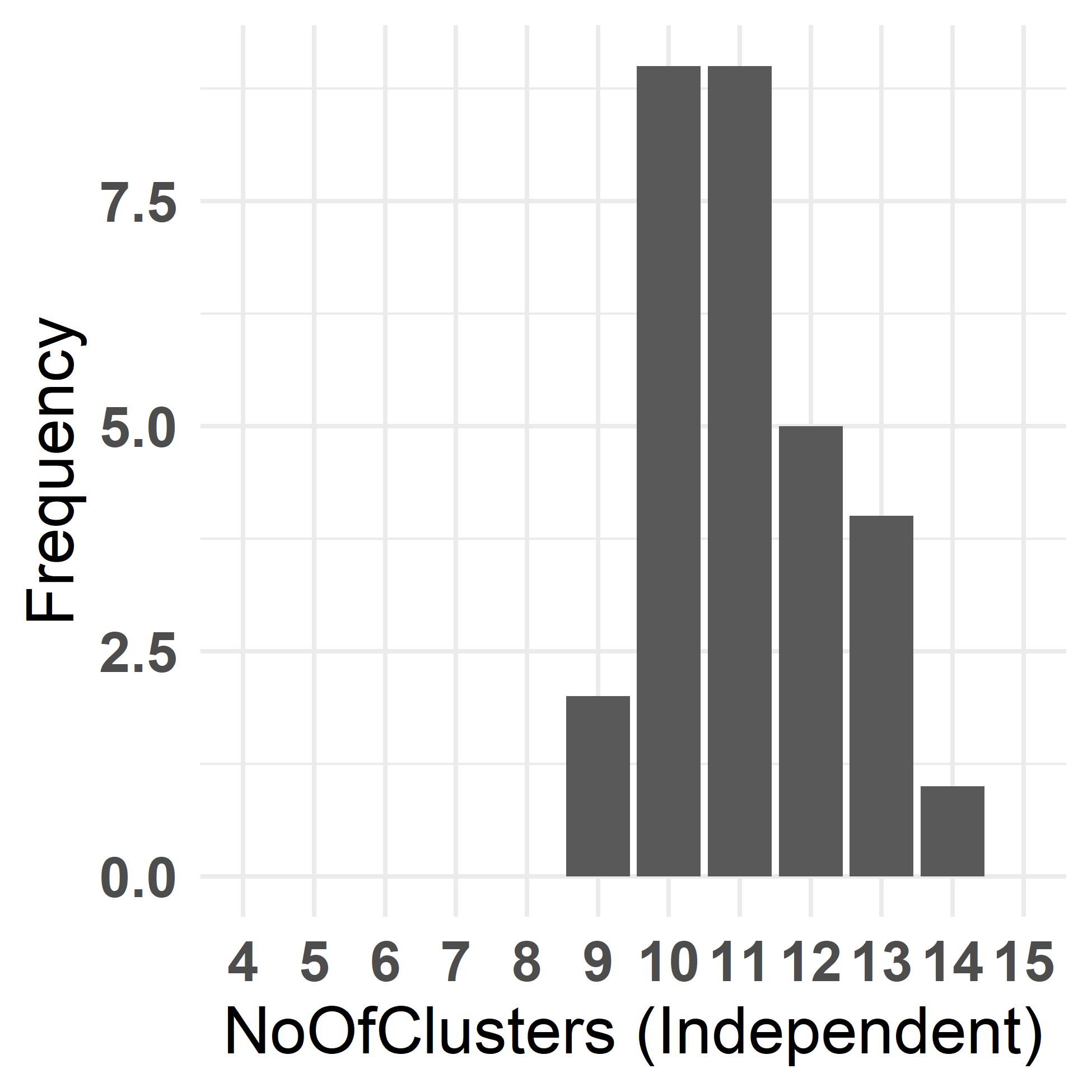} \\
	\vspace{0.5cm}
	\begin{minipage}{0.8\textwidth}
		\caption{Bar plots of the number of identified clusters for 30 different initializations. The prior used is shown within the brackets in the x-axis label.}
		\label{fig:8.3}
	\end{minipage}
\end{figure}

The overall covariance matrix and the benchmark partition are displayed in Figure \ref{fig:8.2}. The first three principle components are used to visualise the data, as they explain 78\% of the total variability. 
We run the sampler repeatedly for $30$ different starting points for each prior, to assess stability, for a burn-in period of $10,000$ iterations, followed by $6,000$ iterations. The induced partitions were stable after the burn-in. Figure \ref{fig:8.3} shows the number of final clusters recognized for each prior given the different initialisations. As the variables are not normally distributed, the number of final clusters identified is usually larger than the benchmark and less stable compared to the simulated data analyses.  

\begin{figure}[h!]
	\centering
	\includegraphics[width=5cm, height=5cm]{pic_main/real-true-2} 
	\includegraphics[width=5cm, height=5cm]{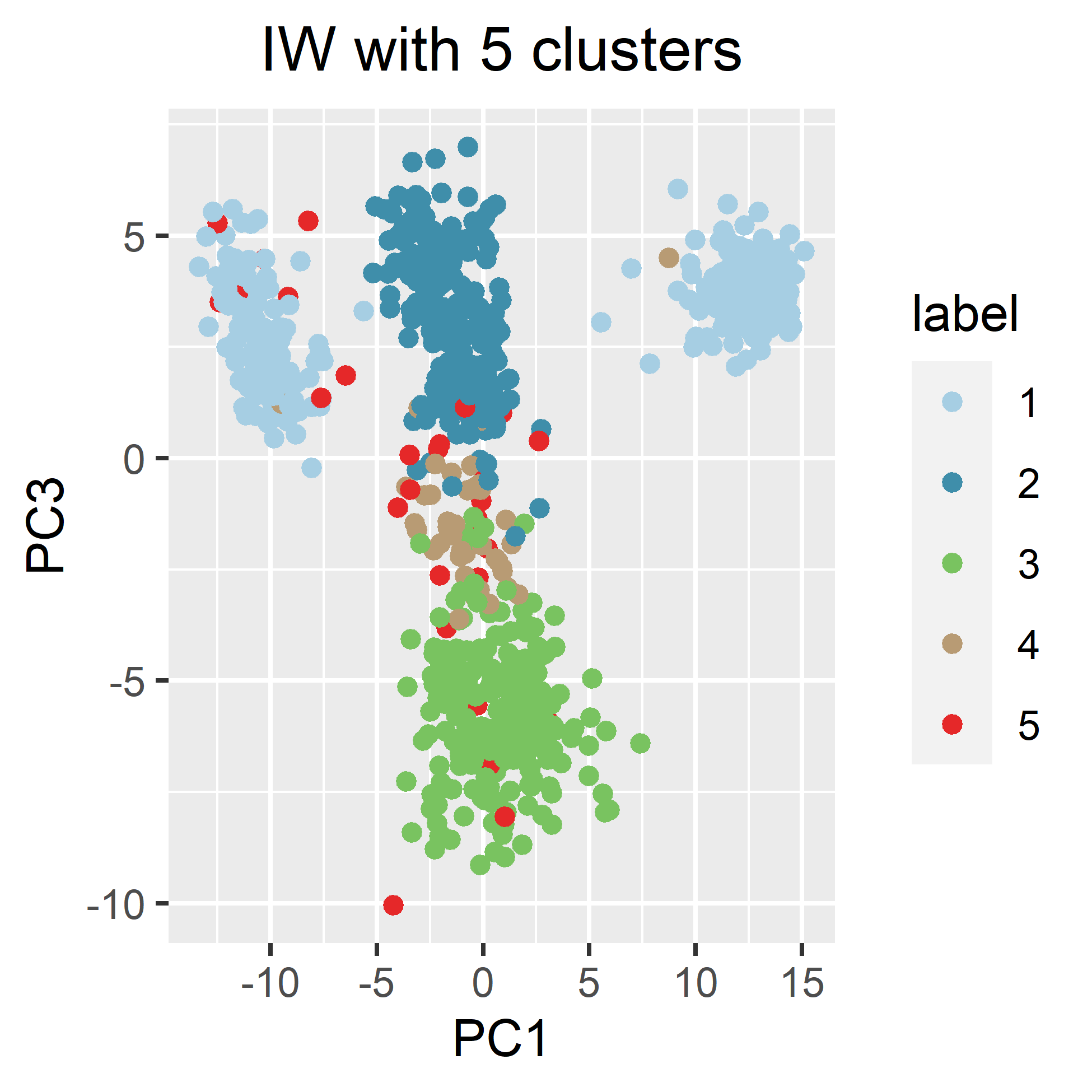} \\
	\includegraphics[width=5cm, height=5cm]{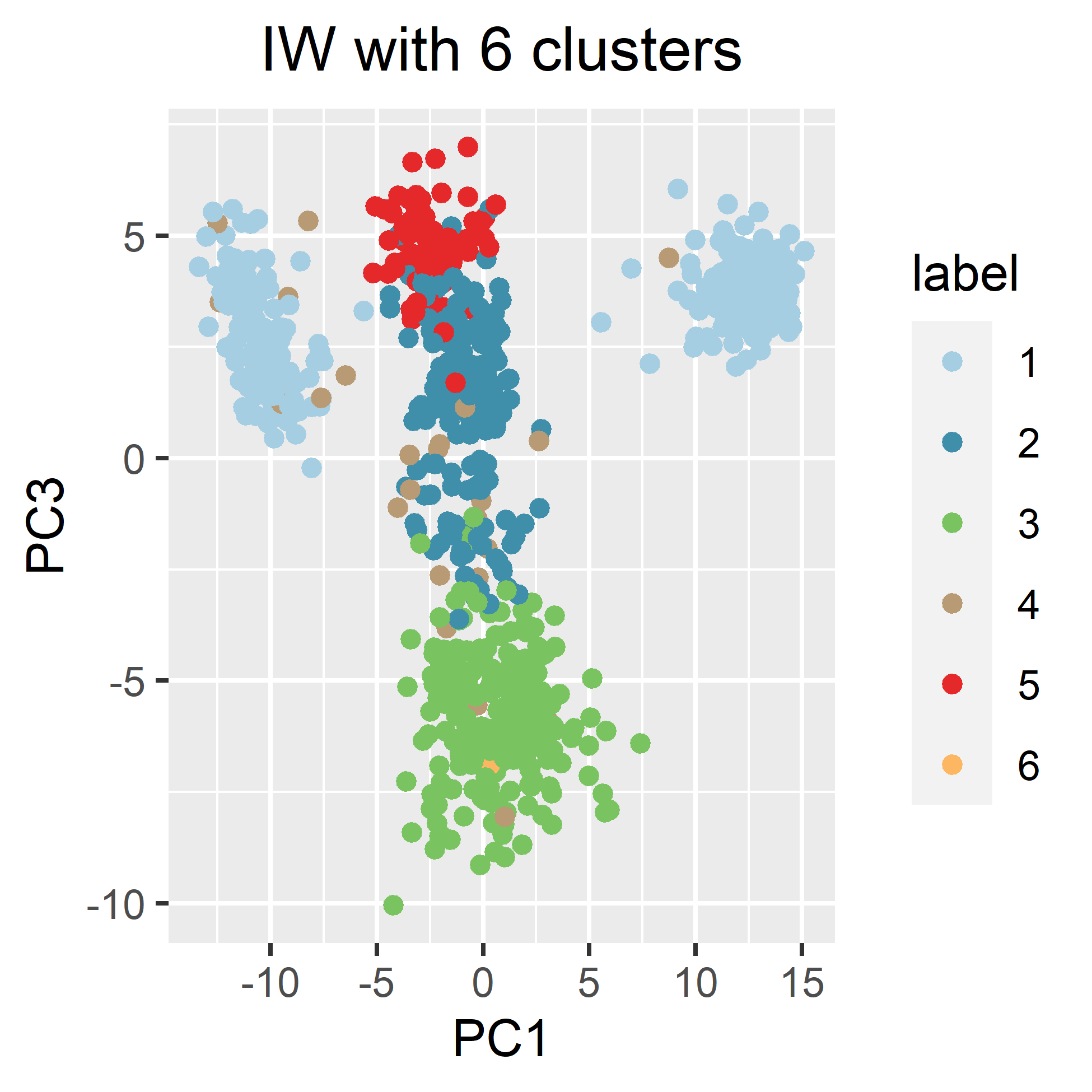} 
	\includegraphics[width=5cm, height=5cm]{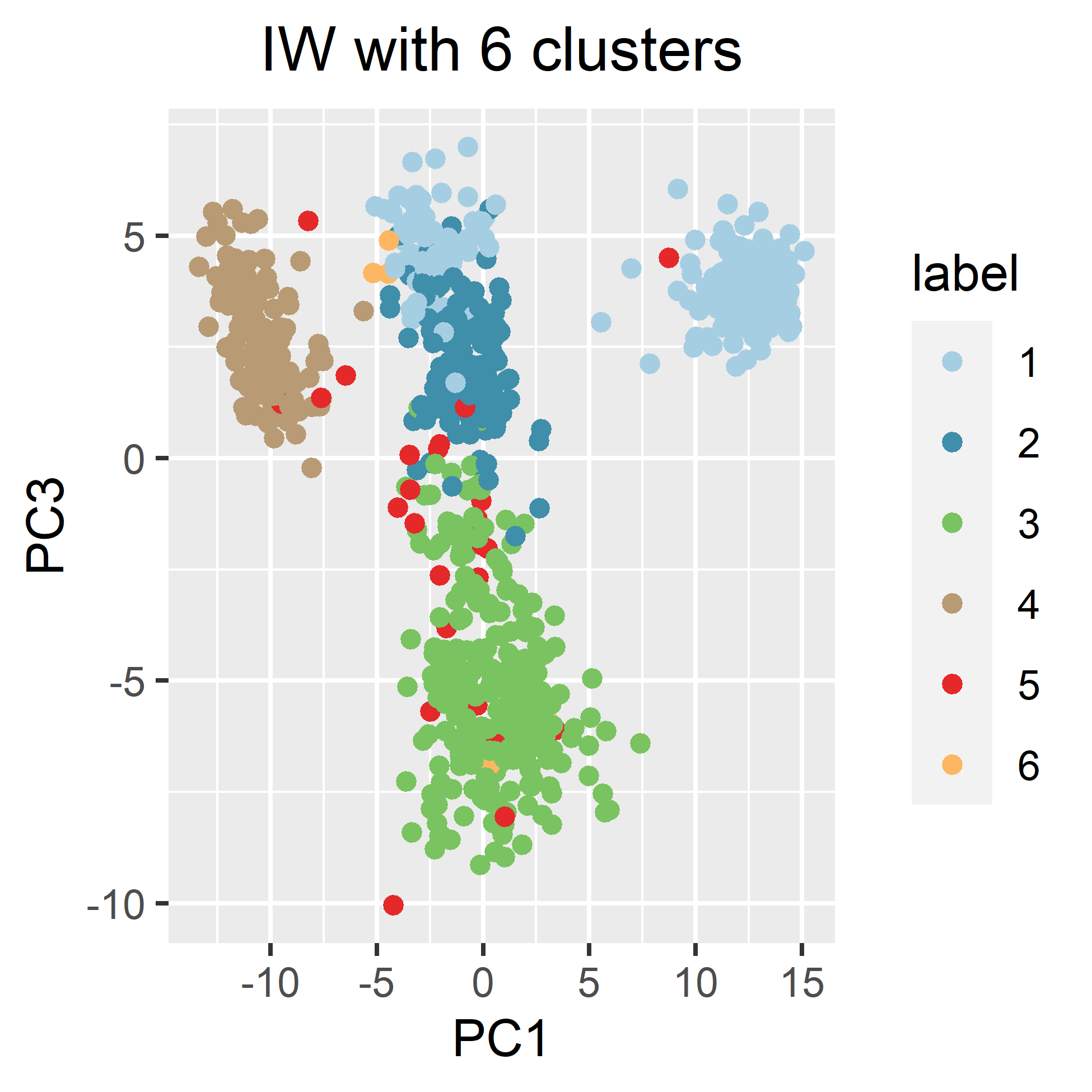} \\
	\includegraphics[width=5cm, height=5cm]{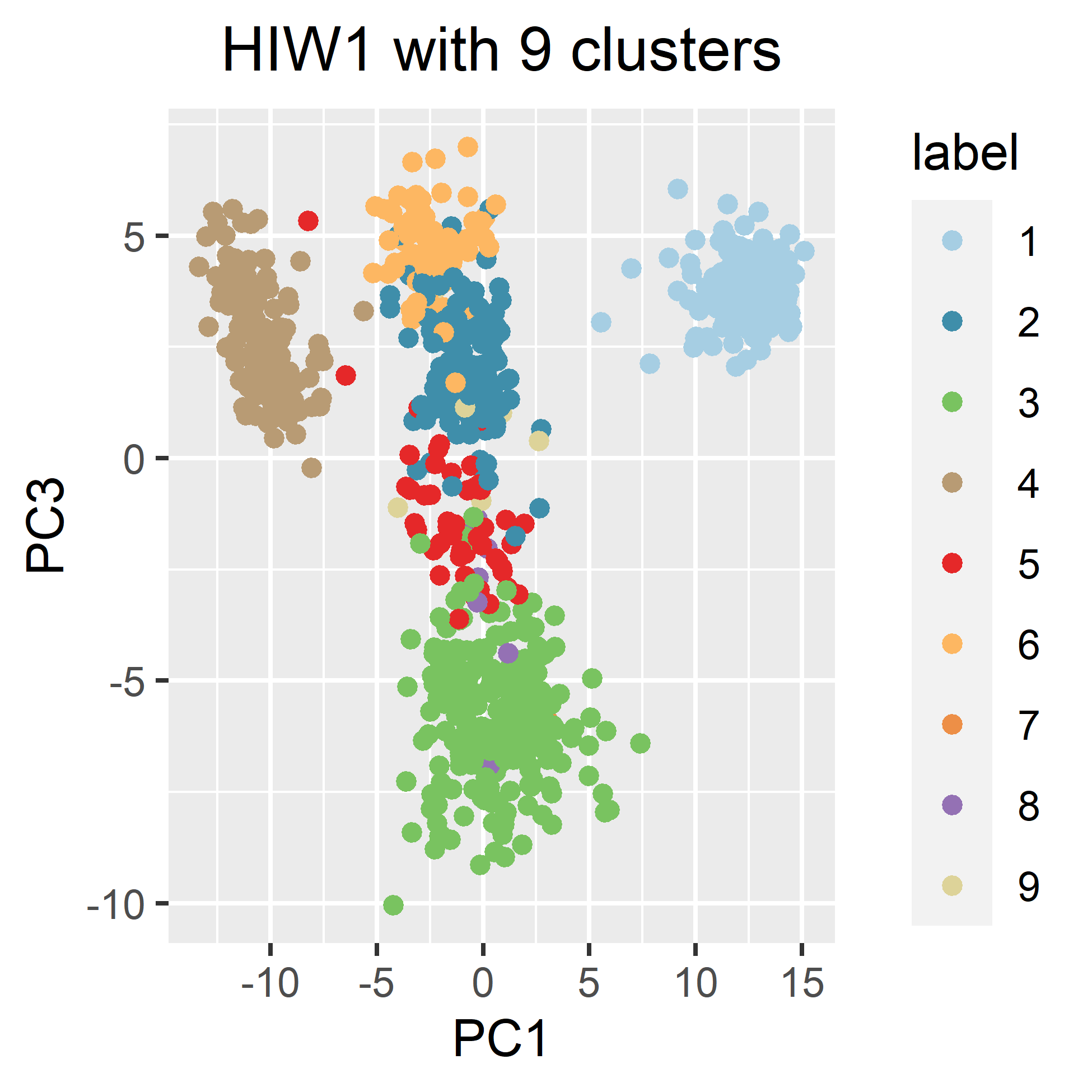} 
	\includegraphics[width=5cm, height=5cm]{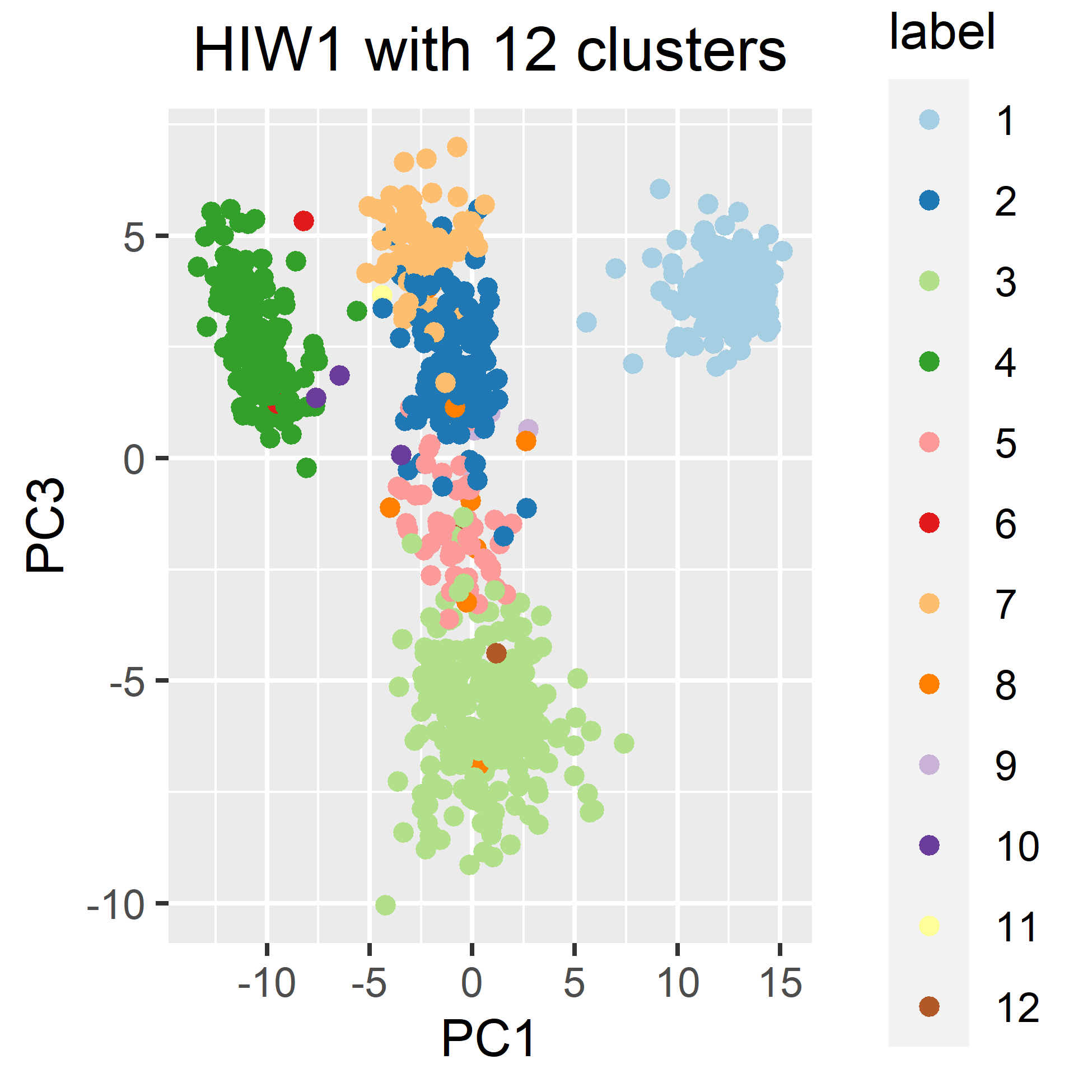} \\
	\includegraphics[width=5cm, height=5cm]{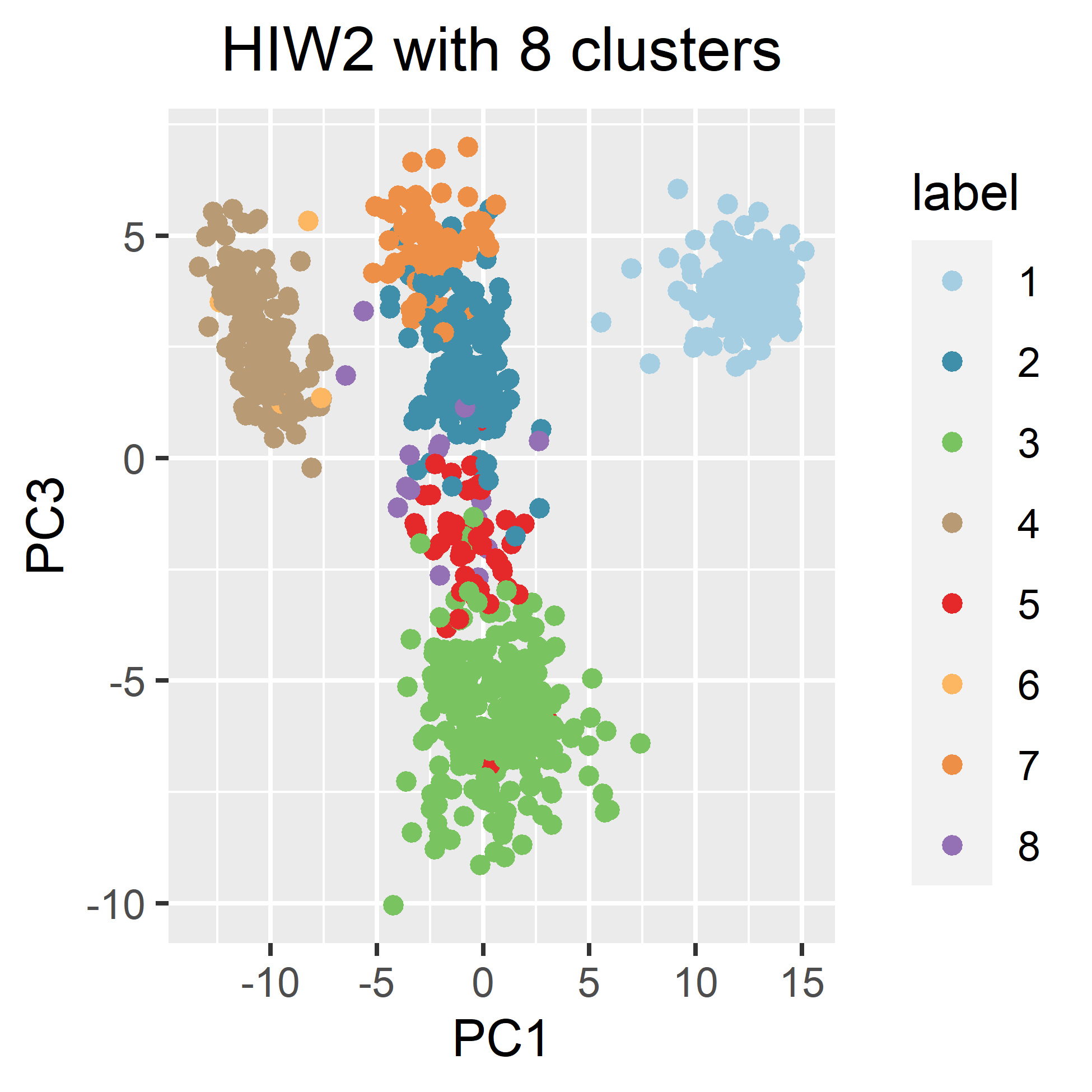} 
	\includegraphics[width=5cm, height=5cm]{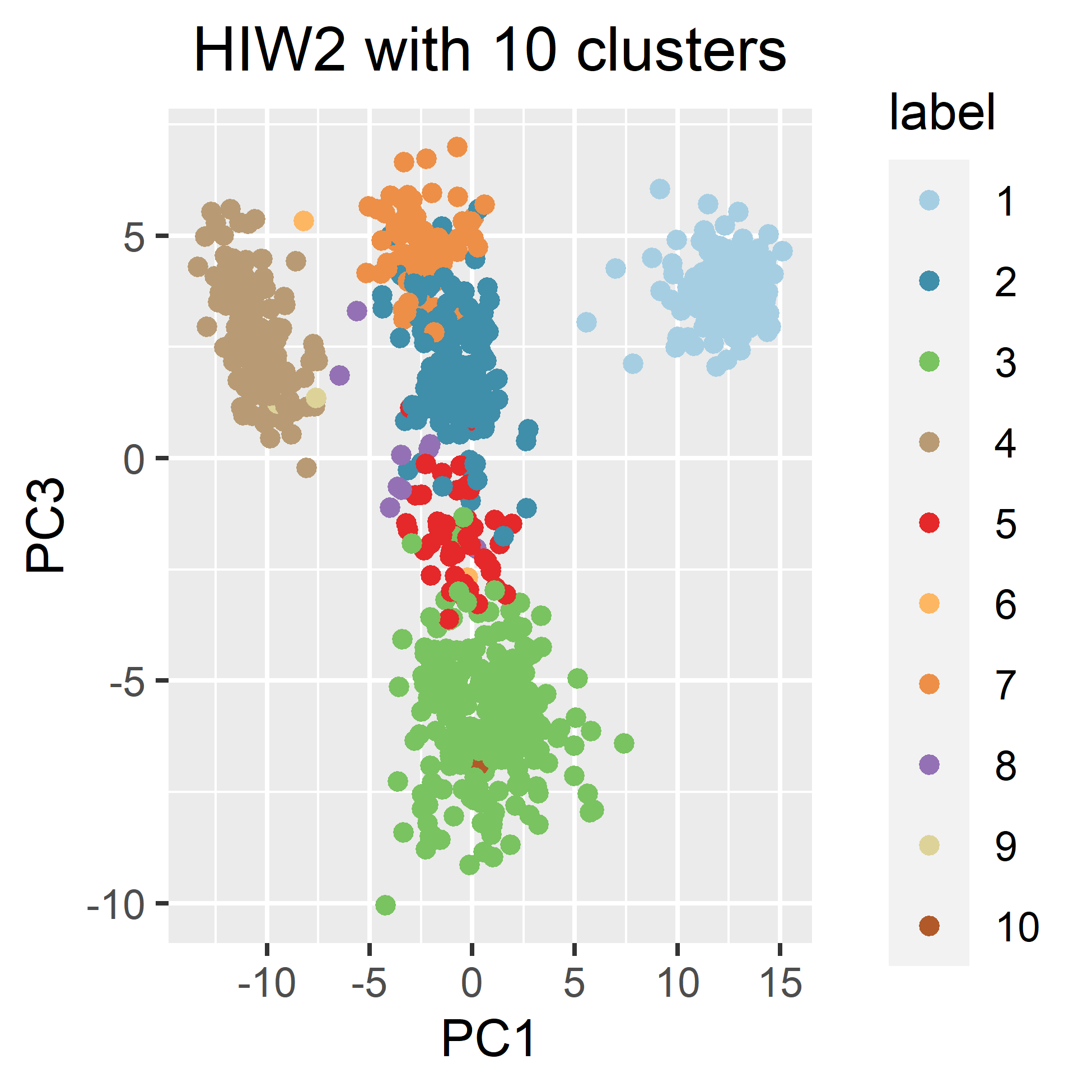} \\
	\begin{minipage}{0.8\textwidth}
		\caption{One representative clustering for a certain number of clusters for priors IW, HIW1 and HIW2.  }
		\label{fig:8.4}
	\end{minipage}
\end{figure}

\begin{figure}[h!]
	\centering
	\includegraphics[width=5cm, height=5cm]{pic_main/real-true-2}  
	\includegraphics[width=5cm, height=5cm]{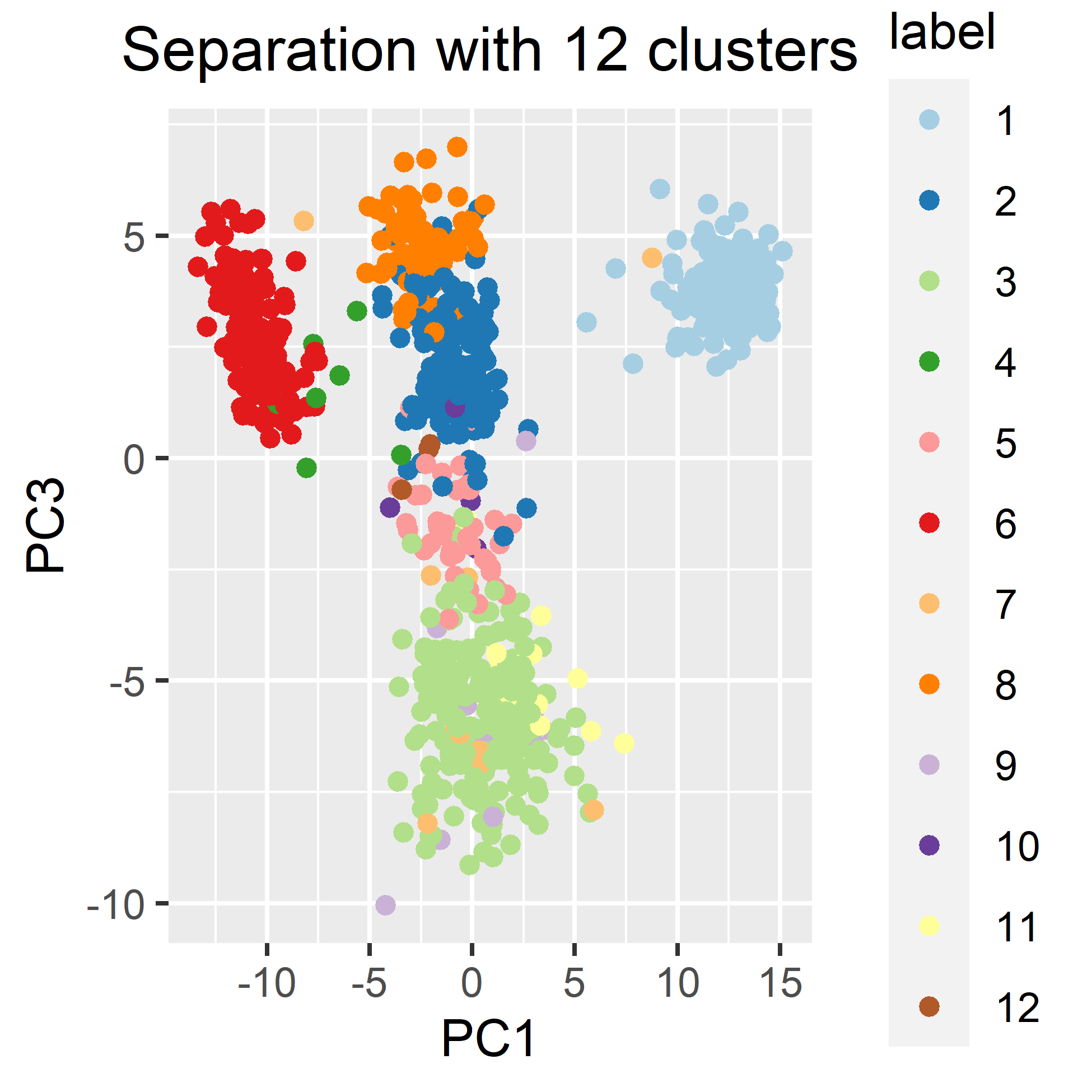} \\
	\includegraphics[width=5cm, height=5cm]{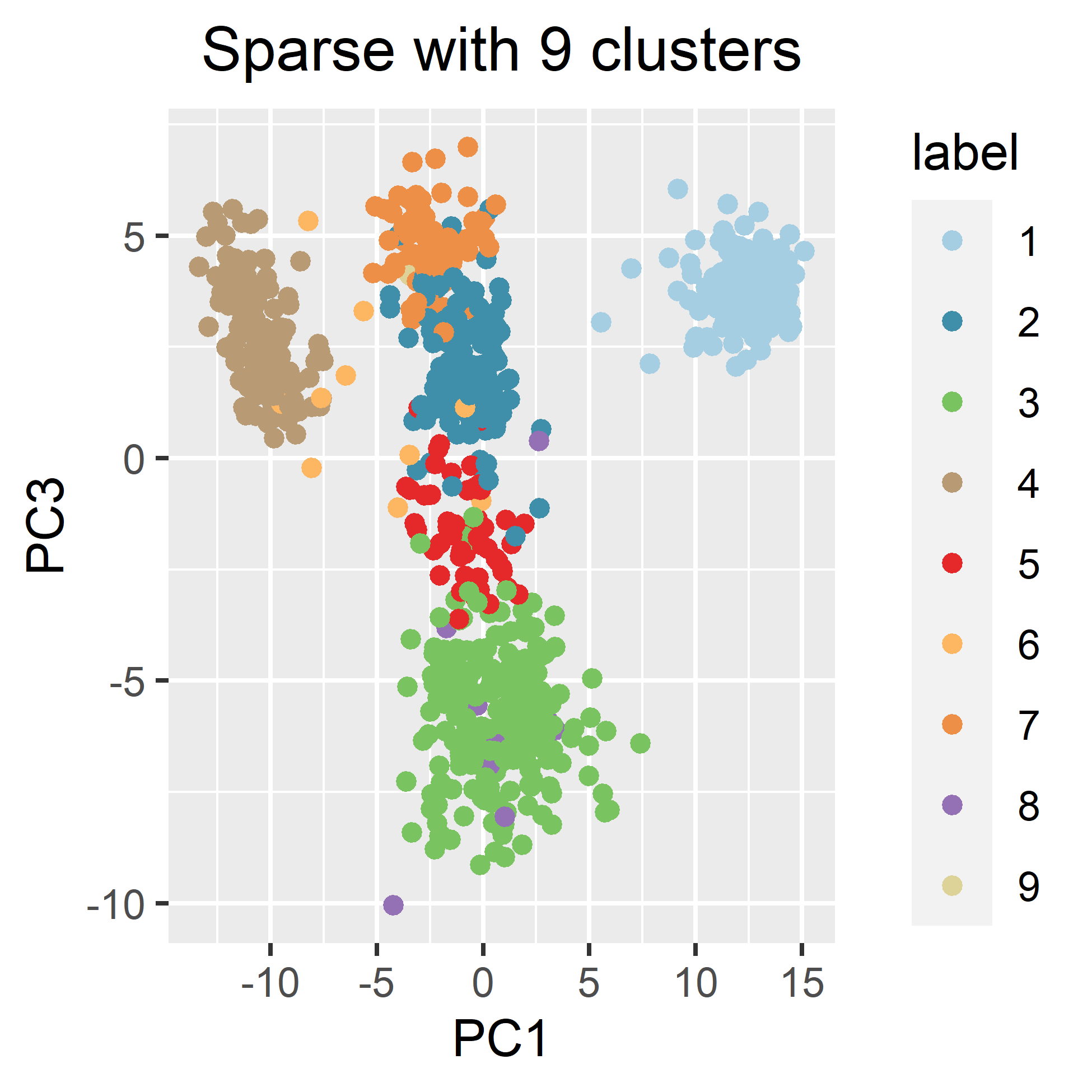} 
	\includegraphics[width=5cm, height=5cm]{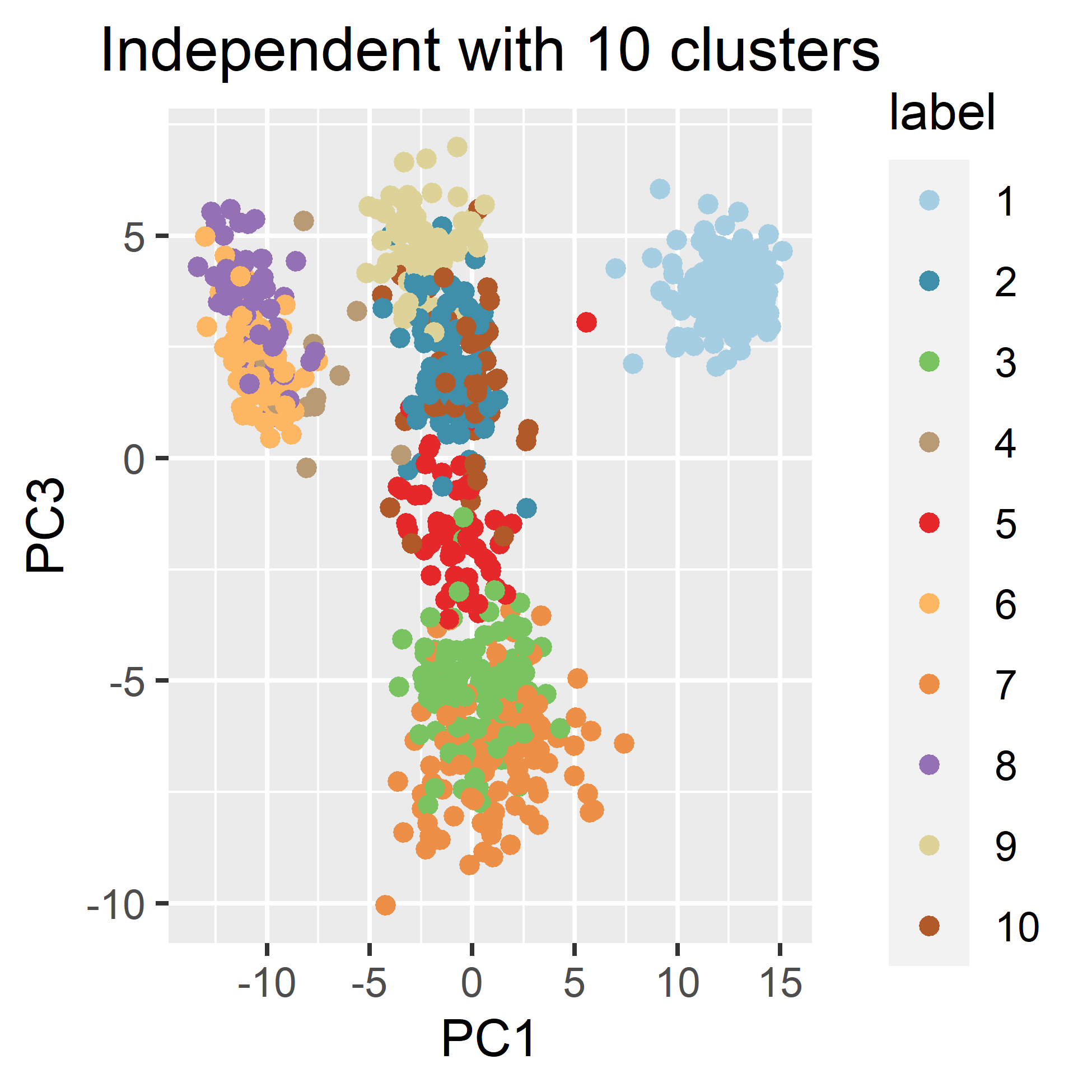} \\
	\begin{minipage}{0.8\textwidth}
		\caption{One representative clustering for a certain number of clusters for the separation,  sparse and independent priors.  }
		\label{fig:8.5}
	\end{minipage}
\end{figure}

\begin{figure}[h!]
	\centering
	\includegraphics[width=5cm, height=5cm]{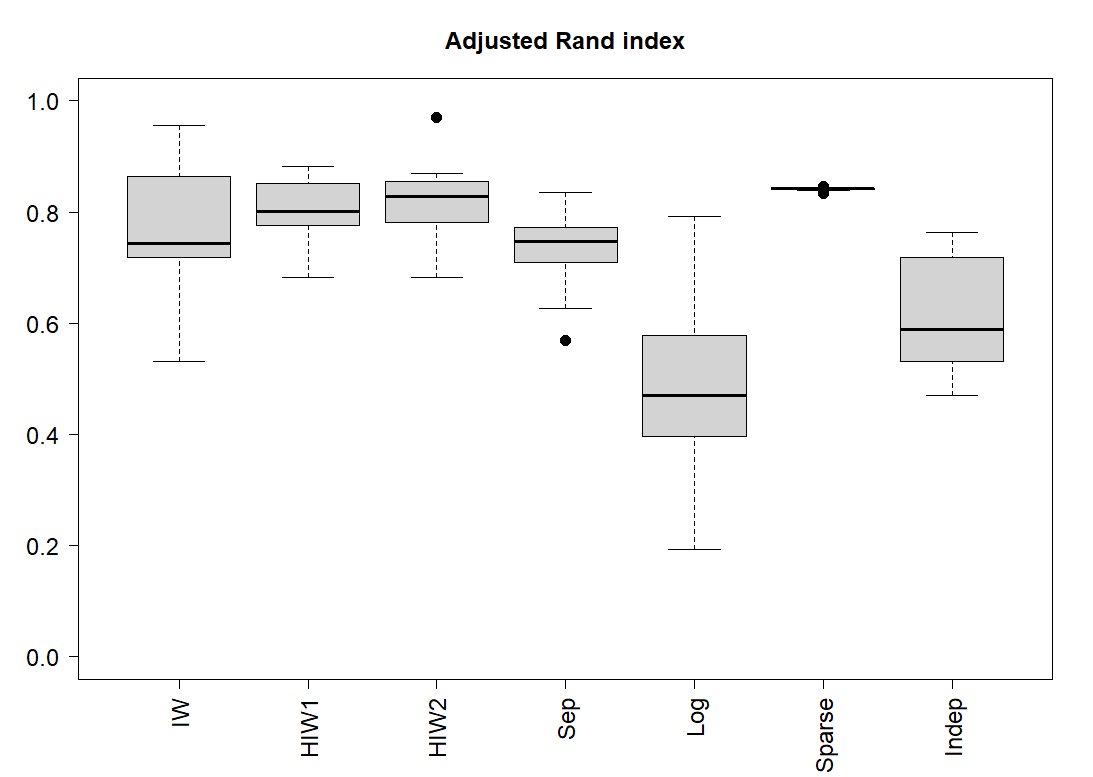} 
	\includegraphics[width=5cm, height=5cm]{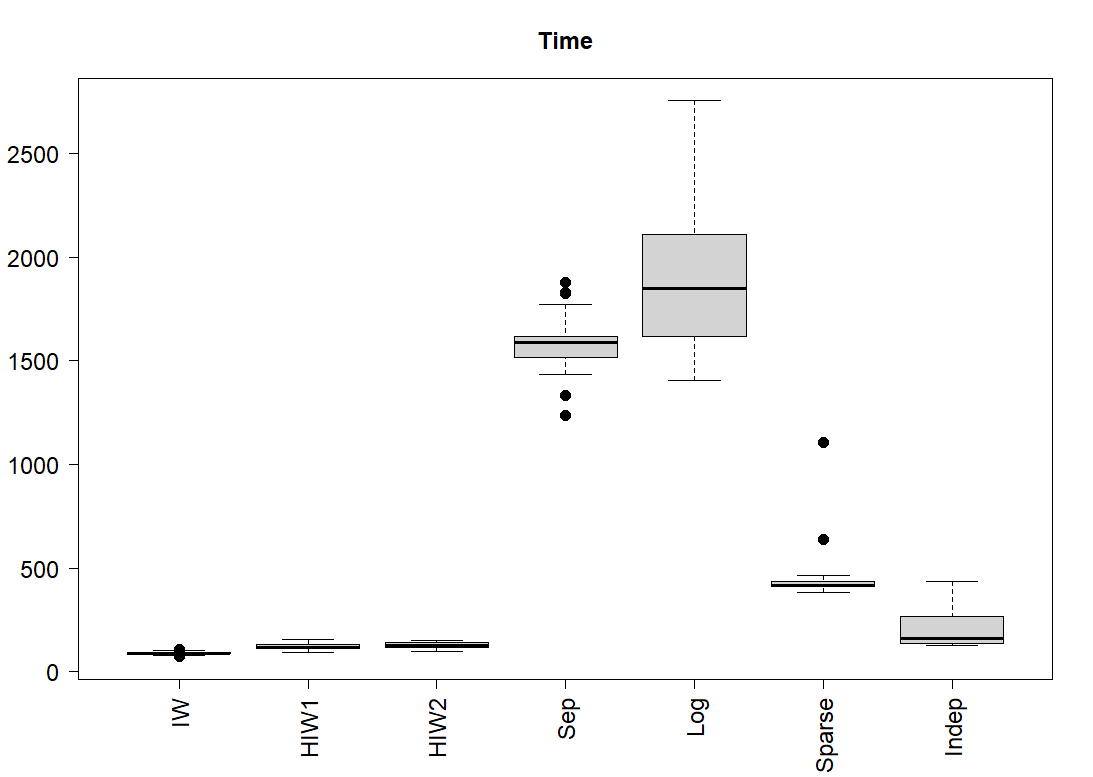} 
	\begin{minipage}{0.8\textwidth}
		\caption{Boxplots of adjusted Rand indices for different priors (left) and computational times after running 30 seeds (right) in seconds. }
		\label{fig:8.6}
	\end{minipage}
\end{figure}

Figures \ref{fig:8.4} and \ref{fig:8.5} demonstrate that the clustering of the subjects may be notably different from the benchmark, even when the number of clusters identified is the same as the target number. On the other hand, the clustering of the subjects may be close to the benchmark when superfluous clusters are small. 
The log prior frequently recognizes a number of superfluous clusters that is larger compared to other priors. It also combines target clusters. Representative clusters for the log prior are not shown, as it is the worst performer in terms of adjusted Rand index. Figure \ref{fig:8.6} (left) displays boxplots of the adjusted Rand indices observed for each prior. Table \ref{tab:8.2} provides means and standard deviations of the adjusted Rand indices. 

The independent prior always recognizes more clusters, as in the simulation studies. True clusters are split, not only for observations at the boundaries of the benchmark clusters, as shown in Figure \ref{fig:8.5}. In terms of Rand index performance, it is only better than the log prior. 

The separation prior generates superfluous clusters too, importantly some not small. Its Rand index performance is not satisfactory compared to other priors.  Figure $\ref{fig:8.5}$ shows an example of $12$ final clusters from the separation prior. 

The performance of the IW is also not satisfactory, as target clusters are often merged.
With the HIW1 and HIW2 priors this problem alleviates, and their performance in terms of Rand index is good, only bettered by the sparse prior. However, small clusters are created at the boundaries of the benchmark groups, and representative partitions are unstable for different initializations in terms of the final number of identified clusters. Figure \ref{fig:8.4} shows representative clustering examples of the $9$ and $12$ final clusters resulted from the HIW1. The partitions with the $9$ and $12$ final clusters are very similar. The difference lies in the small clusters created at the boundaries. Figure $\ref{fig:8.4}$ displays clustering examples of the $8$ and $10$ final clusters resulted from the HIW2. 

The number of clusters identified by the sparse prior is quite stable. The model frequently identifies $9$ groups with no target clusters combined. Figure \ref{fig:8.5} shows one representative clustering of $9$ final clusters. The superfluous clusters mostly exist at the boundaries of target clusters. The sparse prior outperforms all others, with the highest median adjusted Rand index and little variability.

\begin{table}[H] 
	\centering
	\caption{The mean and standard deviation of the Rand index of different priors}
	\begin{tabular}{  m{3.0cm} m{3.0cm} m{3.3cm} } 
		\hline
		Prior type & mean of the adjusted Rand index & standard deviation of the adjusted Rand index  \\ \hline
		IW & 0.771   & 0.098\\ 
		HIW1 &  0.805  & 0.056\\ 
		HIW2 &  0.816 & 0.059  \\ 
		Separation prior &  0.740  & 0.062  \\ 
        Log prior &  0.469   &   0.146 \\ 
        Sparse prior & 0.841 & 0.002  \\ 
		Independent prior & 0.609  & 0.097  \\ \hline
	\end{tabular}
	\label{tab:8.2}
\end{table}


Figure \ref{fig:8.6} (right) shows boxplots of run-times from the $30$ initializations for different priors. Similar to the simulation studies, the log and separation priors are much slower than the rest. The sparse prior is not as quick as the IW, HIW1, HIW2 and independent priors, however, the disparity is relatively small. Information on the posterior distributions of $\alpha$ is given in the Supplemental material, Section S11.


\section{Discussion}

This manuscript provided a thorough overview and comparative investigation into possible priors for covariance matrices within the DPMM context, with detailed guidelines for their implementation, as done within the R package PReMiuM. In addition to investigating the effectiveness of different prior specifications, our aim was to make it easier for investigators to avoid, if desired, some of the standard approaches to fitting mixture models to high-dimensional data such as assuming diagonal covariance matrices; see, for instance, 
\cite{banfield1993} or \cite{galimberti2013}. 

The sparse prior performed well in the different analyses, with the resulting partitions close to the target and stable. Our understanding is that to attain better clustering results, the prior of the within-cluster covariance matrices should be rather informative. Otherwise, it is difficult for both  model and sampler to identify target clusters. Unlike classification problems where the latent labels are observed for part of the dataset, in unsupervised clustering there is no information on the underlying groups. If the sampled  variances for the within-cluster covariance matrices are larger than the truth, it is likely that target clusters are combined. The tuning of $M_{0}$ can be viewed as the process of setting a suitable prior for $\Sigma_{c}$ using the structure of the observations. Note that for the HIW1, HIW2, separation and log priors, we also specify them to be rather informative in the sense that the prior mean variances in $\Sigma_{c}$ are relatively small. This is an acceptable practice within the Bayesian paradigm, where the structure of the data can be used to inform the prior, rather than the individual observations themselves. In \cite{richardson1997bayesian} for instance, weak informative priors are proposed, and it is pointed out that the possible spread of the clusters should be reflected in the prior. 
Sensitivity analyses we performed have shown that clustering results are not detached from the adopted hyperparameter values. This is a mixture modelling characteristic also mentioned by \cite{fruhwirth2018handbook}. In general, it is possible to improve clustering results by fine-tuning the specified hyperparameters to the specific application and dataset under consideration. Further research can be done towards a more systematic process for tuning $M_{0}$, or the hyperparameters for the Separation, HIW1 and HIW2 priors, beyond the guidelines provided in this manuscript. 

We focus on datasets where the number of variables $J$ is beyond a handful (say 2 to 5), with unconstrained within-cluster covariance matrices $\Sigma_{c}$. In our manuscript, $J$ is not of very large size, say $100$ or $1000$. Letting $K$ denote the number of clusters or non-empty components in the mixture likelihood, \cite{chandra2020escaping} proved that as $J$ approaches infinity with a fixed number of subjects, the probability of $K=1$ (all subjects belong to the same group) or $K=n$ (each subject forms its own cluster) tends to one a posteriori. For datasets with very large $J$ or even with $J \gg n$, \cite{chandra2020escaping} advocated to conduct factor analysis on the observed data. Clustering can then be applied to the derived factors. Therefore, results in this manuscript can also be useful for such high-dimensional problems, through the clustering of factors.

Although this manuscript focuses on the conditional samplers of the DPMM, it is also possible to incorporate the studied priors in marginal samplers with similar computational steps given in the Supplemental material, for example, Neal's Algorithm $8$ \citep{neal2000markov}. It may also be of interest to examine the possibility of including the split and merge step \citep{jain2007splitting} in the marginal samplers for some of the priors mentioned in Section 2.  

It is likely the DPMM creates extra small clusters that contain only a few subjects; see \cite{miller2014inconsistency}. In light of this, \cite{miller2017mixture} advocated to use finite mixture models with a prior on the number of components, referred to as the mixture of finite mixtures (MFM), and designed a marginal sampler for the MFM based on the Algorithm 3 of Neal \cite{neal2000markov}. Later, \cite{fruhwirth2020generalized} extended results by \cite{miller2017mixture} and proposed the telescoping sampler for the MFM, which shares similarities with the conditional samplers for the DPMM. It should be possible to incorporate the sparse prior into the MFM using the telescoping sampler with similar steps discussed in the Supplemental material. This is currently under investigation. 

\cite{malsiner2016model} proposed the sparse finite mixture model. The main idea is to specify a finite mixture model with a maximum number of components much larger than the expected number of groups in the data. Then, a sparse prior is placed on the mixture weight $\Psi=\{\psi_{1},....,\psi_{K}\}$ to empty superfluous components. \cite{malsiner2017identifying} stated that the sparse finite mixture model is more suitable for scenarios where the data contain a moderate number of groups and the group number does not increase when more subjects are observed, whereas the DPMM can be more appropriately applied in scenarios where the number of clusters in the data increases with sample size, such as the text mining context. However, a strong correspondence does exist between the sparse finite mixture model and the DPMM. (See \cite{fruhwirth2019here} for a more detailed discussion). We currently investigate the possibility of incorporating the priors for $\Sigma_{c}$ listed in Section 2 to the sparse finite mixture model and study how they can influence clustering outcomes. 

In the real data analysis, we compared the clustering results with the clustering implied by the cancer labels. As we selected genes with expression highly associated with the tumor labels, it was reasonable to expect that the clustering results should closely match the benchmark clustering according to cancer type. To further encourage the clustering of the subjects to be similar to the tumor labels, we could have utilised a profile regression model, as in \cite{molitor2010bayesian}, also implemented within the PReMiuM R package, with tumor labels modelled as an outcome.  However, to achieve a clean comparison between the priors, we decided against directly influencing the clustering of the subjects by including the tumor labels in the modelling. Finally, to identify genes that are important for the clustering, from a larger pool of genes than the 20 we selected, we could have utilised the variable selection approach implemented in PReMiuM \cite{papathomas2012exploring}. However, variable selection is not the focus of this manuscript. 

\bibliographystyle{ba}
\bibliography{thesis}

\begin{acks}[Acknowledgments]
The first author would like to acknowledge the support of the School of Mathematics and Statistics, as well as CREEM, at the University of St Andrews, and the University of St Andrews St Leonard's 7th Century Scholarship.
\end{acks}

\newpage

\begin{figure}[h!]
	\centering
	\includegraphics[width=5cm, height=5cm]{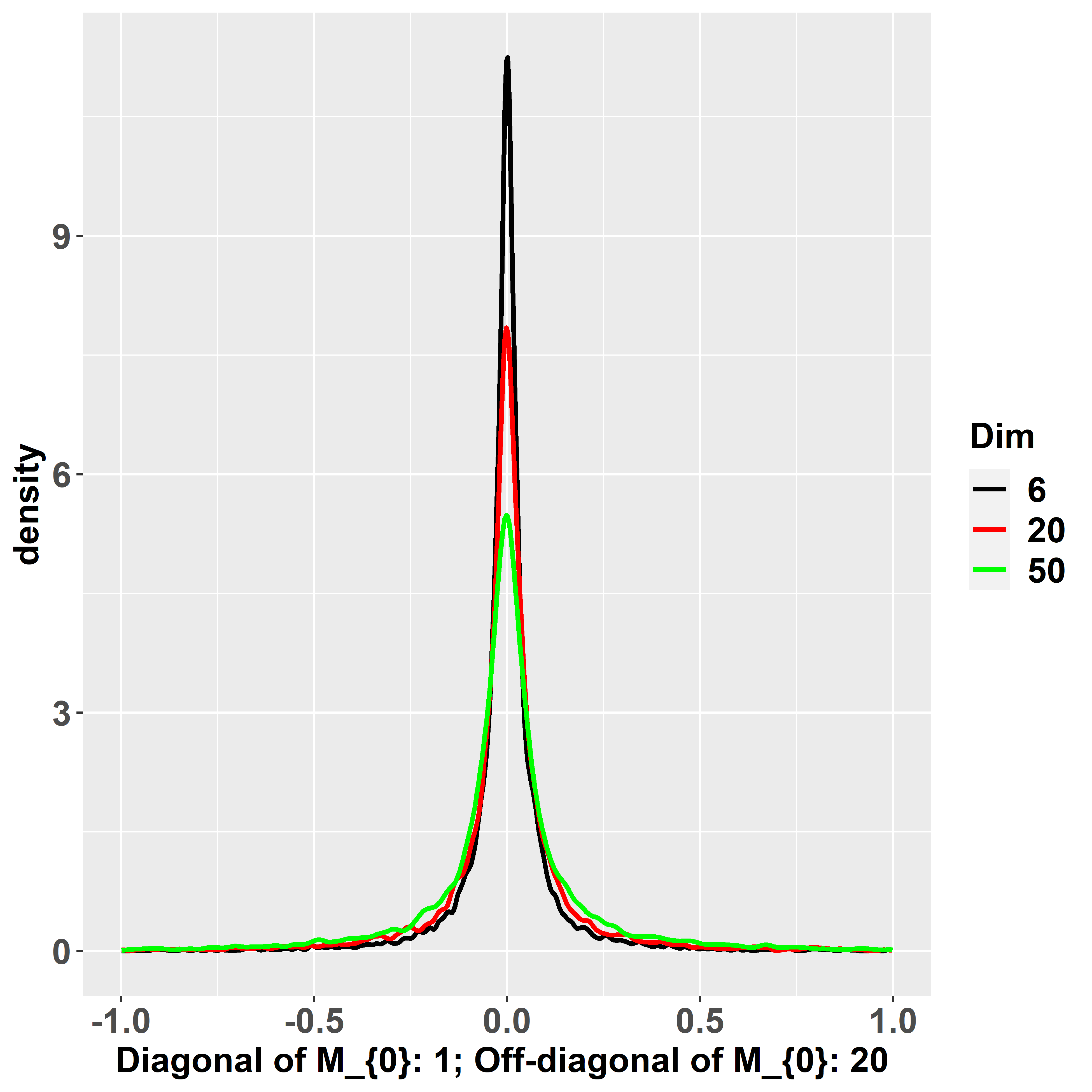} 
	\includegraphics[width=5cm, height=5cm]{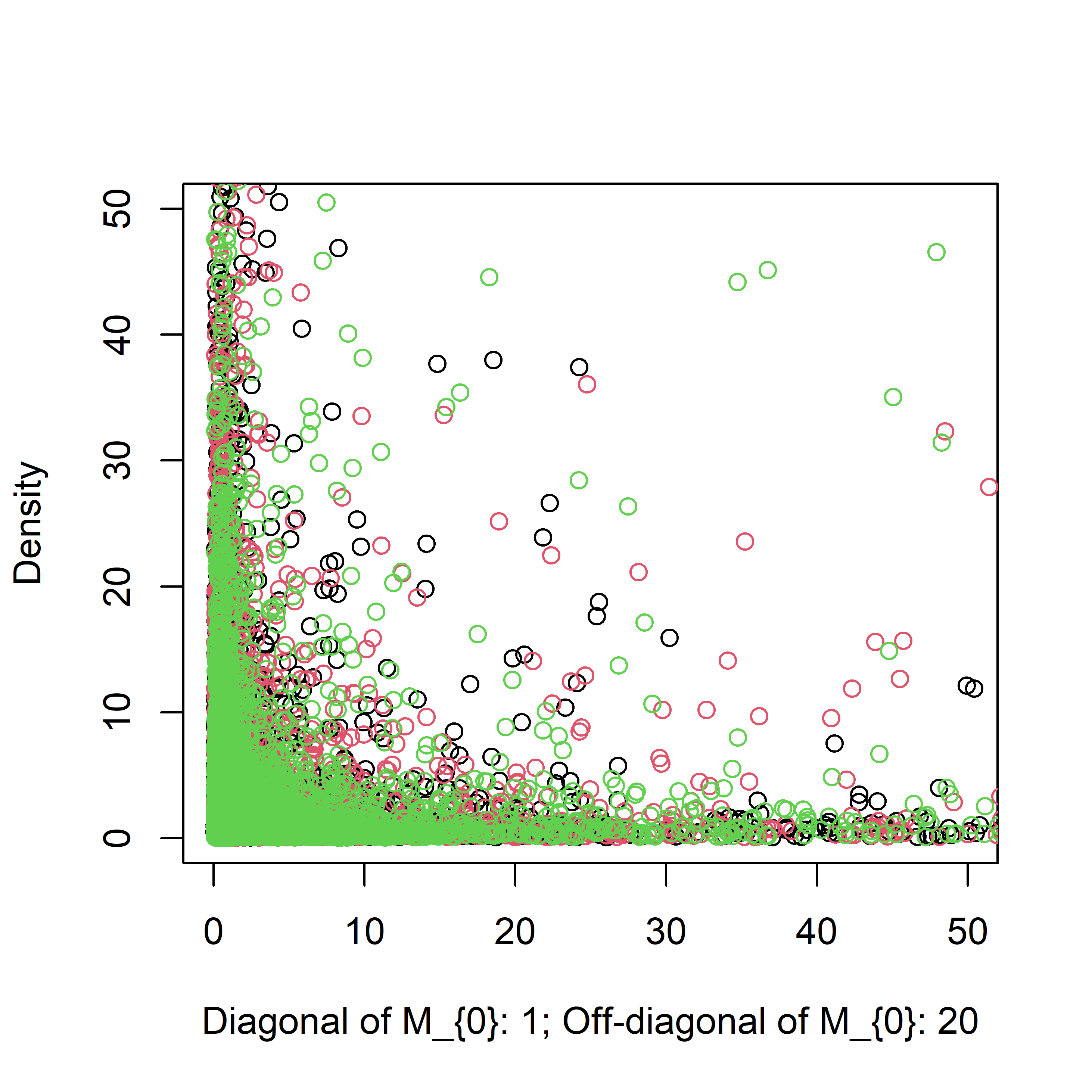} \\ 
	\includegraphics[width=5cm, height=5cm]{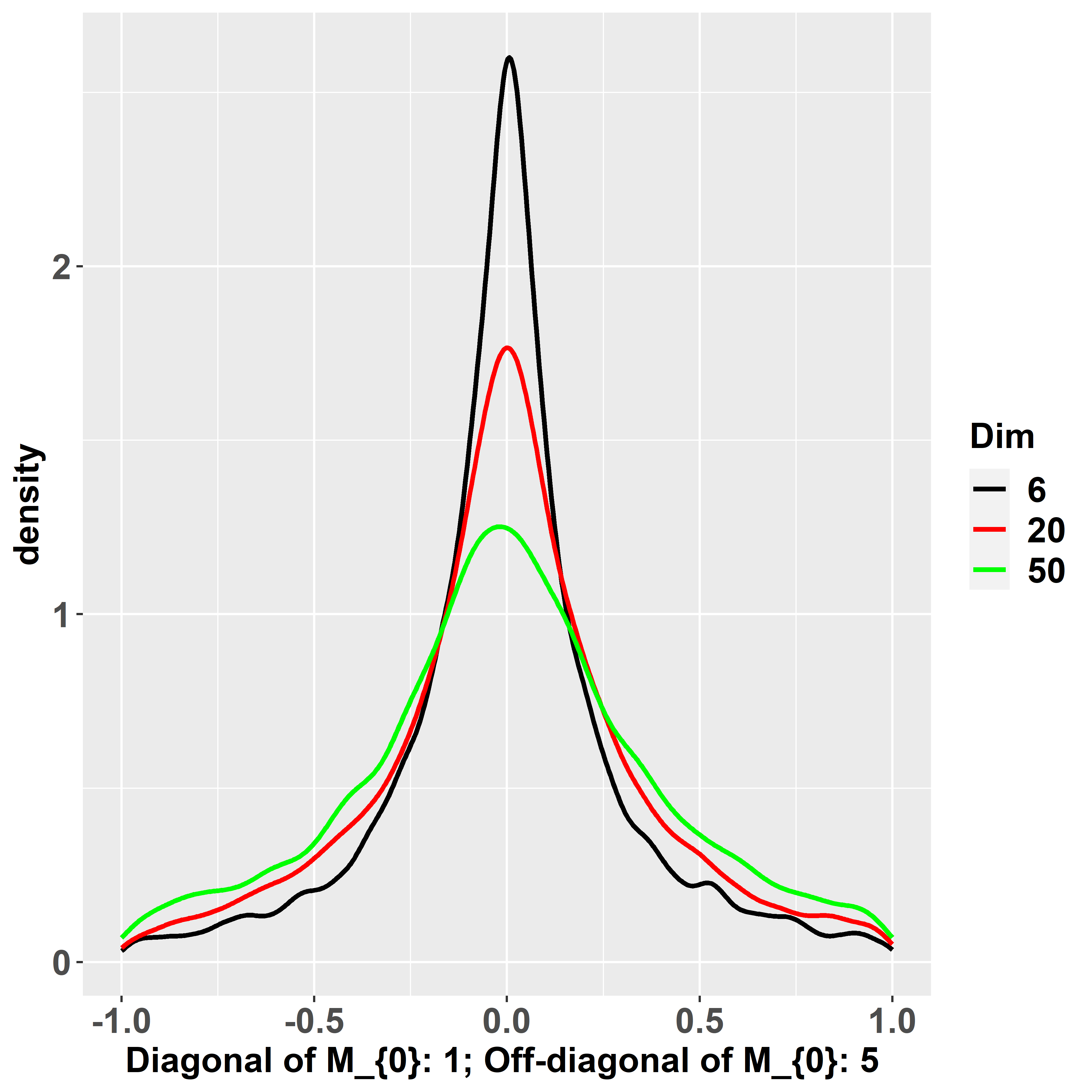} 
	\includegraphics[width=5cm, height=5cm]{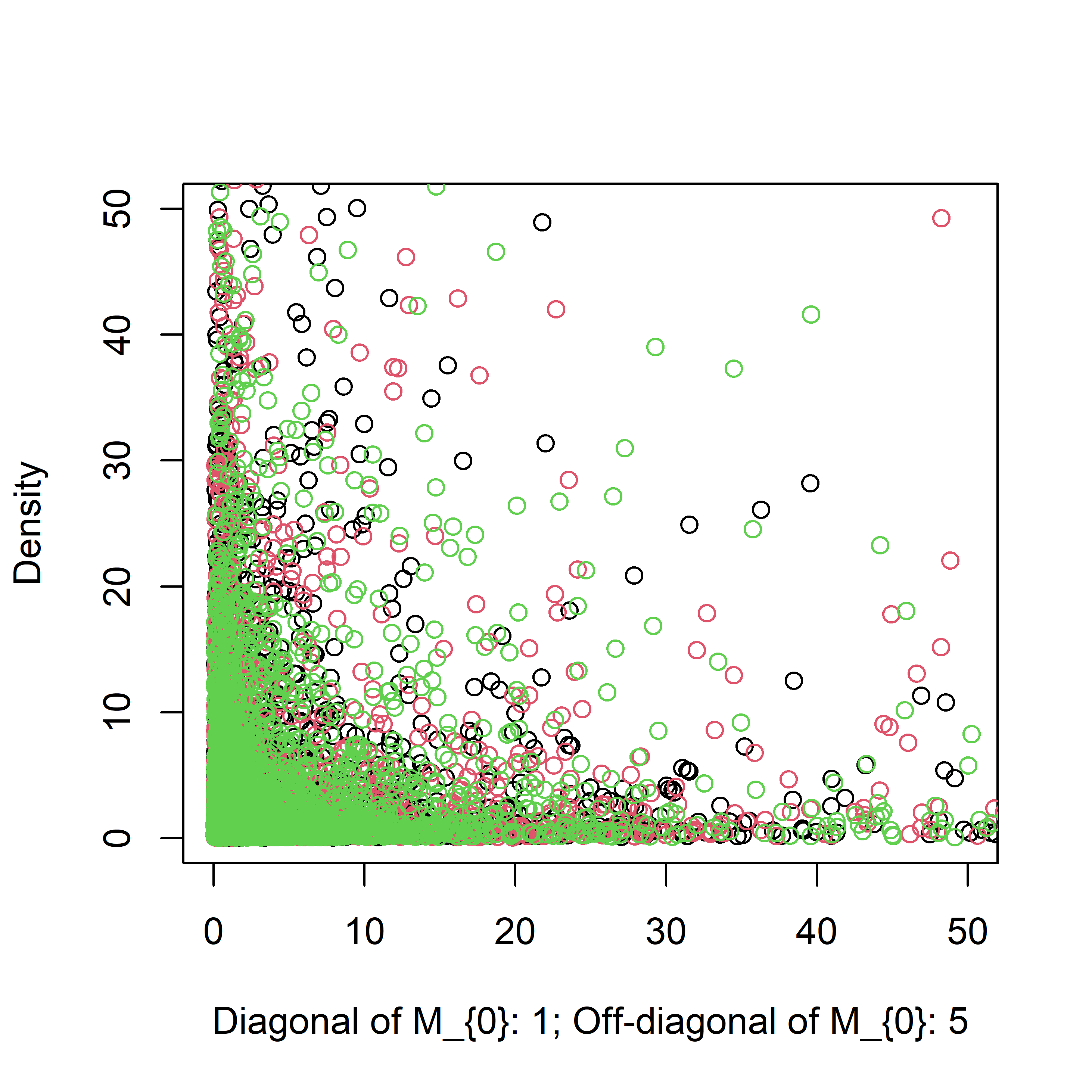} \\ 
	\includegraphics[width=5cm, height=5cm]{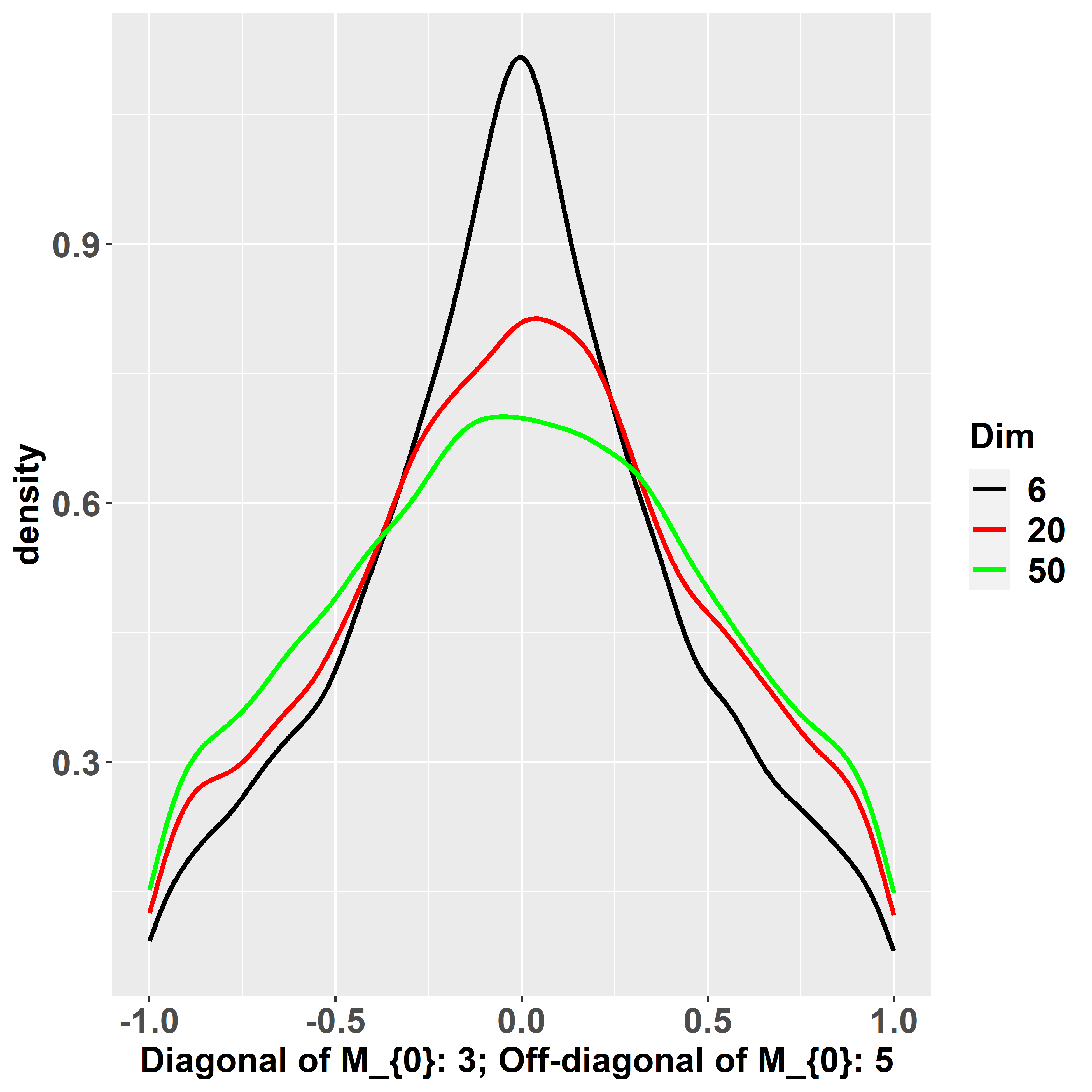} 
	\includegraphics[width=5cm, height=5cm]{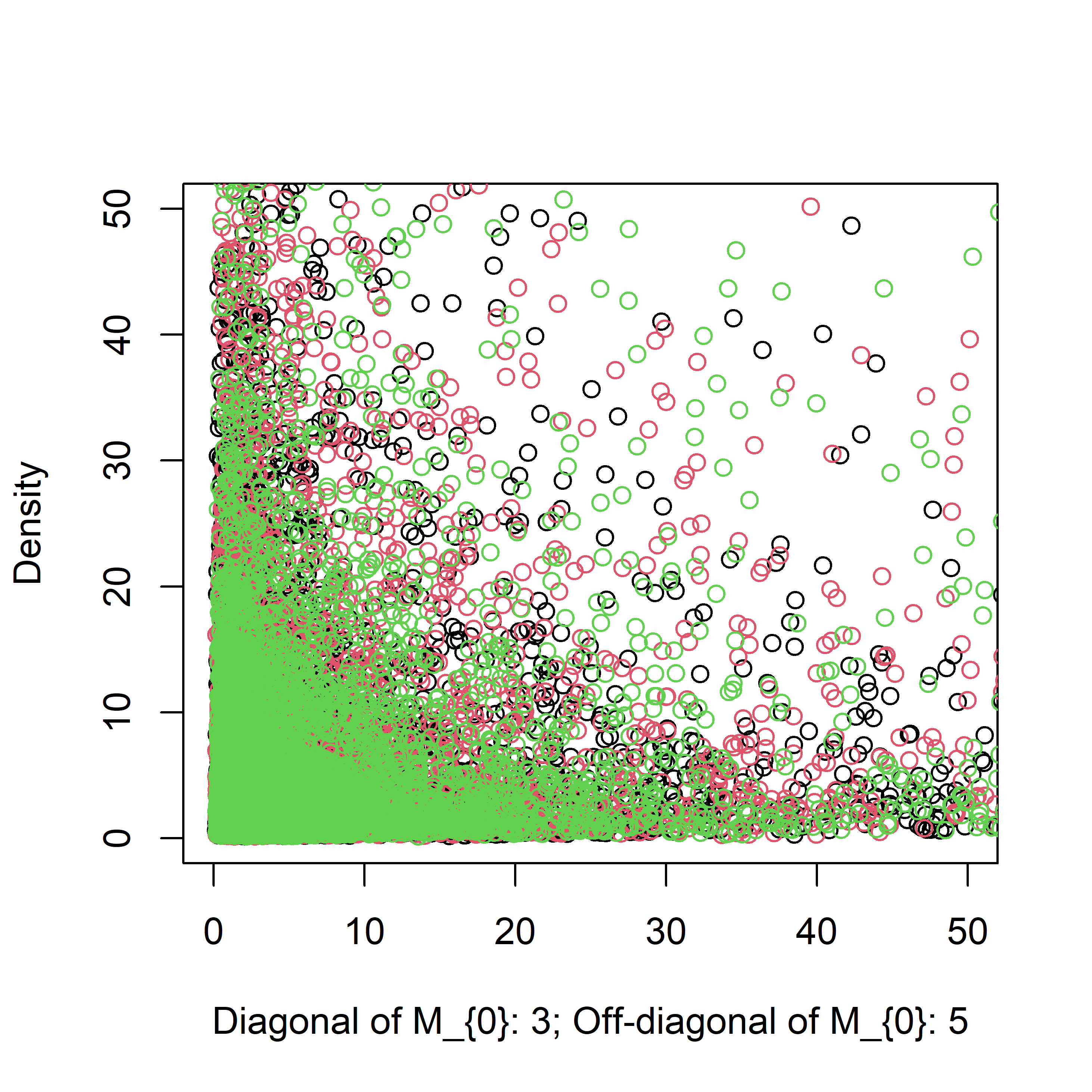} \\ 
	\vspace{0.5cm}
	\begin{minipage}{0.8\textwidth}
		\caption{Empirical prior distributions of correlations when $J=6$, $J=20$ or $J=50$ for different $M_{0}$ (left). Empirical bivariate density plots of variances for $J=6$, $J=20$ or $J=50$ for different $M_{0}$ (right). The black colour corresponds to $J=6$; the red to $J=20$; the green to $J=50$. Diagonal of $M_{0}$ refers to $m_{0,ii}$. Off-diagonal of $M_{0}$ refers to $m_{0,ij}$.}
		\label{fig:4}
	\end{minipage}
\end{figure} 

\end{document}